\documentclass{article}
   \usepackage{graphicx}
   \usepackage{amssymb}
   \usepackage{epstopdf}
   \usepackage{pdfsync}
   \usepackage{color}
   \usepackage[onecolumn]{apjgalley}

             \font\sevenrm=cmr7

          \font\sixrm=cmr6

\def\erg{\varepsilon}
\def\eperp{\vphantom{(}\erg_{\perp}}
\def\lambar{\lambda\llap {--}}
\def\rns{R_{\hbox{\sixrm NS}}}
\def\sigt{\sigma_{\hbox{\sixrm T}}}
\def\sigJL{\sigma_{\hbox{\sixrm JL}}}
\def\sigST{\sigma_{\hbox{\sixrm ST}}}
\def\Thetacol{\Theta_{\hbox{\sevenrm col}}}
\def\thetaBr{\theta_{\hbox{\sevenrm B}r}}
\def\gammadott{{\dot \gamma}_{\hbox{\sixrm T}}}
\def\gammadotres{{\dot \gamma}_{\hbox{\sixrm T,RES}}}
\def\gammadotJL{{\dot \gamma}_{e,\hbox{\sixrm JL}}}
\def\gammadotST{{\dot \gamma}_{e,\hbox{\sixrm ST}}}
\def\gammadotacc{\dot{\gamma}_{\hbox{\sevenrm acc}}}
\def\gammadotCR{\dot{\gamma}_{\hbox{\sixrm CR}}}
\def\lambdaCR{\lambda_{\hbox{\sixrm CR}}}
\def\lambdaacc{\lambda_{\hbox{\sevenrm acc}}}
\def\tauJL{{\tau}_{e,\hbox{\sixrm JL}}}
\def\tauST{{\tau}_{e,\hbox{\sixrm ST}}}
\def\calFST{{\cal F}_{\Delta}}
\def\calGST{{\cal G}_{\Delta}}
\def\calFtauST{{\cal F}_{\tau\Delta}}
\def\calGtauST{{\cal G}_{\tau\Delta}}
\def\calRJL{{\cal R}_{\hbox{\sixrm JL}}}
\def\calRST{{\cal R}_{\hbox{\sixrm ST}}}
\def\tesc{t_{\hbox{\sevenrm esc}}}
\def\fsc{\alpha_{\hbox{\sevenrm f}}}                                
\def\Gammacyc{\Gamma_{\hbox{\sevenrm cyc}}}   
\def\Gammave{\overline{\Gamma}_{1\to 0}}   
\def\dover#1#2{\hbox{${{\displaystyle#1 \vphantom{(} }\over{
   \displaystyle #2 \vphantom{(} }}$}}

{\catcode`\@=11                                                  
\gdef\SchlangeUnter#1#2{\lower2pt\vbox{\baselineskip 0pt\lineskip0pt    
\ialign{$\m@th#1\hfil##\hfil$\crcr#2\crcr\sim\crcr}}}}           
\def\gtrsim{\mathrel{\mathpalette\SchlangeUnter>}}               
\def\lesssim{\mathrel{\mathpalette\SchlangeUnter<}}
\def\rmp{Rev. Mod. Phys.}
\def\asr{Adv. Space Res.}
\def\teq#1{$\, #1\,$}                         
\begin{document} 

\newcommand{\vol}[2]{$\,$\bf #1\rm , #2.}                 
\newcommand{\figureoutpdf}[5]{\centerline{}
   \centerline{\hspace{#3in} \includegraphics[width=#2truein]{#1}}
   \vspace{#4truein} \figcaption{#5} \centerline{} }
\newcommand{\twofigureoutpdf}[3]{\centerline{}
   \centerline{\includegraphics[width=3.8truein]{#1}
        \hspace{0.0truein} \includegraphics[width=3.8truein]{#2}}
        \vspace{-0.2truein}
    \figcaption{#3} }    
\newcommand{\twofigureoutpdfadj}[3]{\centerline{}
   \centerline{\includegraphics[height=3.0truein]{#1}
        \hspace{+0.2truein} \includegraphics[height=2.9truein]{#2}}
        \vspace{-0.1truein}
    \figcaption{#3} }    
\newcommand{\fourfigureoutpdf}[5]{\centerline{}
   \centerline{\includegraphics[width=3.7truein]{#1}
        \hspace{0.0truein} \includegraphics[width=3.7truein]{#2}}
        \vspace{-0.1truein}
   \centerline{\includegraphics[width=3.7truein]{#3}
        \hspace{0.0truein} \includegraphics[width=3.7truein]{#4}}
        \vspace{-0.1truein}
    \figcaption{#5} }    

\title{COOLING RATES FOR RELATIVISTIC ELECTRONS
         UNDERGOING \\ COMPTON SCATTERING
         IN STRONG MAGNETIC FIELDS}  

   \author{Matthew G. Baring, Zorawar Wadiasingh}
   \affil{Department of Physics and Astronomy, MS 108,\\
      Rice University, Houston, TX 77251, U.S.A.\\
      {\it baring@rice.edu, zw1@rice.edu}}

    \and

   \author{Peter L. Gonthier}
   \affil{Hope College, Department of Physics,\\
          27 Graves Place, Holland, MI 49423, U.S.A. \\
      \it gonthier@hope.edu\rm}
\slugcomment{Accepted for publication in \it The Astrophysical Journal\rm .}

\begin{abstract} 
For inner magnetospheric models of hard X-ray and gamma-ray emission 
in high-field pulsars and magnetars, resonant Compton upscattering is
anticipated to be the most efficient process for generating continuum
radiation.  This is due in part to the proximity of a hot soft photon
bath from the stellar surface to putative radiation dissipation regions
in the inner magnetosphere.  Moreover, because the scattering process
becomes resonant at the cyclotron frequency, the effective cross section
exceeds the classical Thomson value by over two orders of magnitude,
thereby enhancing the efficiency of continuum production and the cooling
of relativistic electrons.  This paper presents computations of the
electron cooling rates for this process, which are needed for resonant
Compton models of non-thermal radiation from such highly-magnetized
pulsars.  The computed rates extend previous calculations of magnetic
Thomson cooling to the domain of relativistic quantum effects, sampled
near and above the quantum critical magnetic field of 44.13
TeraGauss.  This is the first exposition of fully relativistic, quantum
magnetic Compton cooling rates for electrons, and it employs both the
traditional Johnson \& Lippman cross section, and a newer Sokolov \&
Ternov (ST) formulation of Compton scattering in strong magnetic fields.
 Such ST formalism is formally correct for treating spin-dependent
effects that are important in the cyclotron resonance, and has not been
addressed before in the context of cooling by Compton scattering. The
QED effects are observed to profoundly lower the rates below
extrapolations of the familiar magnetic Thomson results, as expected,
when recoil and Klein-Nishina reductions become important.
\end{abstract}


\section{INTRODUCTION}
 \label{sec:intro}

Soft Gamma-ray Repeaters (SGRs) and Anomalous X-ray Pulsars (AXPs) are
astrophysical sources, typically associated with supernova remnants
(e.g. AXP 1841-045 associated with the remnant Kes 73: see Vasisht \&
Gotthelf 1997), that are thought to be magnetars, ultra-magnetized
isolated neutron stars.  Observational evidence for supernova
associations of magnetar candidates are discussed in Gaensler (2004) and
Vink (2008). The magnetar model was first proposed by Duncan \& Thompson
(1992) for SGRs and then later extended to AXPs (Thompson \& Duncan
1996). They are defined by their high spin-down rates and long periods
(2-12s) in their X-ray and soft gamma-ray pulsations, suggesting
characteristically short ages $< 10^5$ years and high inferred surface
polar fields, $B_0 \sim 10^{13}-10^{15}$ Gauss (e.g. Vasisht \& Gotthelf
1997). Quiescent X-ray emission has been observed in both AXPs and SGRs
(Tiengo et al 2002) and although counterparts have been observed in the
radio or optical/infrared (Camilo et al. 2006), the bulk of observed
bolometric luminosity is in the X-ray band. In both SGRs and AXPs, high
persistent X-ray luminosity $\sim 10^{35}$ ergs/sec typically exceeds
the classical electromagnetic dipolar rotational energy loss rates by
two to three orders of magnitude, suggesting an energy source not of
rotational spin-down origin, i.e. contrasting the case for the multitude
of young radio and gamma-ray pulsars.

Long timescale flux variability is generally low for quiescent 2--10 keV
band emission in both SGRs and AXPs. However, bursting activity typical
of SGRs was reported by Kaspi et. al (2003) in AXPs.  Such outbursts in
both classes of sources reinforces the magnetar hypothesis, since such
transient events are predicted (e.g. Thompson \& Duncan 1996) as the
consequence of violent rearrangements of superstrong magnetic fields
thredding the neutron star crust. This soft X-ray emission  observed
over the past two decades from AXPs and SGRs has been typically fitted
by blackbody plus power-law components (e.g. Perna et al. 2001), leading
to thermal component temperature estimates of the order of $\sim$ 0.5--1
keV, while the non-thermal tail component is usually steep. More
recently, and surprisingly, pulsed hard X-ray (20--150 keV) tails have
been detected by INTEGRAL along with RXTE, XMM-Newton and ASCA data in
several AXPs (Kuiper et al. 2004, 2006; den Hartog et al. 2008a,b) and
SGRs (Mereghetti et al. 2005a; Molkov et al. 2005; G\"otz et al. 2006;
see also Rea et al. 2009 for the new source SGR 0501+4516). The hard
X-ray tails are characteristically flatter than the power-law tail in
the soft X-ray band, except for SGR 1900+14 which is steeper. Their
extension is, however, constrained by upper limits from the COMPTEL
instrument on the Compton Gamma-Ray Observatory, which imply a spectral
turnover around 200--500 keV (e.g. den Hartog et al. 2008a,b), a feature
whose existence is more or less reinforced by {\it Fermi}-LAT upper
limits to magnetar emission above 100 MeV (Abdo et al. 2010).

In highly-magnetized neutron stars, inverse magnetic Compton scattering
of X-ray surface photons is expected to be highly efficient (e.g. Baring
\& Harding 2007) in generating the X-ray spectra if the real plasma
density is super-Goldreich-Julian. There are various scenarios where
such plasma densities are plausible in magnetars (e.g. Thompson,
Lyutikov \& Kulkarni 2002, or Thompson \& Beloborodov 2005). At the
cyclotron resonance, the cross section exceeds the Thomson cross section
by two or more orders of magnitude (e.g. Daugherty \& Harding 1986;
Gonthier et al. 2000). Recently, the steep tail portions of the soft
X-ray emission of magnetars has been fitted by a resonant Comptonization
model with mildly-relativistic electrons (Lyutikov \& Gavriil 2006; Rea
at al. 2008). The Lyutikov \& Gavriil model uses a nonrelativistic
magnetic Thomson cross section, neglecting electron recoil, and the fits
are of comparable accuracy to a canonical blackbody plus power-law
prescription.  Similiar Comptonization analyses have been presented in
G\"uver, et al. 2007, and G\"uver, \"Ozel \& G{\"o}{\u g}{\"u}{\c s}
2008, for the AXPs XTE J1810-197 and 4U 0142+61, respectively, using
detailed X-ray spectroscopy to constrain the surface magnetic fields of
these magnetars.  Nobili et. al (2008a) have also successfully fitted
the soft X-ray XMM-Newton spectra of AXP CXOUJ1647-4552 using a more
sophisticated Monte Carlo resonant Comptonization model.

For the hard tail X-ray magnetar emission, repeated upscattering by hot,
mildly-relativistic electrons (i.e., Comptonization) may have
difficulties modeling the observed flat spectra due to the relatively
high efficiency of photon escape from the interaction region.  In
contrast, single resonant Compton upscatterings by electrons with
ultrarelativistic Lorentz factors \teq{\gamma_e\gg 1} is potentially a
more viable mechanism, provided such a source of electrons exists in a
scattering locale close to the hot photon bath on the surface.  Previous
 magnetic inverse Compton scattering work applied to pulsar
magnetospheres (Daugherty \& Harding 1989), and also in the context of
neutron star models for gamma-ray bursts (e.g. Dermer 1990) computed
upscattering spectra and electron cooling rates and in the
non-relativistic magnetic Thomson limit (extended from nonmagnetic
Comptonization in Ho \& Epstein 1989).  These magnetic Thomson models
serve as a crucial check in the non-relativistic or weak field limits of
any fully-relativistic QED formalism. Sturner (1995) further extended
Dermer (1990) and presented more sophisticated cooling rate calculations
with consideration of thermal seed photons as well as Klein-Nishina
corrections. These analyses are not necessarily sufficient for
magnetars' hard X-rays in the relativistic quantum domain, where both
fields in excess of the quantum critical value $B_{cr} \approx 4.413$ x
$10^{13}$ G, and ultrarelativistic Lorentz factors for electrons, may be
realized.  Baring \& Harding (2007) and Nobili et al. (2008b) have
presented some basic analyses in this ultrarelativistic QED domain, and
find substantive differences relative to results from magnetic Thomson
cases. Ultimately, the maximum electron energy and emission spectral
characteristics may well be controlled by the cooling rate, so that
accurate computation of such rates is essential for complete models of
inverse Compton spectral formation.  Hence, extending such resonant
Compton cooling analyses to treat the relativistic, quantum domain is
imperative.

Fully relativistic QED magnetic cross sections in the spin-averaged
Johnson \& Lippmann (1949, JL) wavefunction formalism are found in
Herold (1979), Daugherty \& Harding (1986) and Bussard, Alexander \&
M\'esz\'aros (1986), extending earlier non-relativistic quantum
mechanical formulations such as Canuto, Lodenquai \& Ruderman (1971).
These have been the tool of choice for resonant Compton scattering
implementations in neutron star studies.  Yet, in the cyclotron
resonance, the {\it spin-dependent} widths depend on the choice of
electron wavefunctions in a uniform magnetic field, and therefore so
also do the Compton cross sections and scattering rates. The JL
wavefunctions are Cartesian coordinate eigenstates of the kinetic
momentum operator. The Sokolov \& Ternov (1968; ST) ``transverse
polarization'' states constitute a popular alternative, and are
eigenfunctons of the magnetic moment operator that are derived in
cylindrical coordinates. The ST states possess important symmetries
between electron and positron states (e.g. Herold, Ruder \& Wunner 1982;
Melrose and Parle 1983), and under Lorentz transformations along the
magnetic field (Baring, Gonthier \& Harding 2005). Moreover, Graziani
(1993) contended that the ST wavefunctions are the physically-correct
choices for incorporating spin-dependent widths in the resonance of the
QED scattering cross section, since they diagonalize both the magnetic
moment and mass operators. Here, cooling rates are computed for both
sets of states, namely spin-averaged JL scatterings and spin-dependent
ST calculations, highlighting significant differences between the two
when the cyclotron resonance is sampled.  The two wavefunction choices
generate identical Compton cooling rates when the scatterings are
non-resonant.  This is the first time that Sokolov \& Ternov QED
formalism has been deployed in the computation of magnetic Compton rates
for the cooling of relativistic electrons. Since the magnetar
atmospheric thermal photon energies are much less than the electron rest
mass, it is appropriate to specialize to scatterings that leave the
electron in the zeroth Landau level, as noted in Gonthier et al. (2000)
and adopted by  Baring \& Harding (2007); this expedient simplification
is employed here.

The principal thrust of this paper is to serve as a foundation for the
explanation of hard X-ray spectra of magnetars, to enable refinements of
the resonant upscattering model of Baring \& Harding (2007) that
incorporate evolution of the electron distribution due to Compton
cooling.  It presents, for the first time, numerical calculations of the
electron cooling rates, and associated analytical approximations, for
magnetic Compton upscattering in the fully relativistic quantum regime. 
The analysis forms collision integrals for these rates by weighting the
scattering cross section. The principal assumption employed is the
restriction to ultra-relativistic electrons moving along {\bf B}, an
approximation that is generally highly accurate for resonant
interactions in high-field pulsars and magnetars. These rates will
provide important input for future kinetic equation analyses and Monte
Carlo simulations of time-dependent evolution of cooling electron
distributions and concomitant Compton upscattering spectra. Relativistic
reductions profoundly change the rates; classic inverse Thomson
$\dot{\gamma_e} \propto \gamma_e^2$ cooling is not realized in the
ultrarelativistic regime since the cross section is well below the
Thomson value.  The work evaluates rates for both monoenergetic soft
photons, and thermal surface X-rays; the latter case profoundly alters
the kinematic accessibility of the resonance, yielding $\dot{\gamma_e}
\propto \gamma_e^{-1}$ regimes extending to extremely high Lorentz
factors.  The cooling rate calculation  also encapsulates arbitrary
interaction altitudes and colatitudes in neutron star magnetospheres,
via a generalized angular distribution function, under the simplifying
assumption of photons emitted isotropically and uniformly over the
entire stellar surface.  This restriction facilitates greater
tractability of the analysis, and will be relaxed in future work that
accounts for the influence of surface inhomogeneities and anisotropies
(e.g. Zavlin et al. 1994; Thompson \& Beloborodov 2005).  The cooling
rate formalism is expounded upon in Section~\ref{sec:formalism},
specialized results for Thomson regimes and uniform soft photon fields
are discussed in Section~\ref{sec:cooling_Thom}, before the more general
QED cooling calculations for arbitrary locales in the magnetosphere are
addressed in Sections~\ref{sec:cooling_gen}
and~\ref{sec:esoft_anisotropy}. We derive compact analytic asymptotic
approximations to resonance cooling rates that verify the numerics.  In
addition, we observe that the Lorentz factors for the onset of resonant
cooling rates are controlled by simple kinematic constraints that depend
on the surface temperature and the value of the field strength in the
interaction zone.

\section{RESONANT SCATTERING COOLING RATE FORMALISM}
 \label{sec:formalism}

The focus here is specifically for the case of photons initially
propagating along the magnetic field lines.  Such a restriction is
strongly motivated in gamma-ray pulsar models, since photons engaged in
Compton scattering generally collide with ultra-relativistic electrons
moving in the magnetosphere. Virtually all magnetospheric pulsar models,
both of the polar cap or outer gap varieties, inject electrons with
Lorentz factors \teq{\gamma_e\sim 10^4}--\teq{10^7}. This is effectively
mandated by the observation of hard gamma-rays in the EGRET and {\it
Fermi} pulsars (see the {\it Fermi}-LAT Pulsar Catalog, Abdo et al. 2010
for a good synopsis of the status quo). Yet the ensuing cascades
generally fail to cool the electrons down to mildly relativistic
energies so that the \teq{\gamma_e\gg 1} assumption holds (Daugherty \&
Harding 1982; Zhang \& Harding, 2000; Hibschman \& Arons 2001; Rudak,
Dyks \& Bulik 2002; Sturner 1995; Dyks \& Rudak 2000; Harding \&
Muslimov 2002). In cascade development, the components of the electron
momenta perpendicular to {\bf B} are rapidly degraded due to extremely
rapid cyclotron/synchrotron cooling, typically in timescales of
\teq{10^{-20}}--\teq{10^{-16}} seconds.  Hence, the assumption that soft
X-ray photons scatter off electrons that move ultra-relativistically
along the field is generally a secure one, probably extendable to the
magnetar regime.  This \teq{\gamma_e\gg 1} restriction will be adopted
here, with cases of cooling at mildly-relativistic energies being
deferred to future work. For such ultra-relativistic electrons, Lorentz
transformation to the electron rest frame then results in photon
propagation along {\bf B}, regardless of the direction of a low energy
photon in the magnetospheric frame.  This then spawns great
simplification of the relativistic differential cross section for
Johnson \& Lippmann wavefunctions from the fully general form presented
in Daugherty \& Harding (1986), as outlined in Gonthier et al. (2000). 
A similar simplification is afforded for Sokolov \& Ternov formulations
of the cross section (Gonthier et al. 2011, in preparation).  The
compact expressions for these two differential cross sections in the
case of photon propagation parallel to {\bf B} are presented in
Eqs~(\ref{eq:sigma_unpol_JL}) and~(\ref{eq:sigma_unpol_STapprox}) below.

\subsection{Upscattering Kinematics}
 \label{sec:kinematics}

\begin{figure*}[t]
\figureoutpdf{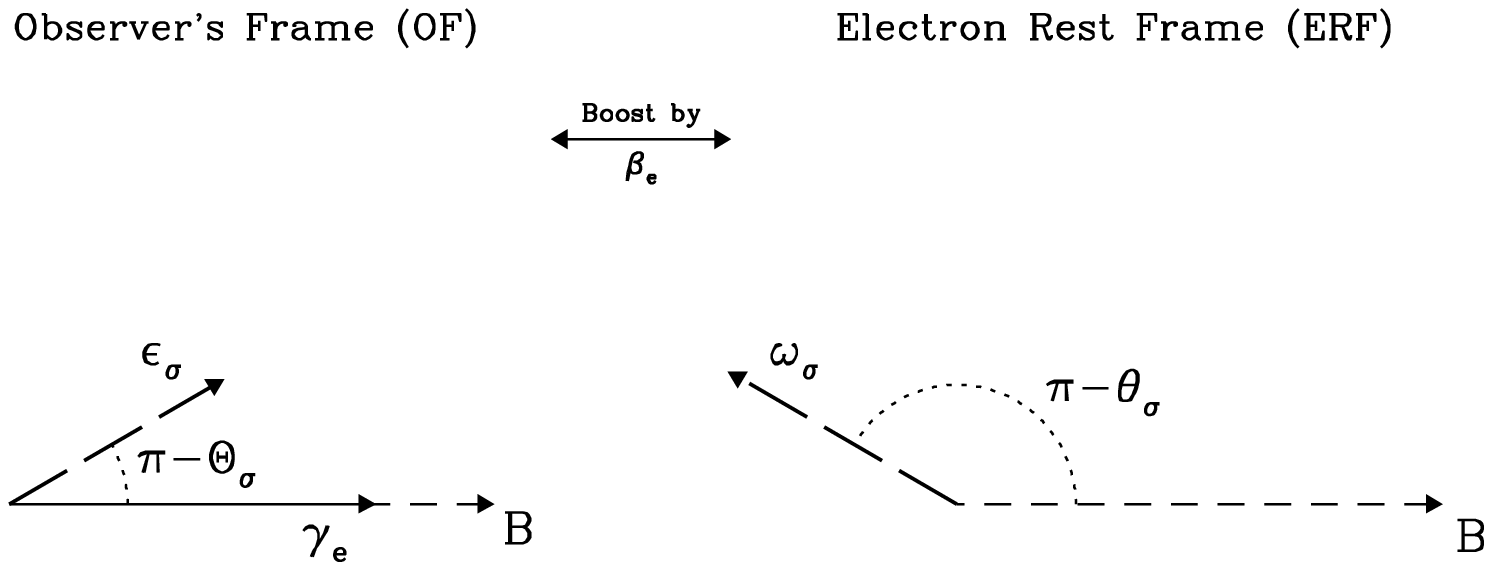}{6.0}{0.0}{0.0}{
Photon momenta and the kinematic geometry for boosts between the
observer's or laboratory frame of the pulsar, and the electron rest
frame, in which the scattering cross section is most simply expressed. 
The boost is along the local magnetic field vector since the electrons
occupy the lowest Landau level.  The label \teq{\sigma =i,f} represents
either the initial (\teq{i}, incoming) or final (\teq{f}, scattered)
photon. The kinematic relationships between photon energies and angles
in the two frames of reference are expressed in
Eq.~(\ref{eq:Lorentz_transform}).
 \label{fig:kinematics} }      
\end{figure*}

To set the scene for the exposition on collision integral calculations
of electron cooling rates, it is necessary to first summarize key
kinematic aspects. Both the Lorentz transformation from the observer's
or laboratory frame (OF) to the electron rest frame (ERF), and the
scattering kinematics in the ERF, are central to determining the
character of resonant Compton upscattering spectra and the cooling
rates. As choices of photon angles in these two reference frames are not
unique, the conventions adopted here are now stated; they follow those
used in Baring \& Harding (2007). Let the electron velocity vector in
the OF be \teq{\vec{\beta}_e}, which is parallel to {\bf B} due to
rampant cyclo-synchrotron cooling perpendicular to the field.  The
dimensionless pre- and post-scattering photon energies (i.e. scaled by
\teq{m_ec^2}) in the OF are \teq{\erg_i} and \teq{\erg_f}, respectively,
and the corresponding angles of these photons with respect to
\teq{-\vec{\beta}_e} (i.e. field direction) are \teq{\Theta_i} and
\teq{\Theta_f}, respectively.  With this definition,
\begin{equation}
   \mu_{i,f}\;\equiv\;
   \cos\Theta_{i,f}\; =\; - \dover{\vec{\beta}_e . \vec{k}_{i,f}}{
        \vert\vec{\beta}_e\vert . \vert \vec{k}_{i,f}\vert }\quad ,
 \label{eq:OF_kinematics}
\end{equation}
and {\it the zero angles are chosen anti-parallel to the electron
velocity}, corresponding to head-on collisions. Here, \teq{\vec{k}_i}
and \teq{\vec{k}_f} are the initial and final photon three-momenta in
the OF, and the \teq{\mu}s denote photon angle cosines hereafter.
Whether \teq{\vec{\beta}_e} is parallel to or anti-parallel to {\bf B}
is irrelevant to the scattering problem.  Boosting by
\teq{\vec{\beta}_e} to the ERF then yields pre- and post-scattering
photon energies in the ERF of \teq{\omega_i} and \teq{\omega_f},
respectively, with corresponding angles with respect to
\teq{-\vec{\beta}_e} of \teq{\theta_i} and \teq{\theta_f}.  The
relations governing this Lorentz transformation are
\begin{eqnarray}
   && \omega_{i,f} \; =\; \gamma_e\erg_{i,f} (1+\beta_e\cos\Theta_{i,f})
   \quad ,\nonumber\\[-5.5pt]
 \label{eq:Lorentz_transform}\\[-5.5pt]
   && \cos\theta_{i,f} \; =\; \dover{\cos\Theta_{i,f} + \beta_e}{
      1 + \beta_e \cos\Theta_{i,f}}\quad ,\nonumber
\end{eqnarray}
the kinematic picture of which is illustrated in
Figure~\ref{fig:kinematics}. The inverse transformation relations are
obtained from these by the interchange
\teq{\theta_{i,f}\leftrightarrow\Theta_{i,f}} and the substitutions
\teq{\omega_{i,f}\to\erg_{i,f}} and \teq{\beta_e\to -\beta_e}.  The form
of Eq.~(\ref{eq:Lorentz_transform}) clearly indicates that head-on
collisions in the OF generate the largest values of \teq{\omega_i}, and
guarantees that for most \teq{\Theta_i}, the initial scattering angle
\teq{\theta_i} in the ERF is close to zero when \teq{\gamma_e\gg 1}. 
Exceptions to this general rule arise when \teq{\cos\Theta_i\approx
-\beta_e}, cases that form a small fraction of the upscattering phase
space, and generally provide only a small contribution to the cooling
rate for ultra-relativistic electrons. The dominance of
\teq{\theta_i\approx 0} interactions motivates the particular laboratory
frame angle convention adopted in Eq.~(\ref{eq:OF_kinematics}).

\newpage

In the electron rest frame (ERF), the scattering kinematics differ from that 
described by the familiar Compton formula in the absence of magnetic fields,
since momentum conservation perpendicular to the field is not mandated.
General forms for the relationship between \teq{\omega_f} and
\teq{\omega_i}, \teq{\theta_i} and \teq{\theta_f} are posited, for example, in
Herold (1979) and Daugherty \& Harding (1986).  In the special case
\teq{\theta_i\approx 0} that is generally operable for the \teq{\gamma_e\gg 1}
scenario here, the kinematic formula for the final photon energy
\teq{\omega_f} in the ERF can be approximated by
\begin{equation}
   \omega_f\; =\; \omega' (\omega_i ,\,\theta_f)\;\equiv\;
     \dover{2\omega_i\, r}{1+\sqrt{1-2\omega_i r^2\sin^2\theta_f}}\qquad , \qquad
     r\; =\; \dover{1}{1+\omega_i (1-\cos\theta_f)}
 \label{eq:reson_kinematics}
\end{equation}
where \teq{r} is the ratio \teq{\omega_f/\omega_i} that would correspond 
to the non-magnetic Compton formula, which in fact does result if 
\teq{\omega_i r^2\sin^2\theta_f\ll 1}.  Eq.~(\ref{eq:reson_kinematics}) 
is a rewriting of Eq. ~(15) of Gonthier et al. (2000; see also Eq. (9) of 
Baring \& Harding 2007), and can be simply rearranged into the 
following convenient form:
\begin{equation}
  (\omega_f)^2\sin^2\theta_f  - 2\omega_i\omega_f (1-\cos\theta_f)
   + 2(\omega_i - \omega_f) \; =\; 0 \quad .
 \label{eq:res_kinematics_alt}
\end{equation}
These kinematic forms are realized for the particular case where
electrons remain in the ground state (zeroth Landau level) after
scattering.  Such forms are relevant for common astrophysical situations
that sample near and below the cyclotron resonance, though well above
the resonance, cases of electron excitation must be entertained (see
Gonthier et al. 2000). Unless otherwise stated, hereafter it will be
assumed that scattered electrons will be produced in the lowest Landau
levels (i.e. spin down to spin down transitions), and the kinematics
described above applies to any spin-state formalism for the collision
cross section.

\subsection{General Cooling Rate Formalism}
 \label{sec:rate_formalism}

To ascertain representative electron cooling rates in pulsar and
magnetar magnetospheres, collision integral calculations of these rates
are performed. For the purposes of this paper, the electrons will be
assumed to be monoenergetic, with Lorentz factor \teq{\gamma_e} and
number density \teq{n_e}, i.e. their distribution function is \teq{n_e
\delta (\gamma - \gamma_e)}. To simplify the formalism for the cooling
rates, monoenergetic, incident photons of dimensionless energy
\teq{\erg_{\gamma}=\erg_s} will first be assumed, with the implicit
understanding that \teq{\erg_s\sim 3kT/m_ec^2} forges the connection
with surface X-ray temperatures.  Clearly values of \teq{\erg_s\sim
3\times 10^{-4} - 3\times 10^{-3}} are commensurate with thermal photon
temperatures \teq{kT\sim 0.1-0.3}keV observed in or inferred for
middle-aged pulsars (see Becker \& Tr\"umper 1997 for a comprehensive
exposition on X-ray pulsar emission) and also radio-quiet, long-period
isolated neutron stars (e.g. Haberl 2007), and hotter surface environs
(\teq{kT\sim 0.5-1}keV) in the more highly-magnetized anomalous X-ray
pulsars (see Perna et al. 2001).   Subsequently, the soft photon
energies \teq{\erg_s} will be distributed via a Planck spectrum.  This
more precise approach provides more than just a smearing out of sharp
cooling function features exhibited in the various figures below that
pertain to monoenergetic soft photons; it provides an extension of the
kinematically accessible phase space for resonant interactions.  It will
become clear that this extension is a profound inclusion, defining the
character of the cooling rates at high Lorentz factors.

The possibility of soft photon anisotropy will be retained in the
formalism.  This is appropriate for moderate and high altitude locales
for scattering interactions, where remoteness from points of origin on
the stellar surface incurs stronger anisotropies in the target X-ray
population.  Accordingly, in the analysis in
Sections~\ref{sec:cooling_Thom} and~\ref{sec:highB_recoil}, the soft
photon number density distribution will take the form
\begin{equation}
   n_{\gamma} (\erg_i ,\, \mu_i )\; =\; n_s\, \dover{f(\mu_i)}{\mu_+-\mu_-}\,
          \delta (\erg_i - \erg_s)\qquad \hbox{for}\qquad 
          \int_{\mu_-}^{\mu_+} \dover{f(\mu)\, d\mu }{\mu_+-\mu_-}\; =\; 1
 \label{eq:ndef}
\end{equation}
prescribing the normalization of the angular portion, which is separable
from the energy dependence.  Observe that the azimuthal dependence is
integrated over in forming \teq{f(\mu_i)}, a step that can be taken from
the outset since the Compton differential cross section does not depend
on such azimuthal angles; this is true regardless of the interaction
locality. Details of how azimuthal dependence is subsumed in
\teq{f(\mu_i)} are expounded in Section~\ref{sec:esoft_anisotropy},
forging a direct connection to the magnetospheric interaction geometry
that is encapsulated in Eq.~(\ref{eq:ang_dist}). To simplify earlier
parts of our analysis, we will first assume photon isotropy within a
cone (or hemisphere), which amounts to setting \teq{f(\mu_i)\to 1}, i.e.
a uniform distribution of angle cosines \teq{\mu_i} in some range
\teq{\mu_-\leq\mu_i\leq \mu_+}. This is often broad enough to encompass
the resonance, i.e. the value \teq{\mu_i=[ B/(\gamma_e\erg_s)
-1]/\beta_e}, but not always so, as will become evident in due course.
In Sections~\ref{sec:therm_esoft},~\ref{sec:mean_energy}
and~\ref{sec:esoft_anisotropy} an alternative normalization protocol
will be adopted, being germane to thermal soft X-ray photon anisotropies
at and above the surface.

The formulation for upscattering spectra appropriate to the resonant
Compton problem was presented in Baring \& Harding (2007). This made use
of generic Compton upscattering formalism presented in Eqs.~(A7)--(A9)
of Ho and Epstein (1989) that is applicable to both Thomson and
Klein-Nishina regimes, and moreover is readily adaptable to incorporate
magnetic kinematics and the QED cross section for fully relativistic
cases of magnetic Compton scattering. The spectrum of photon production
\teq{dn_{\gamma}/(dt\, d\erg_f\, d\mu_f) }, differential in the photon's
post-scattering laboratory frame quantities \teq{\erg_f} and
\teq{\mu_f=\cos\Theta_f}, can be written as
\begin{equation}
   \dover{dn_{\gamma}}{dt\, d\erg_f} \; =\;  \dover{n_e n_s\, c}{\mu_+-\mu_-}
   \int_{\mu_l}^{\mu_u}d\mu_f   \int_{\mu_-}^{\mu_+}d\mu_i    \, f(\mu_i)\,
   \delta \big\lbrack\omega_f -\omega'(\omega_i,\,\theta_f)\, \bigl\rbrack\,
   \dover{1+\beta_e\mu_i}{\gamma_e(1+\beta_e\mu_f)}\,
   \dover{d\sigma}{d(\cos\theta_f) }\quad .
 \label{eq:scatt_spec}
\end{equation}
Observe that in deriving this, the angle convention specified in
Eq.~(\ref{eq:OF_kinematics}) requires the substitution \teq{\beta_e\to
-\beta_e} in Eqs.~(A7--A9) of Ho and Epstein (1989). Here, the notation
\teq{\mu_i=\cos\Theta_i} and \teq{\mu_f=\cos\Theta_f} is used for
compactness.  In Eq.~(\ref{eq:scatt_spec}), the factor
\teq{c(1+\beta_e\mu_i)} expresses the relative velocity in
photon-electron collisions, remembering that \teq{\mu_i=1} represents
head-on impacts.  Also, the \teq{\gamma_e (1+\beta_e\mu_f)} factor in
the denominator arises because of the Lorentz transformation of the
differential cross section between the ERF and the observer's frame. The
rate can then be routinely weighted by the factor
\teq{-(\erg_f-\erg_i)/n_e \approx -\erg_f/n_e} when \teq{\gamma_e\gg 1},
and integrated over all produced energies \teq{\erg_f}, to generate the
required electron cooling rate:
\begin{equation}
   \dot{\gamma}_e \; =\; -  \dover{n_s\, c}{\mu_+-\mu_-}\, \int \erg_f\, d\erg_f
   \int_{\mu_l}^{\mu_u}d\mu_f   \int_{\mu_-}^{\mu_+}d\mu_i    \, f(\mu_i)\,
   \delta \big\lbrack\omega_f -\omega'(\omega_i,\,\theta_f)\, \bigl\rbrack\,
   \dover{1+\beta_e\mu_i}{\gamma_e(1+\beta_e\mu_f)}\,
   \dover{d\sigma}{d(\cos\theta_f) }\quad .
 \label{eq:cool_rate_form}
\end{equation}
Hereafter the specializations \teq{\mu_l\to -1} and \teq{\mu_u\to 1} are
adopted to encompass the full cooling parameter space. Explicit
presentation of the limits on the \teq{\erg_f} integration is suppressed
for the moment, since these are somewhat complicated with this choice of
variables.

It is more convenient to change the angular integration variables
\teq{\mu_{i,f}} at this juncture to \teq{\omega_i} and \teq{\omega_f},
for which the Jacobian indentity for this transformation is
\teq{d\mu_i\, d\mu_f = d\omega_i\, d\omega_f/(\gamma_e^2\beta_e^2
\erg_i\erg_f)}. In addition, the substitution
\teq{(1+\beta_e\mu_i)/(1+\beta_e\mu_f)\to
\omega_i\erg_f/(\omega_f\erg_i)} can be employed.  Then the
\teq{\omega_f} integration is trivial, and the limits on the subsequent
\teq{\omega_i} integration are simply obtained from
Eq.~(\ref{eq:Lorentz_transform}), namely
\teq{\gamma_e(1+\beta_e\mu_-)\erg_s \leq \omega_i \leq
\gamma_e(1+\beta_e\mu_+)\erg_s}.  The limits on the \teq{\erg_f}
integration are less easily derived, since they are formed from the
combination of the Lorentz transformation characteristics in
Eq.~(\ref{eq:Lorentz_transform}) and the Compton scattering kinematic
constraint on \teq{\omega_f} in Eq.~(\ref{eq:reson_kinematics}), in
concert with the domain \teq{\vert\cos\theta_f\vert \leq 1}. The
resulting range is
\begin{equation}
   \gamma_e(1-\beta_e)\, \omega_i \;\leq\;\erg_f\;\leq\; 
   \dover{\gamma_e(1+\beta_e)\, \omega_i}{1+2\omega_i} \quad ,
 \label{eq:kinematic_range_ergf}
\end{equation}
for which the resonant \teq{\omega_i=B} specialization is offered
in Eq.~(15) of Baring \& Harding (2007).   Since the bounds on 
\teq{\erg_f} are dependent on \teq{\omega_i}, it is expedient to 
reverse the order of the integrations.  Then the final nuance in 
the manipulations is to change variables from \teq{\erg_f} to 
\teq{\cos\theta_f} via \teq{\erg_f=\gamma_e\omega_f 
(1-\beta_e\cos\theta_f)}, motivated by the fact that the
differential cross section for Compton scattering is expressed 
most compactly in terms of \teq{\cos\theta_f}.  Note that the 
range in Eq~(\ref{eq:kinematic_range_ergf}) corresponds to 
\teq{-1\leq\cos\theta_f\leq 1}.  The culmination of these steps 
is the final general form for the cooling rate:
\begin{equation}
   \dot{\gamma}_e \; =\; -  \dover{n_s\, c}{\mu_+-\mu_-}\, 
   \dover{1}{\gamma_e^2\beta_e^2\erg_s^2}   \int_{\omega_-}^{\omega_+}
   \omega_i\, d\omega_i   \int_{-1}^1 d(\cos\theta_f )\, [1-\beta_e\cos\theta_f]\,
   \, f(\mu_i)\, \biggl\vert \dover{\partial \erg_f}{\partial (\cos\theta_f) } \biggr\vert\,
   \dover{d\sigma}{d(\cos\theta_f) }
 \label{eq:cool_rate_final}
\end{equation}
for \teq{\mu_i=[ \omega_i/(\gamma_e\erg_s) -1]/\beta_e}, with
\begin{equation}
   \omega_+\; =\; \gamma_e(1+\beta_e\mu_+)\erg_s\qquad ,\qquad
   \omega_-\; =\; \gamma_e(1+\beta_e\mu_-)\erg_s\quad .
 \label{eq:omega_pm_def}
\end{equation}
This double integral form provides the basis for the subsequent 
calculations in this paper.  It is understood that \teq{\omega_f} is 
given by Eq.~(\ref{eq:reson_kinematics}).  The computational 
convenience of this result is underlined by the fact that the 
integration variables are now those embodied in the differential
cross section \teq{d\sigma/d(\cos\theta_f)}.  Using 
\teq{\erg_f=\gamma_e\omega_f (1-\beta_e\cos\theta_f)} and
Eq.~(\ref{eq:res_kinematics_alt}), it can be established that
\begin{equation}
   \dover{\partial \erg_f}{\partial (\cos\theta_f) }\, =\; 
   \gamma_e (1-\beta_e\cos\theta_f)\,  
   \dover{\partial \omega_f}{\partial (\cos\theta_f) } - \gamma_e\beta_e\omega_f
 \label{eq:dergf_dcthetaf}
\end{equation}
for
\begin{equation}
   \dover{\partial \omega_f}{\partial (\cos\theta_f) }\; =\;
   \dover{\omega_f^2\, (\omega_i - \omega_f\cos\theta_f)}{
                2\omega_i - \omega_f [1+ \omega_i (1-\cos\theta_f )\, ]}
 \label{eq:domegaf_dcthetaf}
\end{equation}
In the Thomson limit where \teq{\omega_f\approx\omega_i\ll 1}, the
\teq{\partial \omega_f / \partial (\cos\theta_f) } term contributes
insignificantly to Eq.~(\ref{eq:dergf_dcthetaf}) and the Jacobian
satisfies \teq{\vert \partial \erg_f / \partial (\cos\theta_f) \vert =
\gamma_e\beta_e\omega_f}. Finally, we remark that in the neighborhood of
the resonance, further manipulation of Eq.~(\ref{eq:cool_rate_final}) is
detailed in Section~\ref{sec:cooling_gen}, so as to facilitate both
numerical computations and the derivation of analytic asymptotics for
the rates.

\newpage

\subsection{The Differential Cross Section}
 \label{sec:c-section}

The last remaining ingredient in the formulation is the differential
cross section \teq{d\sigma /d(\cos\theta_f)}.  Here, because the
possibility of near-critical and super-critical fields is entertained,
it is necessary to generalize beyond the magnetic Thomson limit cross
sections employed in the cooling calculations of Daugherty \& Harding
(1989), Sturner (1995) and Harding \& Muslimov (1998). Fully
relativistic, quantum cross section formalism for the Compton
interaction in magnetic fields can be found in Herold (1979), Daugherty
\& Harding (1986), and Bussard, Alexander \& M\'esz\'aros (1986). These
were obtained using the old Johnson \& Lippman (1949) eigenfunctions of
the Dirac equation in a uniform magnetic field. These traditional
wavefunctions are Cartesian coordinate eigenstates of the kinetic
momentum operator that do not diagonalize the magnetic moment operator.
They are conveniently used in cases where information on the spin of the
intermediate electron states is not germane, and can be averaged over;
such is the case for scattering calculations well removed from the
cyclotron resonance.  The cross section for the alternative Sokolov and
Ternov (1968) states is also discussed below, since we employ it as an
upgrade over previous treatments of scattering; this development will
also incorporate full spin dependence of the cross section near the
cyclotron resonance, a desirable advance.

For our  \teq{\gamma_e\gg 1} focus here, where the incoming photons in
the ERF move along the field lines, Gonthier et al. (2000) obtained
algebraic simplification of the results of Daugherty \& Harding (1986). 
Here we use Eq.~(23) of Gonthier et al. (2000), a more compact form of
which is presented in Baring \& Harding (2007). These forms are
specialized to the case of scatterings that leave the electron in the
ground state, the zeroth Landau level that it originates from.  This
expedient choice is appropriate at and below the cyclotron resonance
where ground state-to-ground state transitions provide the only
contribution to the cross section, but must be relaxed well above the
resonance where excitations tend to dominate.  As will become apparent
below, when blackbody soft photon distributions are adopted, the
resonant contribution always overwhelms that above resonance (and below
the first harmonic) when kinematics render both regimes accessible. 
Moreover, astrophysical applications of these cooling rates will most
likely be employed in contexts where the cooling competitively limits
some electron acceleration process so that harmonics of the cyclotron
fundamental are not often encountered in neutron star model applications
involving hard X-rays.  Hence, there is not strong motivation for
addressing excitation transitions: they are expected to introduce only
small corrections to the results presented here because they possess
cross sections near the resonant cyclotron harmonics that are generally
much smaller than that in the neighborhood of the fundamental (e.g. see
Daugherty \& Harding 1986).  Moreover, the decay width is larger in the
harmonics than in the fundamental (e.g. see Eq.~(19) of Herold et al.
1982, for trends with harmonic number \teq{n} in \teq{B\ll 1} cases, and
Fig.~3 of Harding \& Lai 2006, for an illustration that
\teq{\partial\Gamma /\partial n >0} in all field regimes), thereby
reducing the effective cross section at the peak of the resonance. Such
excitation refinements will be deferred to future work.

For the computation of electron cooling rates, it is necessary to sum
over final photon polarizations and also to introduce the effective
(electron spin-averaged) width \teq{\Gamma} associated with the
cyclotron decay lifetime for the intermediate electron states.  Adapting
Eqs.~(12) and~(13) of Baring \& Harding (2007), the differential Compton
cross section in the Johnson and Lippmann (JL) formalism to be used here
takes the form
\begin{equation}
   \dover{d\sigJL}{d(\cos\theta_f)} \; = \; 
   \dover{3\sigt}{16} \, 
   \dover{ (\omega_f)^4\; {\cal T}\,
                  \exp \bigl\{ -\kappa \bigr\} }{ \omega_i\, 
        [2\omega_i -\omega_f - \zeta \, ]}
   \; \biggl\{ \dover{1}{(\omega_i -B)^2 + (\Gamma /2)^2} + 
                \dover{1}{(\omega_i +B - \zeta)^2} \biggr\} 
  \label{eq:sigma_unpol_JL}
\end{equation}
for \teq{\kappa = \omega_f^2\sin^2\theta_f/[2B]} and 
\teq{\zeta = \omega_i\omega_f  (1-\cos\theta_f)}.  The polarization-averaged
factor in the numerator is
\begin{eqnarray}
   {\cal T} &=& 1 + \cos^2\theta_f+ \omega_i (1-\cos\theta_f)^2
                                                      -\omega_f\sin^2\theta_f\quad ,\nonumber\\[4pt]
   \Rightarrow\; \omega_f{\cal T} &=& 2 \omega_i 
                        - (1+\omega_i)\,\omega_f (1-\cos^2\theta_f) \quad ,
 \label{eq:Tpar+Tperp}\\[4pt]
   \Rightarrow\; \omega_f^2 {\cal T} &=& 2 (\omega_i - \omega_f)
                    -  2 \omega_i \omega_f (1+\omega_i) \left\{ 1-  \cos\theta_f\right\}
                   +  2 \omega_i^2 \quad ,\nonumber
\end{eqnarray}
where the kinematic result in Eq.~(\ref{eq:res_kinematics_alt}) has been
used in deriving these identities. Note that for magnetic Compton
scattering, in this particular case of photons propagating along {\bf B}
prior to scattering (i.e. \teq{\theta_i =0}), the differential cross
sections are independent of the initial polarization of the photon
(Gonthier et al. 2000).  This property is a consequence of circular
polarizations forming the natural basis states for \teq{\theta_i =0},
and guarantees that the electron cooling rates computed in this paper
are insensitive to the initial polarization level (zero or otherwise) of
the soft photons.  Note also that the width \teq{\Gamma} should also
appear explicitly in the denominator of the second term inside the
parentheses of Eq.~(\ref{eq:sigma_unpol_JL}), however, since its
contribution to this term is almost negligible for all field strengths,
it is omitted for simplicity.  In addition, only the cyclotron
fundamental appears as a resonance, corresponding to an intermediate
virtual electron state in the first (\teq{n=1}) Landau level.

In this paper, the Sokolov and Ternov (1968) transverse polarization
eigenstate formulation of the magnetic Compton cross section is also
used. Deployment of the JL cross section, while historically convenient,
is unsatisfactory in the cyclotron resonance on physical grounds. Since
the JL states do not diagonalize the spin operator, they do not yield
correct spin-dependent cross sections.  Outside the \teq{\omega_i\approx
B} resonance this is irrelevant, since the spin quantum numbers of the
intermediate electrons are summed, and differential cross sections are
necessarily independent of eigenfunction choices for the intermediate
states.  However, in the cyclotron resonance, the lifetime of the
intermediate electron state is spin-dependent so that sums over electron
propagator spin states no longer generate simple averages, yielding a
feedback between the value of the cross section and the choice of basis
states. The Sokolov \& Ternov (ST) wavefunctions are preferred as
solutions of the magnetic Dirac equation.  Baring, Gonthier \& Harding
(2005) highlighted this by demonstrating that Lorentz transformations
along {\bf B} mix the spin states when JL eigenfunctions are employed as
a basis, while ST states transform unmixed, thereby providing the
inherently attractive simplicity of Lorentz invariance. Moreover they
exhibit symmetry between positron and electron states. Graziani (1993)
observed that they are the physically correct choices for spin-dependent
treatments and for incorporating widths in the scattering cross section.

A full treatment of the ST state formulation for Compton scattering in
strong magnetic fields was presented in Sina (1996), and has been honed
analytically and numerically by Gonthier et al. (2011, in preparation),
for the case of photons incident along the magnetic field lines.  Away
from the fundamental resonance, the differential cross section matches
that of the Johnson \& Lippmann formalism in
Eq.~(\ref{eq:sigma_unpol_JL}), as it should.  In the resonance, both
near the peak and in the wings, specifically in the range \teq{0.95 <
\omega_i/B < 1.05}, Gonthier et al. (2011) demonstrate that the ST
differential cross section can be approximated by the compact analytic
spin-dependent form
\begin{equation}
   \dover{d\sigST}{d(\cos\theta_f)} \; = \; 
   \dover{3\sigt}{64} \, 
   \dover{ \omega_f^2\, e^{-\kappa} }{ \omega_i\,  [ 2\omega_i-\omega_f-\zeta]\, \eperp^3}
                  \sum_{s=\pm 1} \dover{(\eperp - s)^2\, \Lambda_s
                         }{(\omega_i -B)^2 + (\Gamma_s /2)^2}   \quad ,
  \label{eq:sigma_unpol_STapprox}
\end{equation}
for 
\begin{equation}
   \Lambda_s\; =\; \left( 2\eperp + s \right)\, \Bigl\{ \omega_f^2 {\cal T} 
             - [\omega_i-\omega_f] \Bigr\}
             - s\, \eperp^2\, [\omega_i-\omega_f]\quad .
 \label{eq:Lambda_s_def}
\end{equation}
Here the quantum number \teq{s} labels the intermediate electron's 
spin state (either spin up and down), yielding the two Lorentz profiles, 
which are described by the widths
\begin{equation}
   \Gamma_s \; =\; \left( 1 - \dover{s}{\eperp} \right) \,\Gamma
   \quad ,\quad 
   \eperp\; =\; \sqrt{1+2 B}
 \label{eq:ST_widths}
\end{equation}
that emerge from the ST formulation of cyclotron decay rates from the
first Landau level (see Baring, Gonthier \& Harding 2005, including
Eq.~(1) therein, for details). Here \teq{\Gamma} is the same
spin-averaged effective cyclotron width/decay rate as is used in the JL
formulation, namely that in Eq.~(\ref{eq:width_effect}) below. Note that
Eq.~(\ref{eq:sigma_unpol_STapprox}) is an approximation {\it deployed
only in the resonance}, and is generally accurate to better than 0.4\%
whenever \teq{0.95 < \omega_i/B < 1.05}, for a wide range of fields
\teq{B}: see Figure~\ref{fig:ST_csect} in
Appendix~\ref{sec:ST_reduction}.  It contains no non-resonant term
(which has been eliminated for simplicity) like the one appearing in
Eq.~(\ref{eq:sigma_unpol_JL}) that is generated by the second Feynman
diagram (see Gonthier et al. 2000) for the scattering process.  It is
therefore used as a {\it patch} over the interval \teq{0.95 < \omega_i/B
< 1.05}, with Eq.~(\ref{eq:sigma_unpol_JL}) being used for all other
values of \teq{\omega_i/B}, since it represents the spin-averaged
differential cross section appropriate to any eigenstate basis.  A brief
discussion of the derivation of Eq.~(\ref{eq:sigma_unpol_STapprox}) 
from the starting point of Eq.~(3.25) in Sina (1996) is offered in
Appendix~\ref{sec:ST_reduction}.  Noting the identity
\begin{equation}
    \dover{1}{\eperp^3} \sum_{s=\pm 1} (\eperp - s)^2\, \Lambda_s
    \; =\; 4\, \omega_f^2 {\cal T} \quad ,
 \label{eq:ST->JL_identity}
\end{equation}
it becomes clear that the discontinuity in mapping over from the ST
patch in Eq.~(\ref{eq:sigma_unpol_STapprox}) to the spin-averaged JL
differential cross section in Eq.~(\ref{eq:sigma_unpol_JL}) should be
small in the wings of the resonance; numerically, for \teq{0.01 < B <
10^2} it is found to be smaller than 1{\%} at \teq{\vert
\omega_i/B-1\vert =0.05}. The influence of this discontinuity is
negligible on the ST rate calculations, being dominated by the
contribution from much nearer the peak of the resonance at
\teq{\omega_i=B}.

The effective width \teq{\Gamma} does not appear explicitly in the cross
section evaluations of Herold (1979), Daugherty \& Harding (1986),
Bussard, Alexander \& M\'esz\'aros (1986) and Gonthier et al. (2000). 
It must be incorporated in order to account for the fact that the
intermediate electron states effectively have a finite lifetime
\teq{1/\Gammacyc} to cyclotron decay that should formally be embedded in
the resonant denominators that appear in the matrix elements for the
Compton interaction.  Normally this modification is omitted from papers
focusing on cross section calculations.  Yet its inclusion is necessary
in spectral and rate computations since they sample convolutions of the
entire resonance that are integrably divergent without the presence of
natural widths to truncate the cyclotron resonance, thereby generating
Lorentz line profiles. Comprehensive discussions of the incorporation of
such widths in QED scattering formalism via a Breit-Wigner prescription
are provided by Harding \& Daugherty (1991), and later Graziani (1993)
with an extensive formal exposition.  Subtleties appear there in
relativistic regimes (i.e. \teq{B\gtrsim 1}) that mandate an update of
the {\it ansatz} \teq{\Gamma\to \Gammacyc (p_z=0)} that was adopted by
Baring \& Harding (2007) and is predicated on common precedent for use
in non-relativistic Thomson regimes (e.g. see Xia et al. 1985; Daugherty
\& Harding 1989, Dermer 1990; Baring 1994; Liu et al. 2006).  Harding \&
Daugherty (1991; see their Eq. [19]) observed that when incorporating
the cyclotron decay width formally in the denominators, an extra factor
of \teq{{\cal E}_n=\sqrt{1+2 n B + p_z^2}} appears when the matrix
elements are squared.  The upgraded {\it ansatz} then becomes
\teq{\Gamma\to {\cal E}_n\Gammacyc (p_z)}. Here \teq{n=1} is the quantum
number of the intermediate state for the fundamental cyclotron
resonance. The non-zero value of \teq{p_z} that must be inserted in both
\teq{{\cal E}_1} and \teq{\Gammacyc (p_z)} is that obtained by parallel
momentum conservation in the cyclotron interaction when the electron in
the ERF absorbs the incoming photon propagating along {\bf B}, namely
\teq{p_z=\omega_i\approx B} for our case of \teq{\theta_i=0}. Therefore,
in the resonance \teq{{\cal E}_1\to 1+B}.

Values for the width \teq{\Gamma} that is needed for the spin-averaged
JL calculation, and the spin-dependent ST formulation, can be inferred
from the cyclotron decay analysis of Baring, Gonthier \& Harding (2005;
see also Latal 1986; Pavlov et al. 1991; Harding \& Lai 2006).  The
average rates for \teq{1\to 0} transitions at non-zero \teq{p_z} can be
found in their Eqs.~(13) or (23), and are scaled (by \teq{\hbar/m_ec^2})
into dimensionless form here:
\begin{equation}
   \Gammacyc (p_z)\; =\; \Gammave (p_z)\; =\; \dover{\fsc B}{{\cal E}_1} \; I_1(B) \quad ,
   \quad {\cal E}_1\; =\; \sqrt{1 + 2 B + p_z^2}\quad ,
 \label{eq:Gammave_ST_red}
\end{equation}
with
\begin{equation}
   I_1(B)\; = \; \int_0^{\Phi} \dover{d\phi \, e^{-\phi}}{
        \sqrt{(\Phi -\phi )\, (1/\Phi - \phi )}}\;
        \Biggl\lbrack 1 - \dover{\phi}{2} \biggl( \Phi +
                \dover{1}{\Phi}\, \biggr)\, \Biggr\rbrack 
        \quad ,\quad \Phi \; =\; \dover{\sqrt{1+2 B}-1}{\sqrt{1+2B}+1} 
 \label{eq:I1B_BGH05}
\end{equation}
expressing the integration over the angles of radiated cyclotron 
photons.  Observe that right at the peak of the cyclotron resonance,
\teq{\omega_i=B}, the parameter \teq{\Phi} is just that identified in 
the scattering analysis in Appendix A, and the integration variable 
\teq{\phi} is effectively \teq{\omega_f^2\sin^2\theta_f/(2B)}, i.e. the final
angle variable employed in the analytic developments pertaining
to resonant scattering.  The integral for \teq{I_1(B)} can alternatively 
be expressed as an infinite series,
\begin{equation}
   I_1(B) \; =\; \sum_{k=0}^{\infty} \dover{(-1)^k}{k!}\, 
      \Biggl\{ Q_{k} \biggl(1+\dover{1}{B} \biggr) - \Bigl( 1 + \dover{1}{B} \Bigr)
      Q_{k+1} \biggl(1+\dover{1}{B} \biggr) \Biggr\} \quad ,
 \label{eq:I1B_BGH05_alt}
\end{equation}
where the \teq{Q_{\nu}(z)} are Legendre functions of the second kind.
These special functions are finite sums of elementary logarithmic and
polynomial functions of \teq{z} (e.g. see Abramowitz \& Stegun 1965).
Properties of these functions that are germane to their numerical 
evaluation are supplied in the Appendix of Baring, Gonthier \& Harding (2005).
The presence of the \teq{{\cal E}_{1}} factor in Eq.~(\ref{eq:Gammave_ST_red}) 
essentially accounts for time dilation when boosting along {\bf B} from the 
electron rest (\teq{p_z=0}) frame; the Lorentz factor for this boost is simply 
\teq{\gamma = {\cal E}_1/\sqrt{1+2 B}}.  The net {\it ansatz} is that the 
effective width \teq{\Gamma} to be inserted into  Eqs.~(\ref{eq:sigma_unpol_JL}) 
and~(\ref{eq:ST_widths}) can be expressed as
\begin{equation}
   \Gamma\; =\; {\cal E}_1\, \Gammacyc (p_z)
                   \;\equiv\; \sqrt{1+2 B}\, \Gammacyc (p_z=0)
                   \; =\; \fsc B\, I_1(B)\quad ,
 \label{eq:width_effect}
\end{equation}
which is independent of \teq{p_z=\omega_i}.  When \teq{B\ll 1}, this reduces
to the commonly-invoked form \teq{\Gamma\approx 2\fsc\, B^2/3},
which can be deduced using Eq.~(26) of Baring, Gonthier \& Harding (2005),
or the \teq{\Phi\ll 1} limit of the integral in Eq.~(\ref{eq:I1B_BGH05}).
The enhancement of the width by the extra ``relativistic'' factor \teq{\sqrt{1+2 B}} in
supercritical fields is more than offset by the ultra-quantum reductions
in the cyclotron decay rate.  The \teq{\Phi\to 1} limit of Eq.~(\ref{eq:I1B_BGH05}) 
quickly reveals that \teq{\Gamma\approx \fsc\, B\, (1-1/e)} when \teq{B\gg 1}.  
Therefore, \teq{\Gamma/ B\ll 1} always, regardless of the value
of the field strength.  

\section{COOLING RATES IN THE MAGNETIC THOMSON REGIME}
 \label{sec:cooling_Thom}

The exploration of resonant Compton cooling rates focuses first on the
familiar magnetic Thomson regime, commonly treated in the pulsar literature,
before addressing the new territory of the influences of a full QED
treatment of the scattering process.  This Section will restrict considerations
to monoenergetic soft photons that are uniformly distributed in angles within 
a conical sector of the sphere, so as to capture the essence of the magnetic Thomson 
cooling characteristics.  Furthermore, only the Johnson \& Lippmann cross section
will be employed, for simplicity, though with a width choice that essentially maps
over to the spin-dependent Sokolov and Ternov picture.

The Thomson regime is when \teq{B\ll 1} and \teq{\omega_f\approx\omega_i\ll 1},
for which the JL differential cross section in Eq.~(\ref{eq:sigma_unpol_JL}) simplifies to 
\begin{equation}
   \dover{d\sigt}{d(\cos\theta_f )}\; \approx\;\dover{3\sigt}{16}\,
   \Bigl[ 1+\cos^2\theta_f\Bigr]\,
   \Sigma_{\kappa}\Bigl(\dover{\omega_i}{B}\Bigr) \quad ,
 \label{eq:cross_sec_Thom}
\end{equation}
where the subscript T denotes here the magnetic Thomson domain, and 
\begin{equation}
   \Sigma_{\kappa}(\psi)\; =\;\dover{\psi^2}{(\psi -1)^2+\kappa^2}+\dover{\psi^2
}{(\psi +1)^2}\quad ,\quad\kappa\, =\,\dover{2\fsc B}{3}
 \label{eq:Sigma_kappa_def}
\end{equation}
encapsulates both the resonant and non-resonant contributions to the
cross section using \teq{\psi} as the incoming photon energy scaled in
terms of \teq{B}.  Here \teq{\sigt} is the field-free Thomson
cross-section. This form is commonly encountered in past neutron star
radiative applications (e.g. see Dermer 1990; Baring 1994; Liu et al.
2006). The parameter \teq{\kappa =\Gammacyc/(2B)} is a scaling of the
non-relativistic natural line width \teq{\Gammacyc =4\fsc B^2/3} of the
cyclotron harmonic, the only resonance that appears in the Thomson
limit.  However, observe that this choice is precisely twice the
\teq{B\to 0} limit of the spin-averaged width of the Johnson and Lippman
cross section encapsulated in Eq.~(\ref{eq:Gammave_ST_red}).  The reason
for this choice stems from the fact that in this \teq{B\ll 1} domain,
the cyclotron rate is highly spin-asymmetric: spin-flip decays
\teq{n=1\to 0} are profoundly inhibited so that only spin-down to
spin-down transitions are probable (e.g. see Melrose \& Zheleznyakov
1981; Herold, Ruder \& Wunner 1982).  This no-spin-flip cyclotron
transition has a rate given by \teq{\Gammacyc =4\fsc B^2/3}, exactly the
classical rate (e.g. Bekefi 1966), so that the spin-averaged decay rate
is exactly half of this, as outlined in Latal (1986) and Baring,
Gonthier \& Harding (2005).  It is this higher, spin-biased width that
has been adopted in the aforementioned astrophysical applications of
magnetic Thomson scattering. This presumes that one spin state is
prepared by successive, predominantly no-spin-flip cyclotron transitions
from much higher states, a description that is satisfactory for
semi-classical or non-relativistic quantum cyclotron problems.  In the
full, spin-dependent, Sokolov \& Ternov cooling rate offering in
Section~\ref{sec:cooling_gen} below, the scattering process itself
biases the spin preparation of the intermediate state, and this nuance
is taken into account self-consistently in this paper.

\begin{figure*}[t]
\twofigureoutpdf{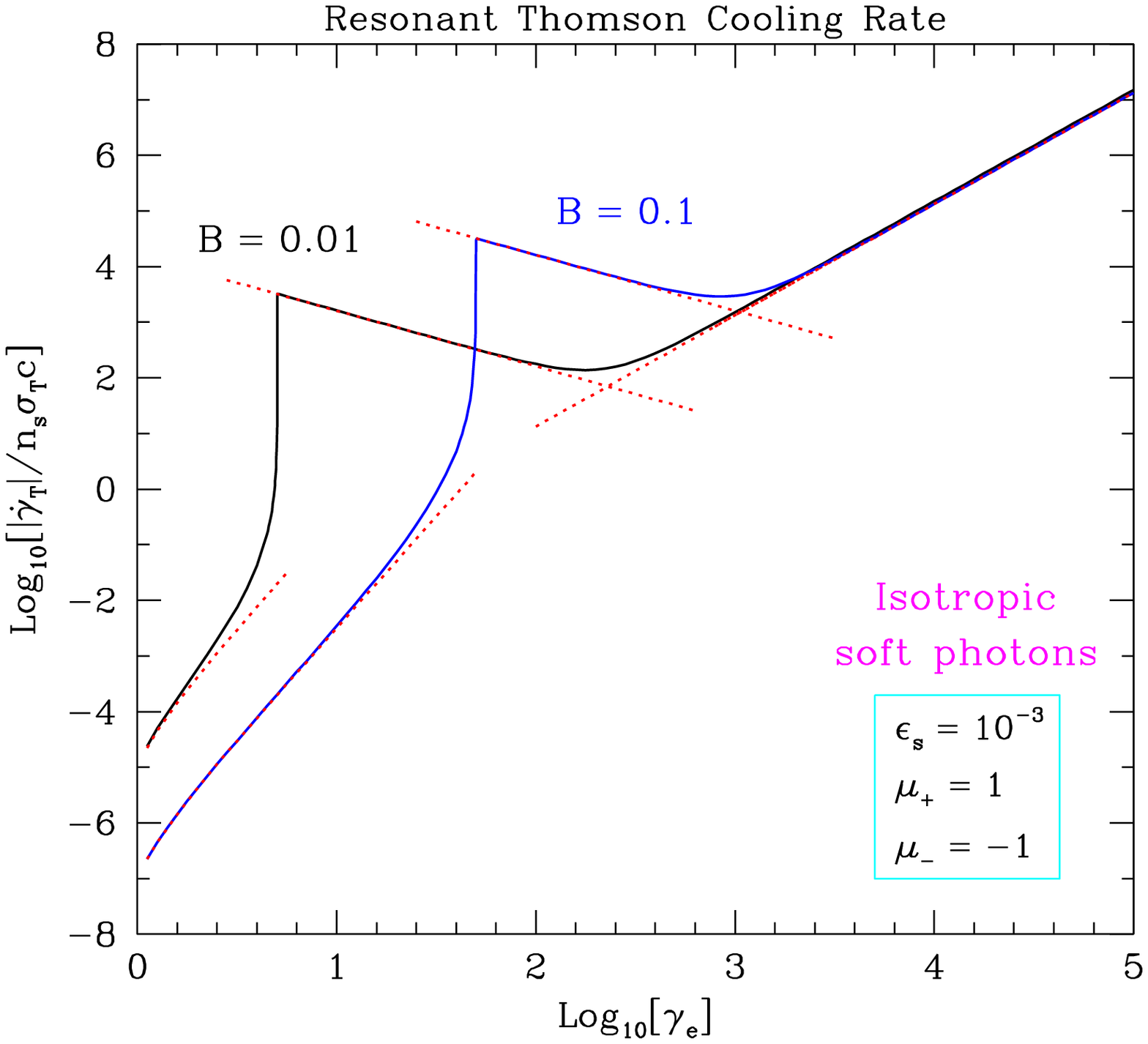}{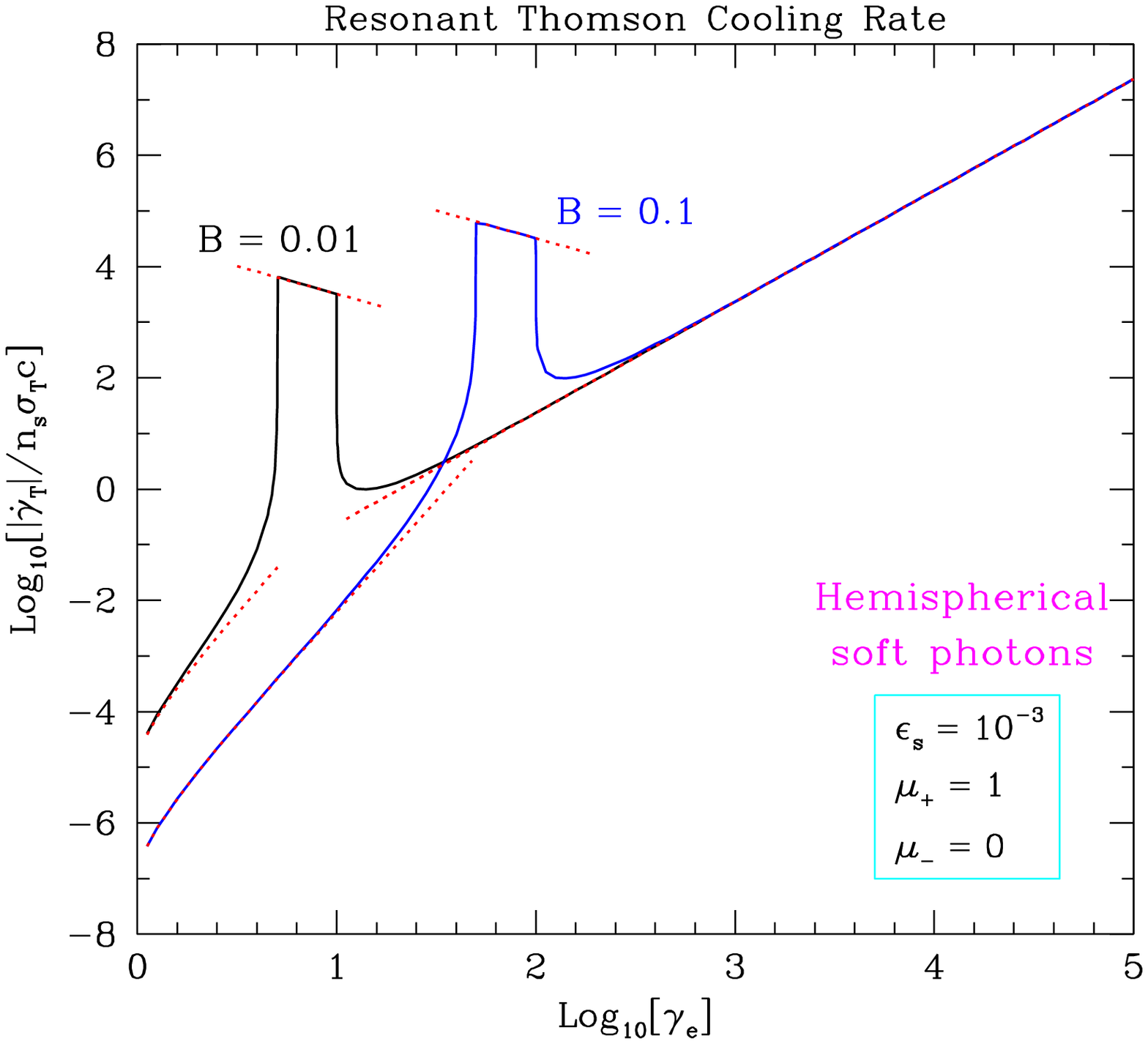}{
Resonant Thomson cooling rates for two different field strengths,
as labelled, in units of \teq{B_{\rm cr} \approx 4.413\times 10^{13}}Gauss,
calculated using Eq.~(\ref{eq:cool_rate_Thom}), i.e. corresponding to
the differential cross section in Eq.~(\ref{eq:cross_sec_Thom}).  The
soft photon energy distribution is monoenergetic, with \teq{\erg_s=10^{-3}},
typical of that for neutron star surface X-rays.  The two cases portrayed
are isotropic soft photons (left panel) and hemispherical \teq{n_s(\erg_s)}
(right panel), the latter being fairly representative of interaction locations
near the stellar surface.  In both panels, the dotted lines encapsulate the 
asymptotic approximations embodied
in Eqs.~(\ref{eq:gammadot_Thom})--(\ref{eq:gammadot_belowres})
[see text].
 \label{fig:resThom_cool} }      
\end{figure*}

For a uniform conical beam of incident photons centered on an axis
coincident with the local magnetic field vector, \teq{f(\mu_i)=1} in the
range \teq{\mu_- \leq \mu_i \leq \mu_+}, and zero outside.  The Thomson
limit then reduces the Jacobian in Eq.~(\ref{eq:cool_rate_final}) to a
simple form \teq{\vert \partial \erg_f / \partial (\cos\theta_f) \vert =
\gamma_e\beta_e\omega_f \approx \gamma_e\beta_e\omega_i}, so that the
cooling rate in Eq.~(\ref{eq:cool_rate_final}) becomes analytically
tractable.  The \teq{\theta_f} integration is almost trivial, and the
resulting compact analytic form, valid for \teq{\gamma_e\gg 1}, is
\begin{equation}
   \dot{\gamma}_e \; =\; -  \dover{n_s\, \sigt c}{2(\mu_+-\mu_-)}\, 
   \dover{B^3}{\gamma_e\beta_e\erg_s^2}  \; 
   \Bigl\lbrack {\cal I}^2_{\kappa}(\psi_+)- {\cal I}^2_{\kappa}(\psi_-) \Bigl\rbrack
   \quad ,\quad \psi_{\pm}\; =\; \dover{\omega_{\pm}}{B}
   \; =\; \dover{\gamma_e\erg_s}{B}\, (1+\beta_e\mu_{\pm})\quad ,
 \label{eq:cool_rate_Thom}
\end{equation}
where the pertinent integrals over \teq{\omega_i} lead to the definitions of
the functions
\begin{equation}
   {\cal I}^p_{\kappa}(\psi )\; =\; \int_1^{\psi}t^p\,\Sigma_{\kappa}(t)\, dt
   + c_p(\kappa )\quad .
 \label{eq:calIkappa_def}
\end{equation}
The constant terms \teq{c_p(\kappa )} are merely introduced to render 
the forms of the \teq{{\cal I}^p_{\kappa}(\psi )} more compact, and cancel 
out in all computations of \teq{\dot{\gamma}_e}.  The specific case
of interest here is \teq{p=2}, for which 
\begin{eqnarray}
{\cal I}^2_{\kappa}(\psi ) &=& \dover{2\psi^3}{3} +(6-\kappa^2)\psi -\dover{1}{1+\psi} 
   -4\log_e(1+\psi ) \nonumber\\[-5.5pt]
 \label{eq:calI2_eval}\\[-5.5pt]
   &+& 2(1-\kappa^2) \log_e\Bigl\lbrack (\psi -1)^2 + \kappa^2\Bigr\rbrack
   +\dover{1-6\kappa^2+\kappa^4}{\kappa}\arctan \Bigl(\dover{\psi -1}{\kappa} \Bigr)
   \quad , \nonumber
\end{eqnarray}
and
\begin{equation}
   c_2(\kappa )\;\equiv\; {\cal I}^2_{\kappa}(1)
   \; =\; \dover{37}{6} - \kappa^2 + 4\log_e\dover{\kappa}{2}
   -4\kappa^2\log_e\kappa\quad .
 \label{eq:c2_eval}
\end{equation}
The combination of Eq.~(\ref{eq:cool_rate_Thom}) and Eq.~(\ref{eq:calI2_eval}) 
agrees algebraically with the leading order (\teq{\gamma_e B\gg \erg_s}) 
contribution for the resonant Thomson cooling rate in Eq.~(22) of Dermer (1990).
Since \teq{\kappa\ll 1}, the mathematical character of \teq{{\cal I}^2_{\kappa}(\psi )}
is that of a step function smoothed on the scale of \teq{\kappa}.
Useful asymptotic forms for \teq{{\cal I}^2_{\kappa}(\psi )} are given by
\begin{equation}
   {\cal I}^2_{\kappa}(\psi )\;\approx\; \cases{
   -\dover{\pi}{2\kappa} \, \Bigl(1-6\kappa^2 \Bigr) -1
   +\dover{2}{5}\psi^5+O(\psi^7)+O(\kappa^2)\;\; , & 
       $\quad \psi\ll 1\;\;$,\cr
   \dover{\pi}{2\kappa} \, \Bigl(1-6\kappa^2 \Bigr) + \dover{2\psi^3}{3} +O(\psi ) 
   +O(\kappa^2)\;\; , & 
       $\quad 1\ll\psi\;\;$.  \cr}
 \label{eq:calI_asymp}
\end{equation}
Only these leading order terms are required for the results subsequently
listed. Cooling rates computed using the magnetic Thomson limit in
Eq.~(\ref{eq:cool_rate_Thom}) are displayed in
Fig.~\ref{fig:resThom_cool}, for both isotropic (\teq{\mu_{\pm}=\pm 1})
and hemispherical (\teq{\mu_-=0}, \teq{\mu_+=1}) monoenergetic soft
photon distributions.  In both cases, the resonant contribution appears
as a sharp ``wall'' signifying its onset at low \teq{\gamma_e},
resolving into a slowly-declining rate.  The hemispherical case is for
roughly head-on collisions between the soft photons and the relativistic
electrons.  In comparing the two panels of the Figure, it becomes
apparent that the removal of those soft photons that chase the electrons
kinematically curtails the resonant ``plateau'' at higher Lorentz
factors \teq{\gamma_e}, so that this feature becomes narrower: its width
is defined by the boundaries \teq{\psi_{\pm} =1}. Other asymptotic
domains for the cooling rate are not qualitatively influenced by such
soft photon collimation.

Various asymptotic cooling rates can now be obtained to establish
correspondence with extant results in the literature, and also to define
the characteristics of general cooling rate numerics in uniform fields. 
The most familiar of these is the non-magnetic cooling rate, which is
generally most salient for  \teq{B\ll 1}. In principal, or magnetic
scattering events this domain is realized when \teq{1\ll
\psi_-\leq\psi_+}, i.e. \teq{B/(\gamma_e\erg_s)\ll 1+\beta_e\mu_-}, for
which the resonance is never sampled kinematically. In practice, this
non-magnetic cooling situation often doesn't arise until \teq{\gamma_e}
is extremely large, or it may not arise at all.  For example, if
\teq{\mu_-\approx -1}, corresponding to electrons and soft photons
chasing each other along \teq{{\vec B}}, then \teq{1+\beta_e\mu_-\approx
1/(2\gamma_e^2)} so that \teq{1\ll \psi_-} cases are not attained in
highly-magnetized systems that satisfy \teq{B\gtrsim \erg_s}.  This
occurrence could be common in high-field pulsars.  However, a sufficient
condition for the realization of a non-magnetic cooling rate is that the
contribution of the resonance is dwarfed by non-resonant cooling. This
arises when \teq{\psi_- < 1\ll \psi_+}, specifically when
\teq{\psi_+^3\gg 3\pi/(4\kappa)}, and this domain is always accessible
for the cross section in Eq.~(\ref{eq:cross_sec_Thom}), regardless of
the value of \teq{\mu_-}.  Then, the second form in
Eq.~(\ref{eq:calI_asymp}) readily yields the classical Thomson cooling rate
\begin{equation}
   {\dot \gamma}_e\; \approx\; \gammadott\; =\; - n_s\, \sigt c\; \gamma_e^2\erg_s  \; 
   \dover{(1+\beta_e\mu_+)^3 - (1+\beta_e\mu_-)^3}{3 \beta_e\, (\mu_+ -\mu_-)} 
   \quad ,\quad \gamma_e\erg_s (1+\beta_e\mu_+) \;\gg\; 
   \biggl( \dover{3\pi B^2}{8\fsc} \biggr)^{1/3}  .
 \label{eq:gammadot_Thom}
\end{equation}
For isotropy with \teq{\mu_+=1} and \teq{\mu_-=-1}, the 
\teq{\beta_e}-dependent factor simply yields \teq{4/3} when \teq{\gamma_e\gg 1},
and this rate then corresponds to the well-known classical form in Eq.~(7.16a)
of Rybicki \& Lightman (1979).  It shall become apparent in 
Section~\ref{sec:cooling_gen} below that in strong fields, 
this regime will never be realized because Klein-Nishina reductions above 
the resonance will reduce the cross section below the Thomson scale, \teq{\sigt}.

As the field is increased, or equivalently \teq{\gamma_e} is reduced,
the phase space for resonant interaction becomes larger, so that domains
\teq{\psi_- < 1 < \psi_+} can be accessed.  For these, the cooling rate is
dominated by the contribution of the cyclotron resonance, and if 
\teq{\psi_-\lesssim 1\ll \psi_+}, with \teq{\psi_+^3\ll 3\pi/(4\kappa)}, then 
both the forms in Eq.~(\ref{eq:calI_asymp}) are employed in arriving 
at the resonant Thomson cooling rate:
\begin{equation}
   {\dot \gamma}_e\; \approx\; \gammadotres\; \equiv\; 
   - \dover{3\pi}{4\fsc}\, \dover{n_s\, \sigt c}{\mu_+ -\mu_-}
   \; \dover{B^2}{\gamma_e\erg_s^2}   \quad ,\quad 
   B \;\ll\; \gamma_e\erg_s (1+\beta_e\mu_+)\; \ll\; 
    \biggl( \dover{3\pi B^2}{8\fsc} \biggr)^{1/3} \quad .
 \label{eq:gammadot_res}
\end{equation}
This is exactly equivalent to the \teq{\gamma_e\gg 1} limit of Eq.~(24) 
of Dermer (1990) in the specific case of monoenergetic soft photons.  
If the ratio \teq{B/(\gamma_e\erg_s)} is increased further, the cooling 
again becomes non-resonant, but this time accessing an energy-dependent 
regime of the cross section, namely \teq{\sigt\propto\omega_i^2}.
This then introduces a stronger dependence on \teq{\gamma_e} than in 
Eq.~(\ref{eq:gammadot_Thom}), principally via the extra factor 
\teq{(\omega_i/B)^2\propto\gamma_e^2\erg_s^2/B^2} that appears 
because \teq{\psi_-\leq\psi_+\ll 1}.  Then
\begin{equation}
   {\dot \gamma}_e\; \approx\; - n_s\, \sigt c\; \dover{\gamma_e^4\erg_s^3}{B^2}  \; 
   \dover{(1+\beta_e\mu_+)^5 - (1+\beta_e\mu_-)^5}{5 \beta_e\, (\mu_+ -\mu_-)} 
   \quad ,\quad   \gamma_e\erg_s (1+\beta_e\mu_+)\; \ll\;  B\quad .
 \label{eq:gammadot_belowres}
\end{equation}
If \teq{B\lesssim \erg_s}, this third asymptotic regime is clearly never
realized. Moreover, because it critically depends on \teq{\omega_i^2}
cross section behavior for incident photons propagating along the field,
relatively small incident photon angles \teq{\theta_i} in the ERF will
modify it profoundly.  When \teq{\sin\theta_i \gtrsim \omega_i/B}, the
scattering cross section below the resonance is approximately
\teq{\sigt\sin^2\theta_i} (for incoming photons of \teq{\parallel}
polarization only; e.g. see Herold 1979), and since \teq{\theta_i\sim
1/\gamma_e}, at low Lorentz factors \teq{\gamma_e\lesssim (B/\erg_s
)^{1/2}} such non-zero ERF \teq{\theta_i} will cause the cooling rate to
saturate at a finite value \teq{{\dot \gamma}_e\sim -n_s\sigt c\,\erg_s}
(a suggestion of the onset of this appears in Figure 2 of Xia, et al.
1985). In other words, the curves in Fig.~\ref{fig:resThom_cool} will be
modified for mildly-relativistic electrons with \teq{\gamma_e\lesssim
3-5}, when taking into account this finite ERF incidence angle
\teq{\theta_i} effect. Finally, observe that as \teq{\gamma_e} decreases
from very high values, these three asymptotic regimes are accessed
sequentially in order from Eq.~(\ref{eq:gammadot_Thom}) to
Eq.~(\ref{eq:gammadot_belowres}).

It is important to note also that for the classical Thomson regime in
Eq.~(\ref{eq:gammadot_Thom}) to be effectively realized in practice, the
neutron star local magnetic field must be relatively low, considerably
lower than the field choices illustrated.  When \teq{B} approaches a
sizeable fraction of unity, the constraint on \teq{\gamma_e} pushes the
scattering into the Klein-Nishina domain \teq{\gamma_e\erg_s\sim 1},
wherein the simplifying assumptions concerning the cross section that
underpin this asymptotic result are strictly not valid.  In order to
realize a truly Thomson regime, one requires \teq{B^2/\fsc\ll 10^{-3}},
specifically \teq{B\lesssim 3\times 10^{-4}} (i.e. \teq{\lesssim
10^{10}}Gauss).  This will occur only at moderately high altitudes in
conventional pulsars, such as ten stellar radii above the surface, or
down near the surface of a millisecond pulsar.  For proximity of
collision locales to the polar cap of normal or high-field pulsars, the
Klein-Nishina reductions and recoil effects embodied in the full
magnetic scattering cross section must be included, thereby motivating
developments in the subsequent Sections, the focal thrust of this paper.

\section{ELECTRON COOLING RATES IN SUPERSTRONG FIELDS}
 \label{sec:cooling_gen}

While the cyclotron resonance to the scattering process is encapsulated
in quantum formulations, it is not an intrinsically quantum aspect: it
is obtained in classical descriptions (e.g. Canuto, Lodenquai and
Ruderman 1971) that introduce it via magnetic modifications to the
dispersion relation. Quantum effects such as Klein-Nishina declines and
electron recoil influences extend beyond the resonance and can be seen a
range of energies in the ERF.  Here fully quantum regimes where
\teq{B\gtrsim 0.1} are explored.  The focus is on surface emission at
the polar cap, where the photon angular distribution is uniform within a
cone whose axis coincides with the neutron star magnetic dipole axis;
the next Section will treat rates appropriate to arbitrary locales in
the magnetosphere. First we explore the particular characteristics of
high-field regimes where electron recoil and Klein-Nishina declines are
paramount.  Then we consider thermally-distributed soft photons, and
highlight how even a modest breadth in the range of \teq{\erg_s}, such
as that present in the Planck spectrum, profoundly alters the resonant
rates.  Then this Section will conclude with a brief exposition on the
mean energy loss in \teq{\gamma}-\teq{e} collisions that isolates the
field domain for which recoil is significant.

\subsection{Characteristics of High Field Regimes: Monoenergetic Soft Photons}
 \label{sec:highB_recoil}

\begin{figure*}[t]
\twofigureoutpdf{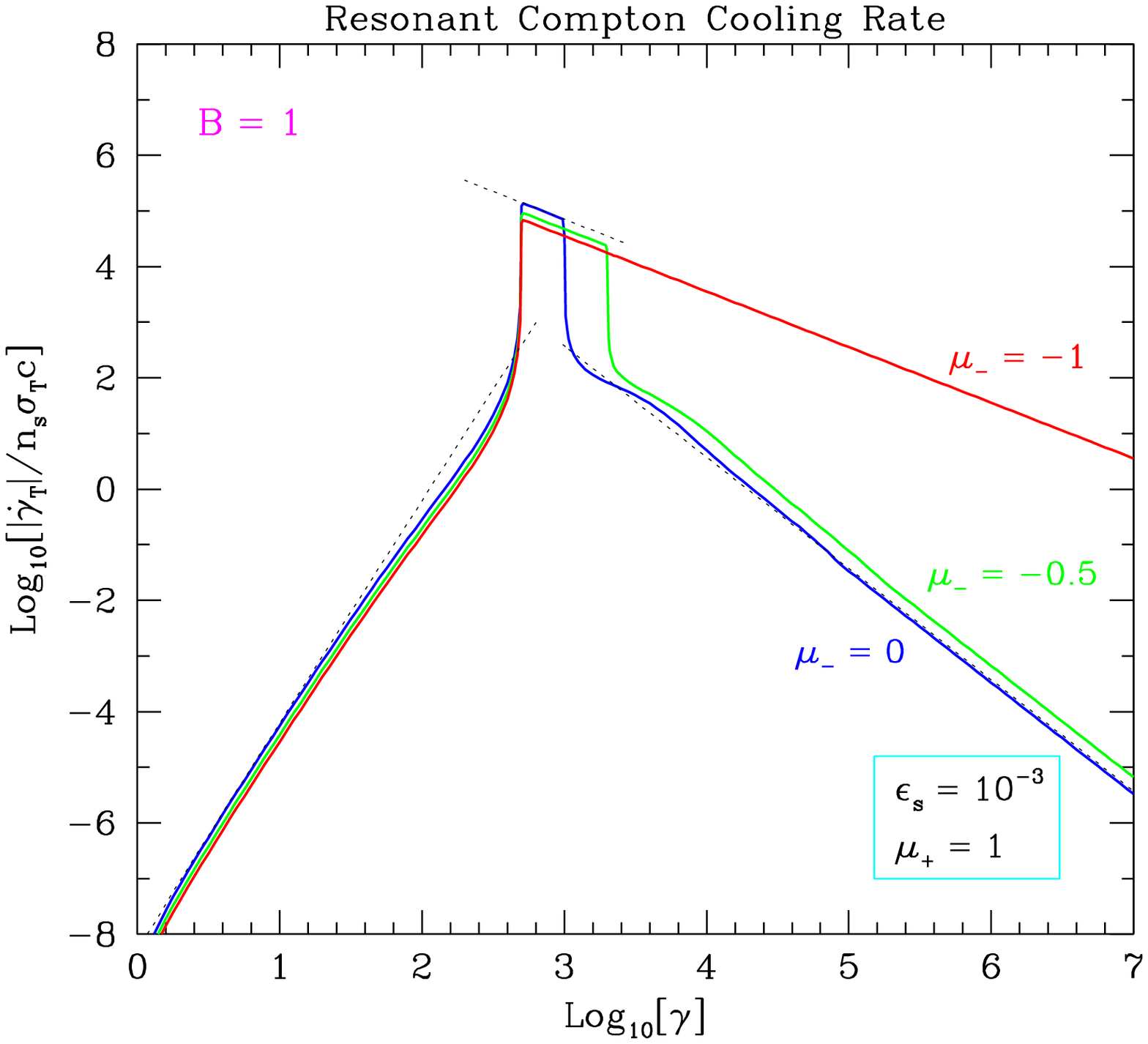}{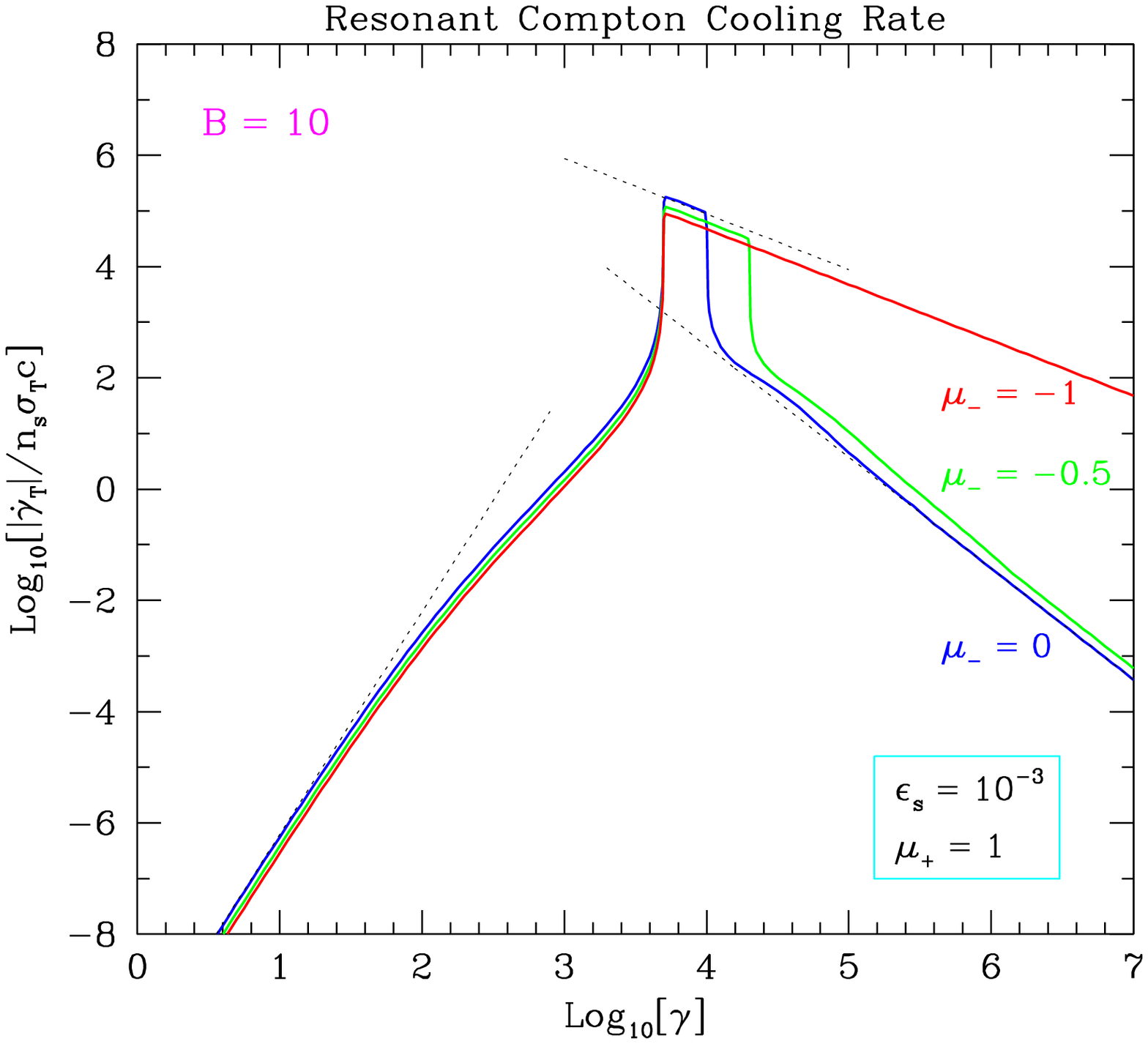}{
Resonant Compton cooling rates for two different field strengths,
\teq{B=1} (left panel) and \teq{B=10} (right panel), again in units of 
\teq{B_{\rm cr} \approx 4.413\times 10^{13}}Gauss.  These are computed
using the QED differential cross section in Eq.~(\ref{eq:sigma_unpol_JL}).  The
soft photon energy distribution is monoenergetic, with \teq{\erg_s=10^{-3}},
corresponding to typical surface X-ray temperatures.  In each panel, three cases 
are depicted: isotropic (\teq{\mu_-=1}) and hemispherical (\teq{\mu_-=0})
soft photons, and an intermediate case, \teq{\mu_-= -1/2}.  In both panels, the 
dotted lines encapsulate the asymptotic approximations listed in
in Eqs.~(\ref{eq:gammadot_belowres}),~(\ref{eq:gammadot_res_Bgg1}) 
and~(\ref{eq:gammadot_aboveres_Bgg1}) for the \teq{\mu_-=0} 
hemispherical soft photon case.
 \label{fig:resComp_cool} }      
\end{figure*}

Cooling rates discussed here are computed using the full formulation in
Eq.~(\ref{eq:cool_rate_final}).  These are double integrals in general,
though we anticipate analytic developments in
Subsection~\ref{sec:therm_esoft} below that will reduce one of these
integrations.  For now, specialization to monoenergetic soft photons is
retained.  For the purposes of illustration, the spin-averaged Johnson
\& Lippmann cross section in Eq.~(\ref{eq:sigma_unpol_JL}) is employed
in computing these rates.  The numerical results, obtained using {\tt
Mathematica} coding, are displayed in Fig.~\ref{fig:resComp_cool}, for
two local field strengths \teq{B=1, 10}.  When the rates are computed
using the spin-dependent Sokolov \& Ternov cross section, similar curve
morphology emerges.  The rates are obtained for soft photons distributed
uniformly within conical sectors with axes coincident with the {\bf B}
vector. All cases have \teq{\mu_+=1}, so that head-on collisions abound;
the curves therefore represent interactions at the magnetic pole near
(or just beneath) the stellar surface for electrons heading into the
neutron star. The cooling rate curves display a steeply-rising portion
below the resonance peak, a \teq{1/\gamma_e} dependence in the resonance
``plateau,' and steeply-declining section at \teq{\gamma_e} above the
resonance.  It is this high Lorentz factor regime that evinces
profoundly different character from that in the Thomson domain
exposition in Fig.~\ref{fig:resThom_cool}: the rapid decline reflects
the reduction of the cross section in the Klein-Nishina regime when
\teq{B\gtrsim 1} (see Gonthier et al. 2000).  Since the Thomson cross
section in Eq.~(\ref{eq:cross_sec_Thom}) does not possess such a
decline, the cooling rates illustrated in Fig.~\ref{fig:resThom_cool}
exhibit the classic \teq{\gamma_e^2} Thomson cooling rate at high
\teq{\gamma_e}. Therefore, when \teq{B\gtrsim 0.03}, the full
relativistic quantum cross section must be used.

It is insightful to generate analytic approximations pertinent to
Fig.~\ref{fig:resComp_cool}.  In supercritical fields, whenever
\teq{\omega_i\gg 1} the differential cross section in
Eq.~(\ref{eq:sigma_unpol_JL}) is extremely sensitive to the amount of
electron recoil in a scattering, by virtue of the exponential factor. 
Specifically, the focus here is on the \teq{\omega_i^2/B\gg 1} domain. 
This spawns a strong dependence of \teq{d\sigJL/d(\cos\theta_f)} on
\teq{\theta_f}, with the dominant contribution arising from narrow
collimation about forward scattering, \teq{\theta_f\approx 0}. The
collimation is controlled by the exponential factor so that outside the
range \teq{(1-\cos\theta_f )\lesssim 3 B/\omega_i^2}, corresponding to
recoil being either moderate or strong, the contributions to the
differential cross section and the total cooling rate are insignificant
when \teq{B\gg 1} and \teq{\omega_i\gg 1}. With this restriction,
\teq{\omega_f\approx\omega_i} and the kinematics are quasi-Thomson.  The
differential cross section in Eq.~(\ref{eq:sigma_unpol_JL}) accordingly
simplifies dramatically to
\begin{equation}
   \dover{d\sigJL}{d(\cos\theta_f )}\; \approx\;\dover{3\sigt}{8}\,
   \exp \Bigl\{ - \dover{\omega_i^2}{B}\, (1-\cos\theta_f )\Bigr\}\,
   \Sigma_{\kappa}\Bigl(\dover{\omega_i}{B}\Bigr) \quad ,
 \label{eq:cross_sec_Bgg1}
\end{equation}
using the definition of \teq{\Sigma_{\kappa}(\psi )} in 
Eq.~(\ref{eq:Sigma_kappa_def}). Furthermore, the Jacobian in 
Eq.~(\ref{eq:cool_rate_final}) again reduces to a simple form 
\teq{\vert \partial \erg_f / \partial (\cos\theta_f) \vert = \gamma_e\beta_e\omega_f
\approx \gamma_e\beta_e\omega_i},
so that the cooling rate in Eq.~(\ref{eq:cool_rate_final}) for \teq{B\gg 1} is analytically
tractable.  The dominant contribution from the relative velocity factor 
\teq{1-\beta_e\cos\theta_f = (1-\beta_e) + \beta_e (1-\cos\theta_f)} is from
the \teq{\beta_e (1-\cos\theta_f)} portion, since \teq{\beta_e} is so nearly unity.
Assuming a uniform beam of incoming photons such that \teq{f(\mu_i)=1}
on \teq{\mu_-\leq\mu_i\leq\mu_+}, the \teq{\theta_f} integration is simple, and 
the resulting compact result is
\begin{equation}
   \dot{\gamma}_e \; \approx \; -  \dover{3 n_s\, \sigt c}{8(\mu_+-\mu_-)}\, 
   \dover{B^3}{\gamma_e\beta_e\erg_s^2}  \; 
   \Bigl\lbrack {\cal I}^{-2}_{\kappa}(\psi_+)- {\cal I}^{-2}_{\kappa}(\psi_-) \Bigl\rbrack
   \quad ,\quad \psi_{\pm}
   \; =\; \dover{\gamma_e\erg_s}{B}\, (1+\beta_e\mu_{\pm})\quad .
 \label{eq:cool_rate_Bgg1}
\end{equation}
The definition of the \teq{{\cal I}^{-2}_{\kappa}} function is given in
Eq.~(\ref{eq:calIkappa_def}), and it can be expressed as 
\begin{equation}
{\cal I}^{-2}_{\kappa}(\psi ) \; =\;  -\dover{1}{1+\psi} 
    + \dover{1}{\kappa}\arctan \Bigl(\dover{\psi -1}{\kappa}\Bigr)
    \quad ,\quad c_0(\kappa )\;\equiv\; {\cal I}^0_{\kappa}(1)
   \; =\; - \dover{1}{2} \quad .
 \label{eq:calI-2_c-2_eval}
\end{equation}
Again, since generally \teq{\kappa\ll 1}, the mathematical character 
of \teq{{\cal I}^{-2}_{\kappa}(\psi )} is that of a step function smoothed 
on the scale of \teq{\kappa}.  Useful asymptotic forms for 
\teq{{\cal I}^{-2}_{\kappa}(\psi )} are given by
\begin{equation}
   {\cal I}^{-2}_{\kappa}(\psi )\;\approx\; \cases{
   -\dover{\pi}{2\kappa} - \dover{\kappa^2}{3}
   + \psi - \psi^2 + O(\psi^3)+O(\kappa^2)\;\; , & 
       $\quad \psi\ll 1\;\;$,\cr
   \dover{\pi}{2\kappa}  - \dover{1}{\psi} 
   + \dover{1}{\psi^2}
   +O\biggl( \dover{1}{\psi^3}\biggr)+O(\kappa^2)\;\; , & 
       $\quad 1\ll\psi\;\;$.  \cr}
 \label{eq:calI-2_asymp}
\end{equation}
Eq.~(\ref{eq:cool_rate_Bgg1}) formally applies only to \teq{B\gg 1} 
regimes.  An alternative formulation that addresses arbitrary field 
strengths is offered in Section~\ref{sec:therm_esoft} below.

As with the Thomson regime considerations, there are three main
asymptotic domains of potential interest here.  At low Lorentz factors,
specifically when \teq{\gamma_e\erg_s\ll 1}, the relevant analytic
approximation is that in Eq.~(\ref{eq:gammadot_belowres}), so that
\teq{{\dot \gamma}_e\propto \gamma_e^4}.  This form is depicted in both
panels of the Figure, and again follows from the \teq{\sigma\propto
\omega_i^2} dependence of the cross section.  When \teq{\gamma_e\erg_s}
exceeds unity, but still is below \teq{B}, Klein-Nishina reductions
emerge below the resonance.  This regime is evident for the \teq{B=10}
case in Figure~\ref{fig:resComp_cool}. The resonant cooling rate for
highly supercritical fields is:
\begin{equation}
   {\dot \gamma}_e\; \approx\; 
   - \dover{3\pi}{4}\, \dover{n_s\, \sigt c}{\mu_+ -\mu_-}
   \; \dover{B^4}{\gamma_e\beta_e\erg_s^2} \;\dover{1}{\Gamma}  \; =\; 
   - \dover{9\pi}{16\fsc}\, \dover{n_s\, \sigt c}{\mu_+ -\mu_-}
   \; \dover{B^2}{\gamma_e\beta_e\erg_s^2}   \quad ,\quad 
 \label{eq:gammadot_res_Bgg1}
\end{equation}
if we use the low-field cyclotron decay width.  Well above the resonance,
recoil influences are profound, Klein-Nishina reductions are dramatic, and
the cooling rate is
\begin{equation}
   {\dot \gamma}_e\; \approx\; 
   - \dover{3}{4}\, \dover{n_s\, \sigt c}{(1+\beta_e\mu_+)\, (1 +\beta_e\mu_-)}
   \; \dover{B^2}{\gamma_e^2\erg_s^3}   \quad ,\quad 
   1\;\ll\; B\;\ll\; \gamma_e\erg_s (1+\beta_e\mu_-)\quad ,
 \label{eq:gammadot_aboveres_Bgg1}
\end{equation}
where the \teq{\psi\gg 1} limit of Eq.~(\ref{eq:calI-2_asymp}) has been
invoked. This limit is depicted in the Figure. Clearly, for isotropic
soft photons with \teq{\mu_-=-1}, it is never realized since the
resonance can always be accessed kinematically for large \teq{\gamma_e},
so that the resonant contribution dominates all others.  It will become
evident below, that distributing the soft photon energies can also
circumvent the appearance of a regime exhibiting this profoundly reduced
cooling rate.

\subsection{Thermal Soft Photons}
 \label{sec:therm_esoft}

Contributions to pulsar and magnetar spectra in the X-ray band from the
neutron star surface are necessarily approximately thermal, due to the
high radiative opacities in their atmospheres.  The spectrum is not
expected to be of a perfect Planck form because there are radial and
colatitudinal temperature gradients; line processes can be involved in
its formation, and therefore Kirchhoff's Third Law seeds the expectation
that absorption lines can be present.  A plethora of theoretical models
of neutron star atmospheres exist, differing in assumptions about
composition, ionization balance, and active radiative transport
processes.  Generally, they now predict non-Planckian forms with at most
modest spectral structure due to the smearing of features by spatial
distribution or vacuum polarization effects (e.g., see Bulik \& Miller
1997; \"Ozel 2001; Ho et al. 2007, and references therein).  Discrete
lines have not been unambiguously observed below 10 keV, but there are
indications of deviations from isothermal Planck forms.  For example,
state-of-the-art spectroscopy of some neutron stars, like RX
J1856.5Ð3754, in the Chandra/XMM Newton era suggests two-component
blackbody models as the best fits to data (e.g., see Burwitz et al.
2003).  In addition, radiative transfer and heat conduction in surface
layers is heavily suppressed perpendicular to the field, so the photon
distribution must be decidedly anisotropic, as discussed by Zavlin,
Shibanov \& Pavlov (1995), P\'erez-Azor'n, Miralles \& Pons (2005) and
Mori, Ho \&  Wynn (2007).  This anisotropy, and departures from Planck
spectral form for the soft photons are higher order influences on
resonant Compton cooling rates.  The focus here is the exploration of
the major features of distributing the soft photons in energy beyond the
delta function approximation in Eq.~(\ref{eq:ndef}). To this end, for
interaction locales on the magnetic polar axis, the Planck spectral form 
\begin{equation}
   n_{\gamma}(\epsilon_s) \; =\; \dover{\Omega_s}{\pi^2 \lambar^3} 
      \, \dover{\epsilon_s^2}{e^{\epsilon_s/\Theta} -1}
 \label{eq:Planck_spec}
\end{equation}
is employed in the computation of Eq.~(\ref{eq:cool_rate_final}) for the
Compton cooling rate.  Here, \teq{\Theta =kT/m_ec^2} is the
dimensionless temperature of the thermal surface photons, and
\teq{\lambar = \hbar/(m_ec)} is the Compton wavelength over \teq{2\pi}. 
Also, \teq{\Omega_s} represents the solid angle of the blackbody photon
population, divided by \teq{4\pi}.  This fractional solid angle is
introduced to accommodate anisotropic soft photon cases, in particular
hemispherical populations (\teq{\Omega_s=1/2}) just above the stellar
atmosphere. The total number density of soft photons is therefore \teq{2
\Omega_s\zeta (3) /(\pi^2 \lambar^3)}, for \teq{\zeta (n)} being the
Riemann {\it zeta} function.  The analysis of this Section will be
restricted to interaction points on the magnetic axis, for which the
angular distribution is flat-topped and within a range specified in Eq.
(\ref{eq:cone_angle}).  In such axisymmetric cases, the solid angle
factor \teq{\Omega_s} is neatly separable from the \teq{\omega_i}
integration. As will soon be apparent, the spreading of seed photon
energies profoundly alters the morphology of the cooling rate curves at
high Lorentz factors \teq{\gamma_e}, a feature that is apparent in magnetic 
Thomson cooling rate expositions (e.g. Dermer 1990; Sturner 1995).

The triple integration in Eq.~(\ref{eq:cool_rate_final}) for photons
with a Planck spectrum is computationally expensive.  Below the
resonance, the integrands are relatively stable and {\tt Mathematica}
was used without undue burden on the integration time.  At the high
Lorentz factors where the resonance is generally sampled, numerical
evaluations of the triple integration are computationally inefficient,
and asymptotic approximations are desirable.  The dominance of the
resonance in determining the rate does in fact lead to analytic
simplification of the integrations, so that two of them can be rendered
tractable, thereby leaving a single numerical integration. Retaining the
full mathematical structure of the Lorentz profile limits the
simplification: if one imposes just a \teq{\gamma_e\gg 1} approximation,
then the cooling rates distill to double integrations.  In either case,
the non-resonant term in the cross section is neglected as a small
contribution whenever the cyclotron resonance is sampled.  The analysis
for these developments is presented in Appendix A; here an outline of
the results is presented.  If one takes the \teq{\gamma_e\gg 1} limit,
the Jacobian factor in Eq.~(\ref{eq:cool_rate_final}) simplifies
according to Eq.~(\ref{eq:dergf_dcthetaf_approx}).  Then a change of
variables for the angle integration is employed.  Consider first the
Johnson \& Lippmann resonant cross section, which constitutes the first
piece in Eq.~(\ref{eq:sigma_unpol_JL}).  For monoenergetic soft photons,
these manipulations yield a JL cooling rate in and near the resonance of
\begin{equation}
   \gammadotJL \; \approx\; - \dover{3}{4}\, \dover{n_s\, \sigt c}{\mu_+-\mu_-}\, 
   \dover{1}{\gamma_e\erg_s^2}   \int_{\omega_-}^{\omega_+}
   \dover{{\cal F}\left(z_{\omega},\, p\right) - {\cal G}\left(z_{\omega},\, p\right)}{
   (\omega_i -B)^2 + (\Gamma /2)^2} 
   \, \dover{\omega_i^4\, d\omega_i}{(1+2\omega_i)^2} 
   \quad ,\quad z_{\omega}\; =\; 1 + \dover{1}{\omega_i}\quad ,
 \label{eq:cool_rate_genJL2}
\end{equation}
for \teq{p=\omega_i/B\approx 1} parametrizing how \teq{\omega_i} maps 
through the peak and wings of the resonance.  As before, \teq{\omega_{\pm} =
\gamma_e\erg_s\, (1+\beta_e\mu_{\pm})} defines the kinematic extrema
of incoming photon energies in the ERF.  The functions \teq{{\cal F}} and \teq{{\cal G}} 
express the integration over \teq{\cos\theta_f}, and take the functional forms
\begin{equation}
   {\cal F}(z,\, p)\; =\; \int_0^{\Phi (z)} \dover{ f(z,\, \phi ) \, e^{ -p\, \phi}\, d\phi }{
         \sqrt{1-2 z \phi + \phi^2}} 
   \quad ,\quad
   {\cal G}(z,\, p)\; =\; \dover{2 (z-1)}{(1+z)^2}
        \int_0^{\Phi (z)} \dover{ g(z,\, \phi ) \, e^{ -p\, \phi}\, d\phi }{ (1-\phi )
         \sqrt{1-2 z \phi + \phi^2}} \quad ,
 \label{eq:calFcalG_zpdef}
\end{equation}
for \teq{f(z,\, \phi )} and \teq{g(z,\, \phi )} being polynomial functions of \teq{z}
and \teq{\phi} as defined in Eqs.~(\ref{eq:calFzpdef}) and~(\ref{eq:calGzpdef}),
respectively.  The incoming photon energy parameter is conveniently expressed via
\begin{equation}
   \Phi (z)\; =\; z-\sqrt{z^2-1} 
          \quad \;\Leftrightarrow\; \quad
   z\; =\; \dover{1}{2} \biggl( \Phi + \dover{1}{\Phi} \biggr) 
 \label{eq:zPhi_rel}
\end{equation}
thereby defining the upper limit to the integrals in Eq.~(\ref{eq:calFcalG_zpdef}),
with \teq{0\leq \Phi (z) < 1} for \teq{z\geq 1}.  Both \teq{{\cal F}} and \teq{{\cal G}} can 
be approximated by analytic functions to the required accuracy, as described in 
Appendix B, sparing the user of a numerical evaluation of the integrals.  Series expansions 
of the integrations in terms of Legendre functions are possible, but do not significantly
expedite the evaluations.  A similar form for the resonant rates is obtained for the 
general Sokolov \& Ternov cooling rate in and near the resonance:
\begin{equation}
   \gammadotST \; \approx\; - \dover{3}{16}\, \dover{n_s\, \sigt c}{\mu_+-\mu_-}\, 
   \dover{1}{\gamma_e\erg_s^2}
   \sum_{s=\pm 1}  \int_{\omega_-}^{\omega_+}
   \dover{(s\eperp + 1)\, \Upsilon_s \left(z_{\omega},\, p\right)}{
   (\omega_i -B)^2 + (\Gamma_s /2)^2} 
   \, \dover{\omega_i^4\, d\omega_i}{(1+2\omega_i)^2} 
   \;\; ,\quad z_{\omega}\; =\; 1 + \dover{1}{\omega_i}\quad ,
 \label{eq:cool_rate_genST1}
\end{equation}
for \teq{p=\omega_i/B\approx 1}, and
\begin{equation}
   \Upsilon_s \left(z,\, p\right)\; =\; \left\{
   \left( \dover{s}{\eperp} +1 \right)\, \Bigl\lbrack{\cal F}(z,\, p) - {\cal G}(z, \, p) \Bigr\rbrack
    + \left( \dover{s}{\eperp} -1 \right) \, \Bigl\lbrack {\calFST}(z, \, p) 
                 - {\calGST}(z, \, p) \Bigr\rbrack \right\}\quad .
 \label{eq:Upsilon_def}
\end{equation}
The label \teq{s=\pm 1} refers to the spin state quantum number 
for the intermediate electron.  The \teq{\calFST} and \teq{\calGST} 
functions are specified by
\begin{eqnarray}
   \calFST (z,\, p) & = & (z+1)
       \int_0^{\Phi (z)} \dover{\phi (1-\phi)\, e^{-p\,\phi}}{\sqrt{1-2 z\phi + \phi^2}}\, d\phi\nonumber\\[-5.5pt]
 \label{eq:calFG_ST_def}\\[-5.5pt]
   \calGST (z, \, p) & = & 
       2\, \dover{z-1}{z+1} \int_0^{\Phi (z)} 
       \dover{\phi^2\, (z+2\phi z - \phi^2)\, e^{-p\,\phi}}{(1-\phi )\, \sqrt{1-2 z\phi + \phi^2}}\, d\phi\quad ,\nonumber
\end{eqnarray}
and, like \teq{{\cal F}} and \teq{{\cal G}}, can be approximated by 
analytic functions (detailed in Appendix B).

Both Eqs.~(\ref{eq:cool_rate_genJL2}) and~(\ref{eq:cool_rate_genST1})
represent single integrations within this construct.  For the cases treated here of 
interaction points on the magnetic axis, the additional integration over the 
Planck spectrum in Eq.~(\ref{eq:Planck_spec}) can be handled in an analytic manner 
by reversing the order of the resulting double integral, and using the definition/identity
\begin{equation}
    \ell_T(\chi ) \; =\;  \int_{\chi}^{\infty} \dover{dx}{e^x -1} \; \equiv\;    \log_e \dover{1}{1- e^{-\chi}} \quad .
 \label{eq:ellT_def}
\end{equation}
With this re-ordering, the \teq{\omega_i} integrations for the two rates are semi-infinite,
i.e. on the interval \teq{0\leq\omega_i\leq\infty}, but the \teq{\erg_s} integration is now 
over a finite interval, defined by 
\begin{equation}
   \chi_-\;\leq\; \chi\;\equiv\; \dover{\erg_s}{\Theta}\;\leq\;\chi_+
   \quad ,\quad
   \chi_{\pm}\; =\; \dover{B}{\gamma_e\Theta\, (1+\beta_e\mu_{\mp})}\quad .
 \label{eq:chi_pm_def}
\end{equation}
One can then quickly write down the result for the neighborhood of the resonance 
in the ST formulation as
\begin{equation}
   \gammadotST \; \approx\; - \dover{3\,\Omega_s}{16}\, 
   \dover{n_s \sigt c}{\gamma_e} \,\Bigl\{ \ell_T(\chi_+) - \ell_T(\chi_-) \Bigr\}
   \sum_{s=\pm 1}  \int_0^{\infty}
   \dover{(s\eperp + 1)\, \Upsilon_s \left(z_{\omega},\, p\right)}{
   (\omega_i -B)^2 + (\Gamma_s /2)^2} 
   \, \dover{\omega_i^4\, d\omega_i}{(1+2\omega_i)^2} \quad ,
 \label{eq:cool_rate_genST_th}
\end{equation}
with a similar factor \teq{\ell_T(\chi_+) - \ell_T(\chi_-)} appearing in
the equivalent JL formulation.  Generally, this factor has approximately
logarithmic dependence on \teq{\gamma_e} and is of the order of unity
for \teq{\chi_-\lesssim 1}, however when \teq{\chi_-\gg 1} and the
angular phase space for access to resonant interactions samples the high
energy tail of the Planck spectrum, the resonant rate is exponentially
suppressed.  Such compact forms were used to compute asymptotic results
for the cooling rates in the resonant domains for the graphical
illustrations in the remainder of this paper.

One remaining analytic refinement is addressed before proceeding to the
display of numerical results.  For all field strengths, the width of the
resonance is comparatively small, in the sense that \teq{\Gamma\ll B}, a
statement that applies to the spin-averaged width in
Eq.~(\ref{eq:width_effect}), but also extends to spin-dependent contexts
when Eq.~(\ref{eq:ST_widths}) is employed.  In such cases, the Lorentz
profile in Eqs.~(\ref{eq:cool_rate_genJL2})
and~(\ref{eq:cool_rate_genST1}) can be approximated by a delta function
in \teq{\omega_i} space of identical normalization:
\begin{equation}
   \dover{1}{(\omega_i -B)^2 + (\Gamma/2)^2}\;\to\;
   \dover{2\pi}{\Gamma}\, \delta (\omega_i-B)\quad .
 \label{eq:resonance_approx1}
\end{equation}
This then trivially evaluates the \teq{\omega_i} integration.  When also 
integrated over the Planck spectrum, the asymptotic expression for the 
resonant Sokolov and Ternov rate, for monoenergetic electrons is
\begin{equation}
   \gammadotST \; \approx\; - \dover{3\, \Omega_s}{4\pi}\, \dover{\sigt c}{\lambar^3}\, 
   \dover{\Theta}{\gamma_e\Gamma}  \, \calRST (B)\, 
   \Bigl\{ \ell_T(\chi_+) - \ell_T(\chi_-) \Bigr\} \quad ,
   \quad \gamma_e\Theta\; \gtrsim\; B\quad ,
 \label{eq:cool_rate_ST_restot_th}
\end{equation}
using the function
\begin{equation}
   \calRST (B)\; =\;  \dover{B^2}{(1+z)^2}\, \Bigl\{  \left\lbrack {\cal F}(z) - {\cal G}(z) \right\rbrack
    +  \left\lbrack {\calFST}(z) - {\calGST}(z) \right\rbrack  \Bigr\}
    \;\approx\; \cases{ \dover{2B^4}{3}\quad &, $\;\;B\,\ll\, 1$,\cr
    B^2 \left( 1 - \dover{2}{e} \right) \quad &, $\;\;B\,\gg\, 1$,\cr}
 \label{eq:calRST_def}
\end{equation}
for \teq{z=1+1/B}, to define the major portion of the magnetic field
dependence of the integrations over the resonance.  Here \teq{{\cal
F}(z)\equiv {\cal F}(z,\, 1)} represents the \teq{p=\omega_i/B=1}
specialization, and similarly for \teq{{\cal G}}, \teq{{\cal
F}_{\Delta}} and \teq{{\cal G}_{\Delta}}. Observe that when
\teq{\mu_-=0}, as is the case for outgoing electrons at the neutron star
surface, \teq{\ell_T(\chi_+)\to 0}. The equivalent result to
Eq.~(\ref{eq:cool_rate_ST_restot_th}) for the JL formulation is simply
obtained by replacing \teq{\calRST (B)} by the function \teq{\calRJL
(B)} that is given by
\begin{equation}
   \calRJL (B)\; =\;  \dover{2B^2}{(1+z)^2}\, 
      \Bigl\{  \left\lbrack {\cal F}(z) - {\cal G}(z) \right\rbrack \Bigr\}
    \;\approx\; \cases{ \dover{4B^4}{3}\quad &, $\;\;B\,\ll\, 1$,\cr
    B^2 \left( 1 - \dover{2}{e} \right) \quad &, $\;\;B\,\gg\, 1$,\cr}
 \label{eq:calRJL_def}
\end{equation}
The Johnson \& Lippmann \teq{B\ll 1} limit, when substituted into
Eq.~(\ref{eq:cool_rate_ST_restot_th}), reproduces exactly the thermal
cooling rate in Eq.~(54) of Dermer (1990) for choices of
\teq{\Omega_s=1/2}, namely hemispherical soft photons at the surface,
and \teq{\Gamma = 4\fsc B^2/3}, the standard classical cyclotron decay
width. Comparing Eqs.~(\ref{eq:calRST_def}) and~(\ref{eq:calRJL_def}),
the ratio of the resonant cooling rates in the two formulations is just
described by the simple function \teq{\calRST (B)/\calRJL (B)} of the
magnetic field strength:
\begin{equation}
   \dover{\calRST (B)}{\calRJL (B)} \; =\; \dover{1}{2} \left\{ 
   1 + \dover{{\calFST}(z) - {\calGST}(z)}{{\cal F}(z) - {\cal G}(z)} \right\}
   \quad ,\quad z\; =\; 1 + \dover{1}{B}\quad .
 \label{eq:calRST_calRJL_rat}
\end{equation}
This function ranges from \teq{1/2} in the magnetic Thomson domain,
where \teq{\Gamma_{+1}\ll\Gamma_{-1}} in Eq.~(\ref{eq:ST_widths}), to
unity in the ultra-quantum, \teq{B\gg 1} regime, where the choice of
wavefunctions is immaterial since the spin-dependent widths
\teq{\Gamma_{\pm 1}} collapse to the spin-averaged one \teq{\Gamma}. 
Note that the resonant cooling rates scale roughly as
\teq{\Theta/\gamma_e}, modulo the logarithmic factors.  If the electron
Lorentz factor drops too low, the leading order mildly-relativistic
correction is of the order of \teq{1/\gamma_e^2}; see for example,
Eq.~(35) of Harding \& Muslimov (1998).  The derived asymptotic rates at
resonance (i.e. Eq.~(\ref{eq:cool_rate_ST_restot_th}) for the ST
formulation, and its JL equivalent) provide useful checks on the ensuing
numerical evaluations, and expressions that can be used with facility in
resonant Compton cooling models for X-ray and gamma-ray emission in
neutron star systems.

\begin{figure*}[t]
\twofigureoutpdf{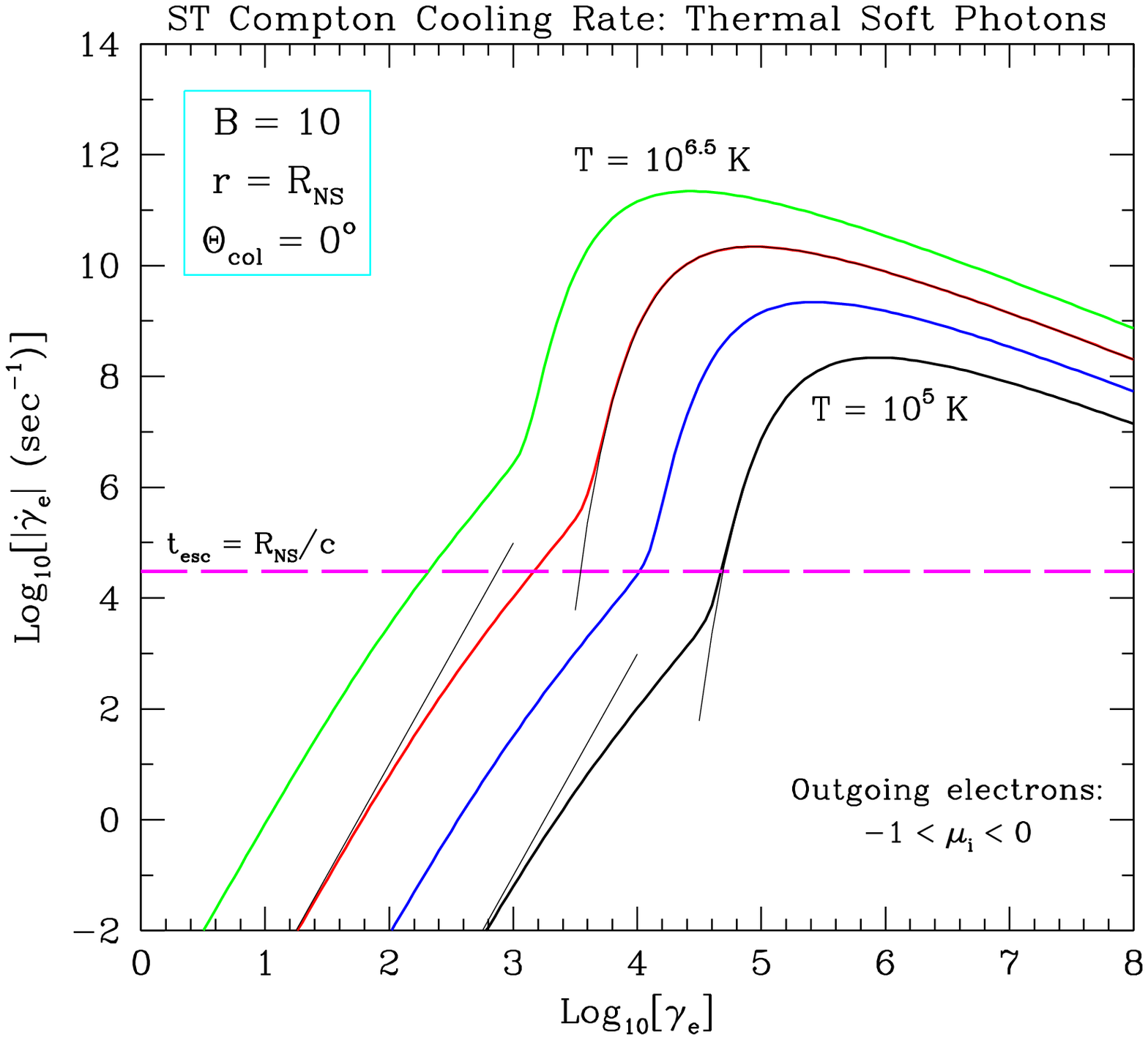}{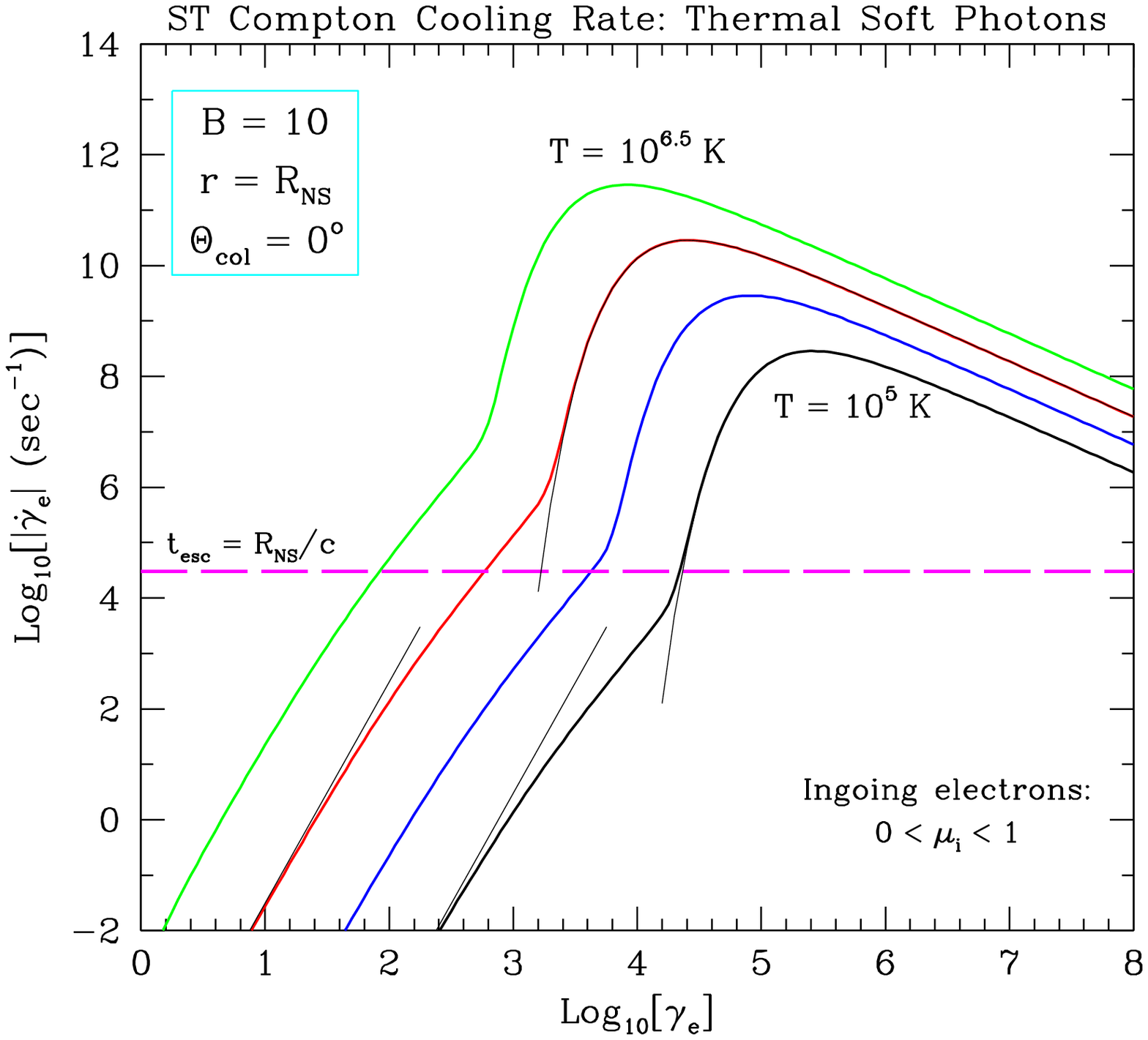}{
Resonant Compton cooling rates for field strength \teq{B=10}, in units of 
\teq{B_{\rm cr} \approx 4.413\times 10^{13}}Gauss, for interaction locales at the 
surface (\teq{r=\rns}) and the magnetic pole (\teq{\Thetacol =0^{\circ}}) of the neutron star.  
These are computed using the Sokolov \& Ternov differential cross section in 
Eq.~(\ref{eq:sigma_unpol_STapprox}) in the resonance, and the 
spin-averaged Johnson \& Lippmann form in Eq.~(\ref{eq:sigma_unpol_JL}) at all 
other ERF energies \teq{\omega_i}.  
The two cases depicted are for outgoing electrons (upward; left panel) and ingoing 
(downward; right panel) electrons.
The soft photon energy distribution is thermal, with surface X-ray temperatures as labelled, 
and is isotropic within a hemisphere.  In both panels, the 
lightweight lines encapsulate the asymptotic approximations for the resonant and 
non-resonant regimes, embodied
in Eqs.~(\ref{eq:cool_rate_ST_restot_th}) and~(\ref{eq:gammadot_belowres_th}),
respectively.  The horizontal dashed line denotes the light escape timescale 
\teq{\tesc} corresponding to a stellar radius.
 \label{fig:resComp_th_ST_B10} }      
\end{figure*}

\newpage

\begin{figure*}[t]
\figureoutpdf{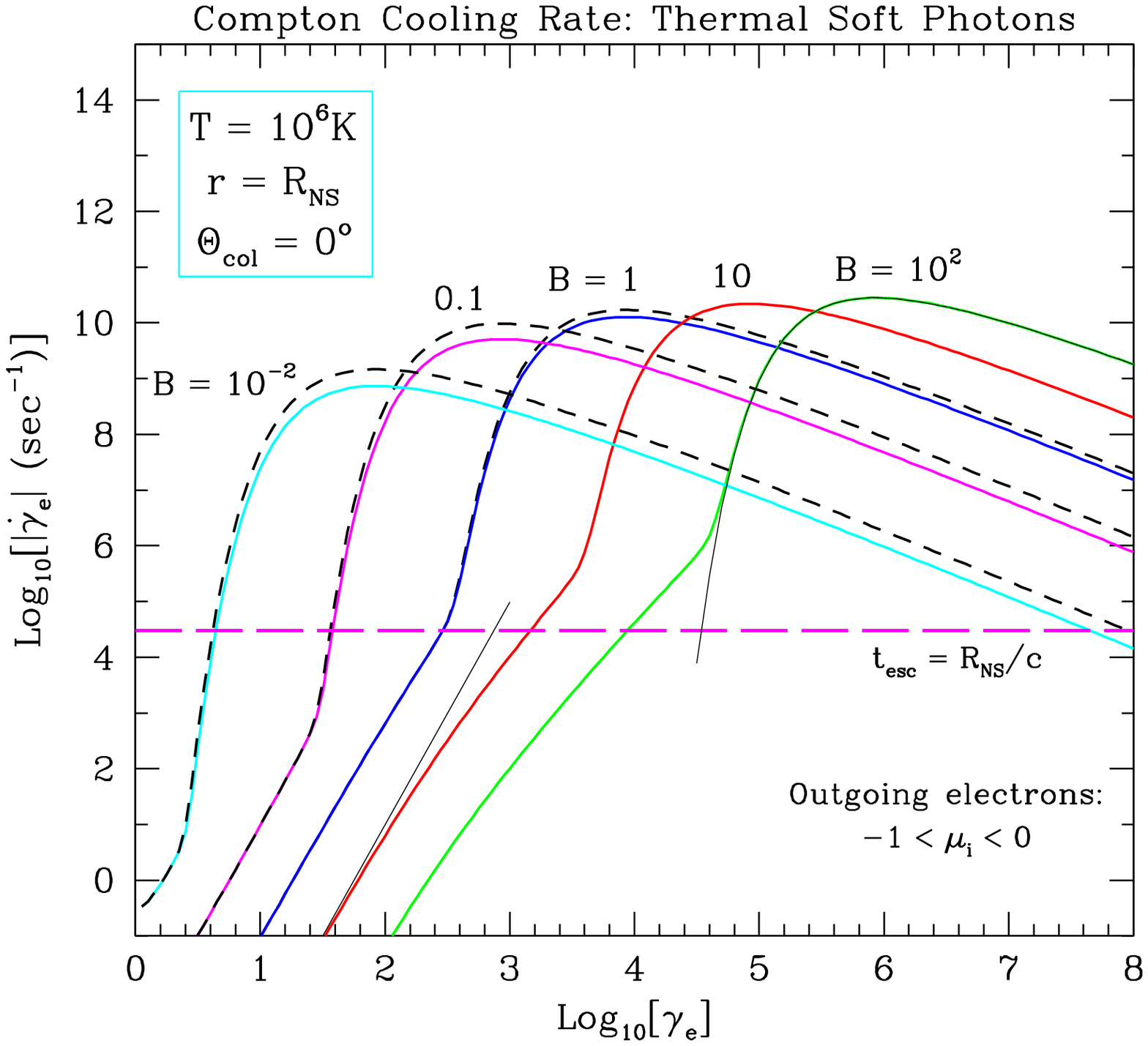}{3.8}{0.0}{-0.2}{
Resonant Compton cooling rates for five different field strengths, as
labelled, and in units of \teq{B_{\rm cr} \approx 4.413\times
10^{13}}Gauss. All curves are for surface thermal temperatures
\teq{T=10^6}K, for scatterings at the polar surface with colatitude
\teq{\Thetacol =0^\circ}, and for outgoing (upward moving) electrons. 
Solid curves are for Sokolov \& Ternov (ST) cases, and the dashed curves
for \teq{B=10^{-2}, 0.1, 1} are Johnson \& Lippman (JL) determinations.
The \teq{B=10,100}, JL evaluations are essentially indistinguishable on
this plotting scale from their ST counterparts, and so were not
exhibited.  The light solid curve in the resonant regime is the
\teq{B=100} asymptotic ST result in
Eq.~(\ref{eq:cool_rate_ST_restot_th}). The light solid straight line for
the low \teq{\gamma_e}, non-resonant regime is the \teq{B=10} asymptotic
result in Eq.~(\ref{eq:gammadot_belowres_th}). As in
Fig.~\ref{fig:resComp_th_ST_B10}, the horizontal dashed line denotes the
light escape timescale \teq{\tesc} corresponding to a stellar radius.
 \label{fig:resComp_th_T6_Bvar} }      
\end{figure*}

It is also possible to derive asymptotic approximations to the rate in 
the non-resonant regime at much lower Lorentz factors. By integrating 
Eq.~(\ref{eq:gammadot_belowres}) over the Planck spectrum, in the 
low \teq{\gamma_e} quasi-Thomson regime, all values of \teq{\erg_s} 
are sampled, and an analytic approximation is obtainable:
\begin{equation}
   {\dot \gamma}_e\; \approx\; - \dover{8\pi^4\Omega_s}{315}\, \dover{ \sigt c}{\lambar^3} \; \dover{\gamma_e^4}{\beta_e}
   \, \dover{\Theta^6}{B^2}  \; \Bigl\{ (1+\beta_e\mu_+)^5 - (1+\beta_e\mu_-)^5 \Bigr\} 
   \quad ,\quad   \gamma_e\Theta (1+\beta_e\mu_+)\; \ll\;  B\quad .
 \label{eq:gammadot_belowres_th}
\end{equation}
This result is applicable to both JL and ST formulations, since the
differential cross section is independent of electron spin-state choice
outside the resonance. It coincides precisely with the result one would
generate by taking the leading order, \teq{\gamma_e^4}, term of Eq.~(20)
of Dermer (1990) and inserting the Planck spectrum of
Eq.~(\ref{eq:Planck_spec}) therein, for any choice of soft photon solid
angle \teq{\Omega_s}.  For the on-magnetic axis illustrations in
Figs.~\ref{fig:resComp_th_ST_B10} and~\ref{fig:resComp_th_T6_Bvar}, the
hemispherical choice \teq{\Omega_s=1/2} is adopted, corresponding to
near-surface electon-photon collisions. It should be emphasized that
when \teq{\gamma_e\lesssim 10}, this formula should be corrected using
more general forms of the differential cross section than are employed
here, namely those applying to non-zero incoming photon angles
\teq{\theta_i} in the ERF.  The mathemematical involvement of such
generalizations is substantial (e.g. see Daugherty \& Harding 1986;
Bussard, Alexander \& M\'esz\'aros 1986), and are beyond the scope of
the current work, whose focus is mainly on higher Lorentz factor,
resonant domains.

Cooling rates for soft photons with the Planck distribution are
exhibited in Fig.~\ref{fig:resComp_th_ST_B10}.  These constitute surface
polar collisions for the two cases of outward-propagating (left panel)
and inward-moving (right panel) electrons.  While the outgoing case
mimics the situation invoked on models of conventional gamma-ray
pulsars, returning pair currents in such pulsars and twisted
magnetosphere models for magnetars include the possibility of
inward-propagating electrons. Of course, the Doppler beaming of
scattered photon angles guarantees that such downward electron cases
spawn upscattered photons that mostly impact and thereby heat the
neutron star surface, if the interaction altitude is low.  The
morphology of the cooling rate curves in the Figure is very similar for
the two electron directions, with modest numerical differences incurred
due to the differing angular phase space in scatterings for the two
cases.  Notably, greater logarithmic curvature is evident in the
resonant domain for outgoing electrons since then \teq{\chi_+\gg\chi_-}.
 The key feature of these plots is that resonant interaction parameter
space extends now to arbitrarily high Lorentz factors, contrasting the
strong drops at the upper end of the ``resonant plateaux''  evinced in
the monoenergetic soft photon cases.  This extension of the resonant
contribution, also realized in the magnetic Thomson studies of Dermer
(1990), Sturner (1995) and Harding \& Muslimov (1998), is borne in the
fact that as \teq{\gamma_e} increases, there is always a low soft photon
energy \teq{\erg_s} available to access the cyclotron resonance at
\teq{\omega_i=B}.  Therefore the shapes of the cooling curves in the
resonant contribution serve as inverted images of the Planck spectrum:
the exponential tail is sampled in the sharp rises on the left, the peak
being controlled by the \teq{\erg_s\sim \Theta = kT/(m_ec^2)} regime,
and the high \teq{\gamma_e} portion where \teq{{\dot \gamma}_e \propto
1/\gamma_e} samples the Rayleigh-Jeans tail of the Planck distribution.

For each surface temperature \teq{T}, the cooling curves depicted were
calculated using the Sokolov \& Ternov differential cross section in
Eq.~(\ref{eq:sigma_unpol_STapprox}) in the resonance, with the
spin-averaged Johnson \& Lippmann form in Eq.~(\ref{eq:sigma_unpol_JL})
being employed at all other ERF energies \teq{\omega_i}.  The numerical
results were obtained as full two-dimensional integrations based on
Eq.~(\ref{eq:cool_rate_final}), but employing an integration by parts
over the \teq{\erg_s} variable so as to eliminate the \teq{\omega_i}
integration, and the weaken sensitivity of the integrand to
\teq{\erg_s}.  This computational protocol can be simply adopted only
for cases where \teq{f(\mu_i)} is constant over the integration range;
hemispherical angular distributions satisfy this constraint. At this
high field strength of \teq{B=10}, resonant JL cooling rates are
indistinguishable on the scale of this plot from the displayed ST ones.
Also depicted in the plots for \teq{T=10^5}K and \teq{T=10^6}K are the
asymptotic resonant rates resulting from
Eq.~(\ref{eq:cool_rate_ST_restot_th}), which are just single
integrations.  The precision of this asymptotic form is considerably
better than 1\% when compared with the full numerical integrations. 
Hence, it is expedient to use the asymptotic approximation whenever
resonant cooling rates are needed in pulsar and magnetar models
incorporating Compton upscattering interactions.

An illustration of the magnetic field dependence of the cooling rates at
polar surface locales is offered in Fig.~\ref{fig:resComp_th_T6_Bvar},
for outgoing electrons, and for Planckian soft photons of temperature
\teq{T=10^6}K possessing a hemispherical distribution
(\teq{\Omega_s=1/2}).  Close agreement with the Sokolov \& Ternov
asymptotic forms in Eqs.~(\ref{eq:cool_rate_ST_restot_th})
and~(\ref{eq:gammadot_belowres_th}) is evident for the highlighted
choices, and is numerically obtained for all field strengths.  For
\teq{B=10} and \teq{B=100}, the resonant JL cooling rates are
indistinguishable on the scale of this plot from the displayed ST ones,
deviating by less than around 2 percent for \teq{B \gtrsim 10}. 
However, for \teq{B=1}, the dashed curve depicts the JL cooling rate
that differs in the resonant domain from the ST one by a factor of 1.32,
according to Eq.~(\ref{eq:calRST_calRJL_rat}), a departure that
increases to a factor of \teq{\approx 2} when \teq{B=10^{-2}}. A notable
feature of this Figure is that at the ``peak'' of the resonance, defined
by \teq{\gamma_e\sim B/\Theta}, the value of the cooling rate is almost
independent of the field strength. This property can readily be deduced
from Eq.~(\ref{eq:cool_rate_ST_restot_th}) using the supercritical field
proportionality \teq{\calRST (B)\propto B^2} in
Eq.~(\ref{eq:calRST_def}), and also the width dependence
\teq{\Gamma\propto B} deducible from Eq.~(\ref{eq:width_effect}) in the
same field regime.  This behavior changes for subcritical fields. It is
clear from Eqs.~(\ref{eq:cool_rate_ST_restot_th})
and~(\ref{eq:calRST_def}) that when \teq{B\ll 1}, the cooling rates
scale with field as \teq{B^2} at fixed \teq{\gamma_e}, which indicates a
decline of the peak rate as subcritical fields are accessed. Moreover,
the difference between the \teq{{\dot \gamma}_e\propto B^2} behavior at
\teq{B\ll 1} and the weaker dependence (\teq{{\dot \gamma}_e\propto B})
in supercritical regimes is noteworthy.  Previous Thomson regime
expositions on cooling rates (e.g. Daugherty \& Harding 1989; Dermer
1990; Sturner 1995), essentially employ the \teq{B\ll 1} limit of the
Johnson \& Lippman field dependence listed in Eq.~(\ref{eq:calRJL_def}).
 If extrapolated to the \teq{B\gtrsim 1} domain, such a protocol would
introduce an overestimate of the resonant cooling rate by the factor
\teq{4 B^4/(3\calRST )}, which is approximately \teq{4 B^2 (1-2/e )/3}
when \teq{B\gg 1}.  This large inaccuracy motivates the deployment of
the relativistic quantum magnetic scattering formalism offered here: a
principal conclusion of this paper is that the QED cooling rates
profoundly improve upon the older magnetic Thomson formulations when
\teq{B\gtrsim 0.1}.

\begin{figure*}[t]
\figureoutpdf{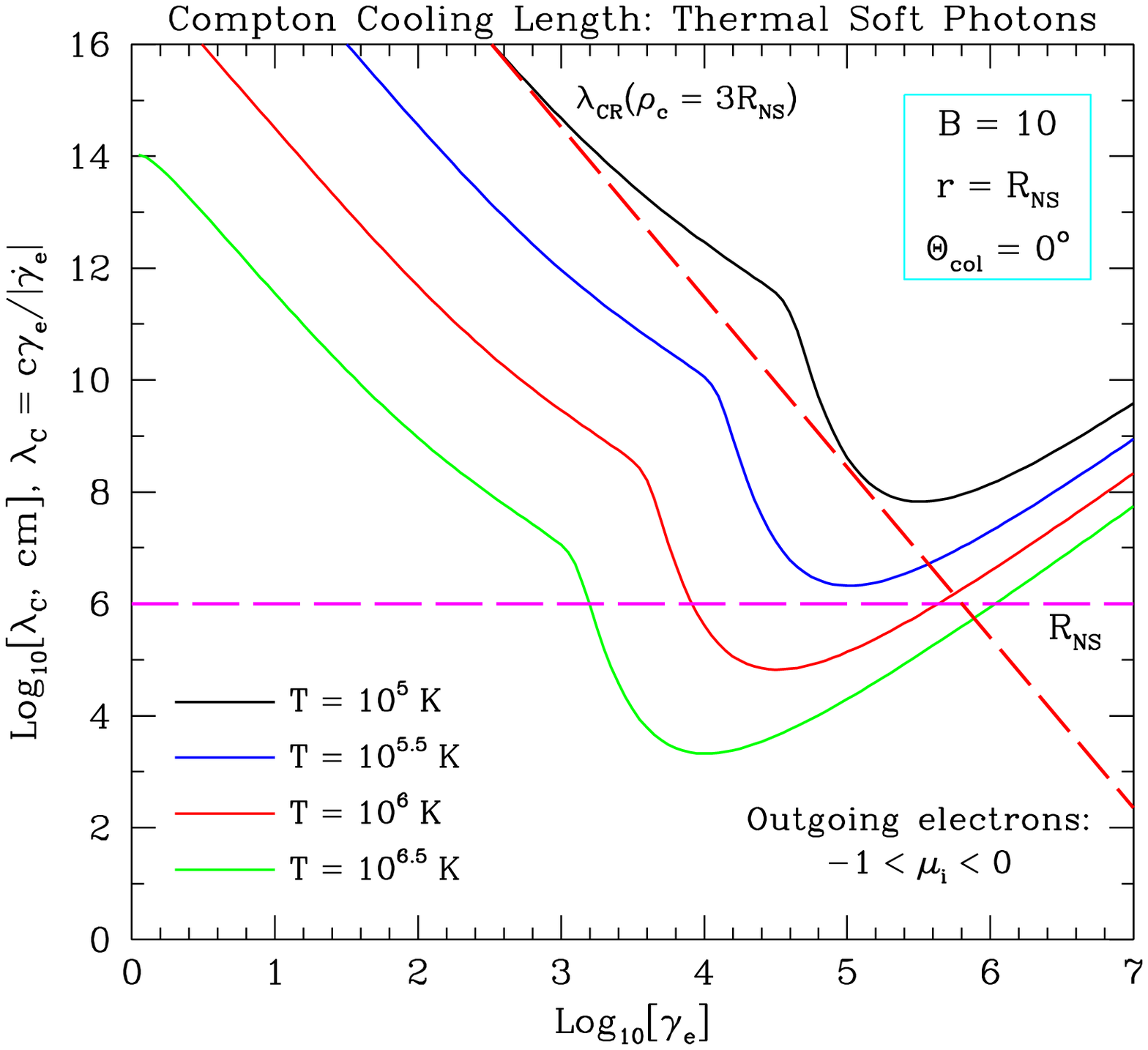}{3.8}{0.0}{-0.2}{
Resonant Compton cooling lengths for \teq{B=10} corresponding to the
cooling lengths in the left panel of Fig.~\ref{fig:resComp_th_ST_B10},
for different X-ray soft photon temperatures, as marked.  The outgoing
electrons are scattered just above the polar cap (\teq{\Thetacol
=0^\circ}), and at the stellar surface (\teq{r=\rns}), so that the
X-rays are defined by a hemispherical angular distribution. Only ST
calculations are exhibited, with JL evaluations virtually coinciding
with these curves because \teq{B\gg 1}.  The diagonal, dashed line
represents the curvature radiation cooling length near the surface,
according to Eq.~(\ref{eq:CR_coolrate}).  When \teq{T > 10^{5.5}}K,
cooling arises on scales shorter than \teq{\rns} for a range of
\teq{e^-} Lorentz factors.
 \label{fig:resComp_th_T6_mfp} }      
\end{figure*}

One measure of the effectiveness of resonant Compton cooling is whether
its lengthscale is shorter than the neutron star radius.  One naturally
anticipates that this may be the case near the neutron star surface,
where the field is high, the X-ray photon bath is intense, and
accordingly the scattering is very efficient.  To assess this
quantitatively, we form resonant Compton cooling lengths \teq{\lambda_c
= \gamma_ec/\vert {\dot\gamma}_e\vert} from the cooling rates, and plot
them in Fig.~\ref{fig:resComp_th_T6_mfp}. For illustrative purposes, the
\teq{\lambda_c} are displayed only for ST evaluations of the cooling
rates.  The neutron star radius \teq{\rns} is highlighted to benchmark
the cooling scales.  It is evident that for this \teq{B=10}, polar
colatitude example, when \teq{T\gtrsim 3\times 10^5}K,
\teq{\lambda_c\lesssim \rns} and cooling is very efficient near the
magnetic poles.  This is true also near the equator, as will become
evident when colatitudinal influences on the cooling rates are explored
in Section~\ref{sec:esoft_anisotropy}. The shortest cooling lengths
arise at the onset of the resonant contribution, which is defined by
setting \teq{\gamma_e\Theta\sim B} in
Eq.~(\ref{eq:cool_rate_ST_restot_th}), and approximating the logarithmic
\teq{{\ell}_T} terms by unity.  This generates the scaling
\begin{equation}
   \lambda_c\Bigl\vert_{\gamma_e\sim B/\Theta}
   \;\sim\; \dover{4\pi}{3\Omega_s}\ \dover{1}{\Theta^3}
        \, \dover{\lambar^3}{\sigt}\, \dover{B^2\Gamma}{\calRST}
   \;\equiv\; \dover{1}{2\Omega_s\Theta^3} 
        \, \dover{\lambar}{\fsc}\, \dover{B^2\Gamma}{\fsc\calRST}
 \label{eq:cool_mfp_scaling}
\end{equation}
for the Sokolov \& Ternov computation.  Using
Eq.~(\ref{eq:width_effect}) for the width \teq{\Gamma}, and
Eq.~(\ref{eq:calRST_def}) for \teq{\calRST}, it becomes clear that the
\teq{B^2\Gamma/(\fsc\calRST )} factor is \teq{1/2} when \teq{B\ll 1},
i.e. is independent of \teq{B}. For surface polar interaction locales,
when \teq{\Omega_s=1/2}, this sets \teq{\lambda_c\sim \lambar
/(2\fsc\Theta^3)} at the onset of resonant cooling in the magnetic
Thomson domain.  In contrast, for highly supercritical fields, the
\teq{B^2\Gamma /\calRST} factor yields a field dependence proportional
to \teq{B}, a signature of the Klein-Nishina reduction of the cross
section in this ultra-quantum regime. Then \teq{\lambda_c\sim 2\lambar
B/(\fsc\Theta^3)} for \teq{B\gg 1} at the magnetic pole on the stellar
surface.  Observe that the shapes of the cooling length curves are
qualitatively similar to those computed in Daugherty \& Harding (1989)
and Sturner (1995), who both employed the magnetic Thomson cross section
rather than the full magnetic QED forms that are the focus here.

It is natural to compare these results with the key lengthscale in polar
cap models of gamma-ray pulsars.  In these energetic neutron stars, the
principal mechanism for primary photon production is curvature emission
in the presumed dipolar field morphology. The cooling rate
\teq{\gammadotCR} for classical curvature radiation in pulsar models
(e.g. Ruderman \& Sutherland 1975; Daugherty \& Harding 1982) can be
employed to express the curvature emission cooling length
\teq{\lambdaCR}:
\begin{equation}
   \gammadotCR\; =\; -\dover{2}{3}\, \dover{r_0 c}{\rho_c^2}\;\gamma_e^4
   \quad \Rightarrow\quad 
   \lambdaCR \; =\; \dover{\gamma_ec}{\vert \gammadotCR \vert} 
                         \; =\; \dover{3}{2\gamma_e^3}\, \dover{\rho_c^2}{r_0}\quad .
 \label{eq:CR_coolrate}
\end{equation}
Here \teq{r_0=\fsc \lambar =e^2/(m_ec^3)} is the classical electron
radius.  Also, \teq{\rho_c} is the local radius of field curvature,
which generally scales as the altitude \teq{r} in a pulsar magnetosphere, 
except above the pole, where it scales as \teq{r^{1/2}}.   
In general, \teq{\rho_c > r}, and the curvature
radius decreases with colatitude.  As a representative case, the
curvature cooling length \teq{\lambdaCR} obtained from
Eq.~(\ref{eq:CR_coolrate}) for the (optimal) near-surface interaction case
\teq{\rho_c/\rns =3} is depicted in the right panel of
Fig.~\ref{fig:resComp_th_T6_mfp} as the steep, dashed red line. 
Clearly, if \teq{\gamma_e\lesssim 10^{5.5}}, resonant Compton cooling is
much more efficient than that due to curvature emission when
\teq{T\gtrsim 10^{5.5}}K. The origin for this is obviously that the
scattering process masquerades in some sense as stimulated cyclotron/synchrotron
emission at the resonance, and so its efficiency far exceeds that for
curvature radiation.

A different assessment of the efficiency of resonant Compton cooling is defined by 
the conditions under which it quenches electron acceleration in the magnetosphere.
The rate of electrostatic acceleration in the magnetospheres of either gamma-ray 
pulsars or magnetars is largely an unknown commodity.  The simplest assumption
is that the parallel electric field \teq{E_{\parallel}} invoked in an electrostatic gap
due to departures from Goldreich-Julian (Goldreich \& Julian 1969)
current flow (e.g. Shibata 1995; Takata et al. 2006) is a sizable fraction 
of the co-rotation {\bf v}\teq{\times}{\bf B} electric field, which scales as
\teq{\sim r\Omega B/c}.  Such is approximately the case in twisted magnetosphere 
models for magnetar energization/dissipation (e.g. see Thompson \& Beloborodov 2005;
Zane, Nobili \& Turolla 2011).  Defining an acceleration efficiency parameter 
\teq{\eta} via \teq{E_{\parallel} = 2\pi \rns \eta B/(Pc)}, the electrostatic acceleration rate
\teq{\gammadotacc} and lengthscale \teq{\lambdaacc} can quickly be written down:
\begin{equation}
   \gammadotacc\; =\; \dover{2\pi\eta\rns\, B}{P\lambar}
   \quad \Rightarrow\quad 
   \lambdaacc \; =\; \dover{\gamma_ec}{\vert \gammadotacc \vert} 
                         \; =\; \dover{\gamma_e}{2\pi B}\, \dover{\lambar \, P c}{\eta\rns}\quad ,
 \label{eq:accel_rate}
\end{equation}
where the magnetic field \teq{B} in this equation is expressed in units
of \teq{B_{\rm cr}}. Evaluation of this for conditions typical of
magnetars yields acceleration length scales of the order of
\teq{\lambdaacc\sim 10^{-2} - 10^{0}}cm when \teq{\gamma_e\sim 10^4};
these short scales are consequences of the extremely high
\teq{E_{\parallel}} assigned by a Goldreich-Julian construct near the
surface of a magnetar. Such values are clearly inferior to the cooling
lengths by at least 3--4 orders of magnitude; a similar situation arises
for normal pulsars -- see Fig.~3 of Harding \& Muslimov (1998).
Accordingly, for resonant Compton to effect a radiation-reaction-limited
acceleration (RRLA) in a magnetar, the accelerating fields must be weak
enough to set \teq{\eta\lesssim 10^{-4}}. Then, the maxiumum Lorentz
factor is controlled by the onset of resonant interactions, and is
proximate to the minima exhibited in Fig.~\ref{fig:resComp_th_T6_mfp}
near \teq{\gamma_e\sim B/\Theta}.  RRLA is commonly invoked in models of
young and middle-aged gamma-ray pulsars using curvature emission at
fairly high altitudes where the magnetic field is much weaker and
\teq{\lambdaCR} is much longer than illustrated in the Figure.  In such
circumstances, higher values of \teq{\eta} can be tolerated and the
equality \teq{\lambdaacc = \lambdaCR} is realized for \teq{\gamma_e\sim
10^6 - 10^7}.  Then, cooling due to resonant Thomson scattering can
intervene to temporarily slow, but not halt, the acceleration (e.g.
Daugherty \& Harding 1996; Harding \& Muslimov 1998).  Returning to
magnetars, if \teq{\eta} exceeds around \teq{10^{-4}}, then the
accelerating electric fields must be quenched by some process other than
resonant Compton cooling.  Screening of the fields by magnetic pair
creation \teq{\gamma\to e^{\pm}} is an obvious candidate, given its
invocation in models of conventional pulsars. If the electrons can
acquire modest pitch angles or populate sufficiently high Landau levels
in the strong magnetic field, then RRLA spawned by cyclo-synchrotron
radiation is also a possibility.

\subsection{Mean Energy Losses for Electrons}
 \label{sec:mean_energy}

The cooling rates computed so far are immediately useful for kinetic
equation analyses where the cooling is {\it continuous}, i.e. changes in
the electron energy incurred by resonant Compton interactions are small
or infinitesimal: this is the Thomson scattering regime.   In Compton
cooling problems where Klein-Nishina domains are sampled and electron
recoil is significant, kinetic equation formulations of the evolution of
the electron distribution function require a more complicated treatment
involving collisions integrals (e.g. see Blumenthal \& Gould 1970),
whose differences do not collapse to Fokker-Planck type differential
constructs where \teq{{\dot \gamma}_e} explicitly appears.  Therefore,
it is instructive to assess when the resonant Compton process is in
quasi-Thomson regimes, or when electron recoil is substantial. 
Intuitively, Klein-Nishina cases are expected to correspond to
\teq{\gamma_e\erg_s\gtrsim 1} or \teq{B\gtrsim 1}, and this turns out to
be the case.  To quantify this parameter space, here the {\it mean
energy loss per collision} is calculated, as a function of neutron star
parameters \teq{B} (localized) and surface temperature \teq{\Theta}. 
This requires an additional derivation of the Compton {\it collision
rate} to augment the cooling rates computed so far, which are
essentially averages of the dimensionless electron energy loss,
\teq{-\Delta\gamma_e =\erg_f-\erg_i}, weighted by the differential cross
section.  The incoming photon energy can be neglected, so that the
\teq{{\dot\gamma}_e} rates approximately average the value of
\teq{\Delta\gamma_e}. If the reaction rate of the collisions per
electron is \teq{{\dot n}_e/n_e=1/\tau_e}, i.e. the inverse collision
timescale, then the mean energy exchange per collision is
\begin{equation}
   - \dover{\langle \Delta\gamma_e\rangle}{\gamma_e} 
   \; =\; \dover{\tau_e\, \vert {\dot\gamma}_e\vert }{\gamma_e}\quad .
 \label{eq:mean_energyloss}
\end{equation}
The reaction rate can be assembled simply by replacing the 
\teq{\erg_f=\gamma_e\omega_f (1-\beta_e\cos\theta_f) \equiv \omega_f/\gamma_e/(1+\beta_e\mu_f)} 
factor in the integrand of the cooling rate expression in Eq.~(\ref{eq:scatt_spec}) by unity.
The number density rate per electron can then be written down by applying this routine
modification to Eq.~(\ref{eq:cool_rate_final}), which then yields
\begin{equation}
   \dover{1}{\tau_e} \; =\;  \dover{n_s\, c}{\mu_+-\mu_-}\, 
   \dover{1}{\gamma_e^3\beta_e^2\erg_s^2}   \int_{\omega_-}^{\omega_+}
   \omega_i\, d\omega_i   \int_{-1}^1 d(\cos\theta_f )
   \, \dover{f(\mu_i)}{\omega_f}\, \biggl\vert \dover{\partial \erg_f}{\partial (\cos\theta_f) } \biggr\vert\,
   \dover{d\sigma}{d(\cos\theta_f) }\quad ,
 \label{eq:coll_rate}
\end{equation}
a minus sign being introduced to render the rate positive.  
The development of this integration is identical to that for the cooling 
rates, and follows the procedures outlined in Appendix A.
The collisional reaction rate can then be written in a form that
parallels Eq.~(\ref{eq:cool_rate_genJL2}), namely
\begin{equation}
   \dover{1}{\tauJL} \; \approx\; \dover{3}{4}\, \dover{n_s\, \sigt c}{\mu_+-\mu_-}\, 
   \dover{1}{\gamma_e^2\erg_s^2}   \int_{\omega_-}^{\omega_+}
   \dover{{\cal F}_{\tau}\left(z_{\omega},\, p\right) - {\cal G}_{\tau}\left(z_{\omega},\, p\right)}{
   (\omega_i -B)^2 + (\Gamma /2)^2} 
   \, \dover{\omega_i^3\, d\omega_i}{1+2\omega_i} 
   \quad ,\quad z_{\omega}\; =\; 1 + \dover{1}{\omega_i}\quad ,
 \label{eq:coll_rate_JL}
\end{equation}
for \teq{p=\omega_i/B\approx 1}, for the Johnson \& Lippmann
formulation; the partner result for ST collisional rates is given in
Eq.~(\ref{eq:coll_rate_ST_app}). These forms can be routinely integrated
over the Planck spectrum for the soft photons in
Eq.~(\ref{eq:Planck_spec}). Numerical computations of these collisional
rates are not explicitly illustrated in this paper.  The general forms
for these rates are summarized in Appendix A via approximations
appropriate to the resonant regime in
Eqs.~(\ref{eq:coll_rate_JLdelta_app})
and~(\ref{eq:coll_rate_STdelta_app}), together with low and high-field
asymptotic results.

\begin{figure*}[t]
\figureoutpdf{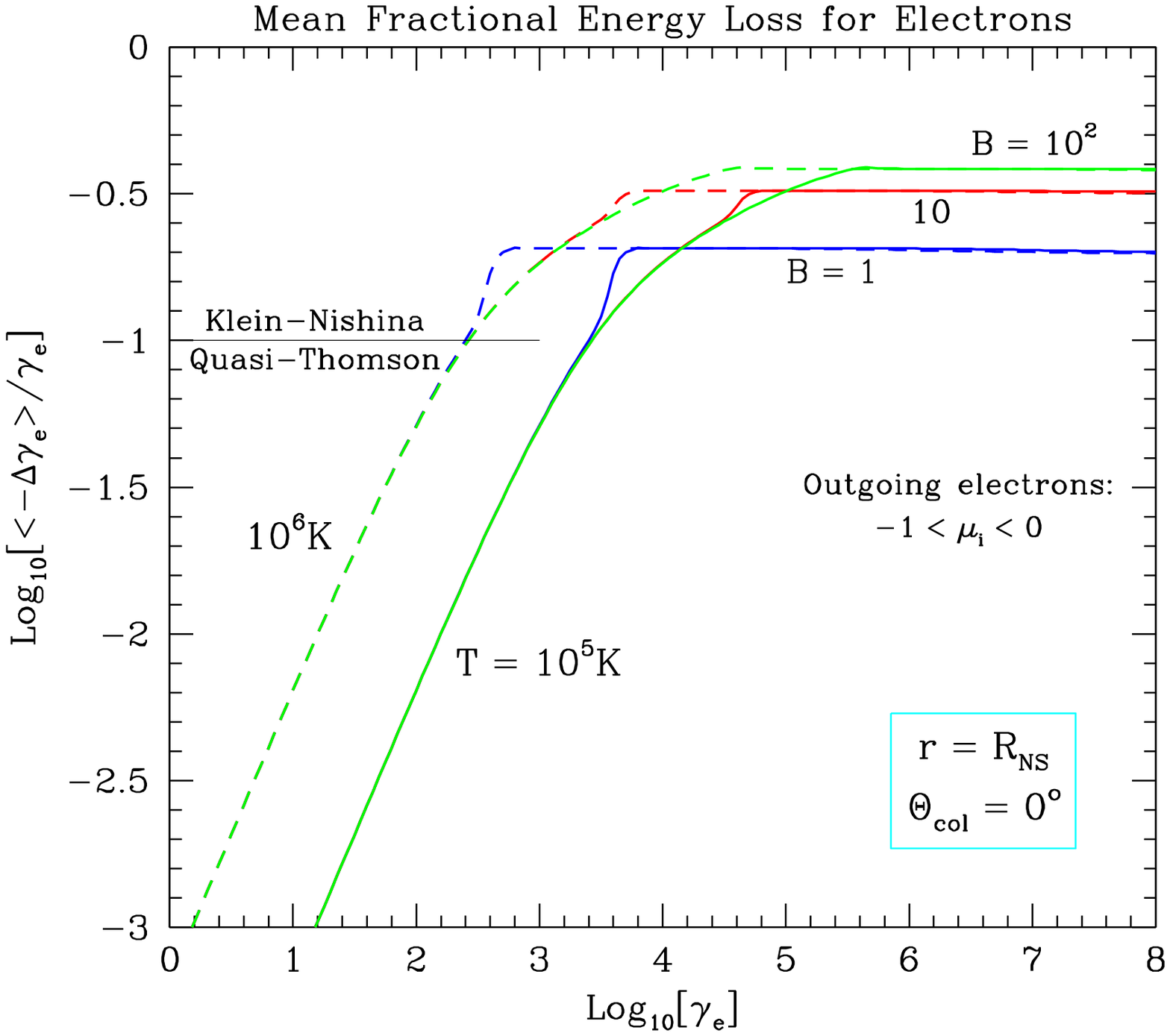}{3.8}{0.0}{-0.2}{
The mean fractional energy loss for electrons in magnetic Compton
interactions, as defined in Eq. (\ref{eq:mean_energyloss}), with thermal
photons at the neutron star surface and at the magnetic pole
(\teq{\Thetacol =0^\circ}). The six cases correspond to two
temperatures, \teq{T=10^5}K (solid curves) and \teq{T=10^6}K (dashed
curves), and for each three magnetic fields, \teq{B=1,10, 100}, as
labelled. They constitute Sokolov \& Ternov (ST) evaluations for
outgoing (upward moving) electrons. At low Lorentz factors, \teq{\langle
-\Delta \gamma_e \rangle \ll \gamma_e}, and the scatterings are in the
quasi-Thomson regime, with the mean fractional energy loss being given
by Eq.~(\ref{eq:delgamdgam_nonres}). At high \teq{\gamma_e}, a magnetic
Klein-Nishina regime delineated approximately by \teq{\langle -\Delta
\gamma_e \rangle / \gamma_e > 0.1} is realized, where the upscattered
photons acquire a sizable fraction of the incident electron energy.  As
\teq{\gamma_e\to\infty}, the fractional energy loss depends only on
\teq{B}, and not temperature \teq{T}, as given by
Eq.~(\ref{eq:delgamdgam_res}).
 \label{fig:resComp_th_delgam} } 
\end{figure*}

Combining the cooling and collisional rates, the ratio in
Eq.~(\ref{eq:mean_energyloss}) for the mean energy exchange per
collision is illustrated in Fig.~\ref{fig:resComp_th_delgam} for the
Sokolov and Ternov cross section. These were computed for surface
interactions at the magnetic pole, where \teq{\Omega_s=1/2}, and soft
photon temperatures \teq{T=10^5}K and \teq{T=10^6}K. We note that since
both the cooling and collisional rates scale as \teq{\Omega_s}, the
solid angle dependence drops out of the problem for on-magnetic axis
locales. The curves evince a clear delineation between two regimes.  At
low Lorentz factors, \teq{\gamma_e\lesssim 1/(10\Theta )}, the
fractional energy loss by the electrons is much less than unity, and a
quasi-Thomson regime emerges where the interactions are not resonant. 
This is the portion of phase space where cooling modifications to the
evolution of electron distributions can be modeled using Fokker-Planck
forms for kinetic equations.  Using the analysis in Appendix A, the
ratio of the cooling rate in Eq.~(\ref{eq:gammadot_belowres_th}) to the
reaction rate in Eq.~(\ref{eq:collrate_belowres_th}) for thermal soft
photon spectra is independent of \teq{B}, and leads to
\begin{equation}
   - \dover{\langle \Delta\gamma_e\rangle}{\gamma_e} 
   \; =\; \dover{\tau_e\, \vert {\dot\gamma}_e\vert }{\gamma_e}
   \;\approx\;  \dover{ 4\pi^6\, \gamma_e \Theta }{945\, \zeta(5)} \,
   \dover{(1+ \beta_e \mu_+)^5- (1+\beta_e \mu_-)^5}{(1+ \beta_e \mu_+)^4- (1+\beta_e \mu_-)^4} 
   \quad ,\quad \gamma_e\Theta\;\ll\; 1\quad , 
 \label{eq:delgamdgam_nonres}
\end{equation}
where \teq{\zeta (n)} is the Riemann {\it zeta} function. In the case
where \teq{\gamma_e} is moderately large so that \teq{\beta_e \approx 1}
and \teq{\mu_+ = 0, \mu_- = -1} at the surface on the magnetic axis, the
numerical factors yield a ratio of \teq{3.924 \gamma_e \Theta}.  This
analytic form nicely describes the curves in
Fig.~\ref{fig:resComp_th_delgam} for \teq{\gamma_e\lesssim 10^2}. Since
the scattering in the ERF generates \teq{\omega_f\approx\omega_i\propto
\gamma_e\erg_s}, i.e. Thomson kinematics, the upscattered photon energy
in the OF scales as \teq{\gamma_e^2\Theta} so that one naturally
anticipates a fractional energy change proportional to
\teq{\gamma_e\Theta}, and independent of the field strength.  As
expected, since the resonance is not sampled in this \teq{\gamma_e}
range, Eq.~(\ref{eq:delgamdgam_nonres}) applies to both JL and ST cases.

At the higher \teq{\gamma_e}, the mean energy exchange saturates at
sizable fractions of unity that are dependent on the field strength. 
This Klein-Nishina domain of large electron recoil is coincident with
resonant interactions, and for such Lorentz factors, the evolution of
electron populations due to magnetic Compton cooling must be treated
using full Boltzmann collisional integrals. The mean energy losses for
\teq{\gamma_e\gtrsim 1/\Theta } are assembled in Appendix A; using
Eqs.~(\ref{eq:coll_rate_JLdelta_asymp})
and~(\ref{eq:coll_rate_STdelta_asymp}), together with the equivalent JL
and ST asymptotic limits for the cooling rates earlier in Appendix A, it
can be inferred that they possess the limiting forms
\begin{equation}
   - \dover{\langle \Delta\gamma_e\rangle}{\gamma_e} 
   \; =\; \dover{\tau_e\, \vert {\dot\gamma}_e\vert }{\gamma_e}
   \;\approx\; \cases{ \dover{e-2}{e-1}\;\approx\; 0.418\quad &, $\;\;B\,\gg\, 1$,\cr
    \vphantom{\Bigl(} B\quad &, $\;\;B\,\ll\, 1$.\cr}
 \label{eq:delgamdgam_res}
\end{equation}
Note that these limiting ratios apply to both JL and ST ratios, which are identical;
in other words, the spin-dependent nuances in each of the cooling and collisional
rates cancel in both the \teq{B\ll 1} and \teq{B\gg 1} domains.  It should be noted 
that for field strengths in between these limits there is, in general, a modest 
difference between the JL and ST predictions for the mean energy loss,
maximized at just under 9\% in the \teq{B\sim 2-3} range.
Clearly when \teq{B\lesssim 1}, the resonant interactions do not sample 
Klein-Nishina reductions, and a magnetic Thomson description suffices.
The insensitivity to the value of \teq{\gamma_e} is evident from a comparison 
of Eqs.~(\ref{eq:cool_rate_genJL2}) and~(\ref{eq:coll_rate_JL}), from which
one can also infer that integrating over the Planck spectrum will yield 
\teq{\tau_e {\dot\gamma}_e/\gamma_e} also independent of temperature,
as is observed.  In strong recoil regimes, information on the initial photon
energy is obscured as the electron loses much of its energy in a scattering 
event.  Note that the morphology of the curves in Fig.~\ref{fig:resComp_th_delgam}
is similar for other interaction altitudes and colatitudes.

\newpage

\section{ALTITUDINAL AND COLATITUDINAL INFLUENCES}
 \label{sec:esoft_anisotropy}

The focus on isotropic photon distributions of the previous section
needs to be extended to include geometric  influences at arbitrary
locations in pulsar magnetospheres. The electron cooling rate
determinations depend on the altitude \teq{r} above the neutron star 
surface and the colatitude \teq{\Thetacol} away from the magnetic axis, 
due to the changing photon angle with respect to the magnetic field 
and because of a curtailed sampling of photons from the surface at 
high altitudes.  We now assume target X-ray photons are emitted 
uniformly and radially from the star surface, and thus at a given 
altitude are collimated in a cone of angle $\theta_c$ given by 
\begin{equation}
   \cos \theta_c \; =\; \sqrt{1- \Bigl( \dover{\rns}{r}\Bigr)^2}
   \quad ,\quad
   r\;\geq\; \rns
 \label{eq:cone_angle}
\end{equation}
(e.g. Rybicki \& Lightman 1979; Dermer 1990). The solid angle of the
soft photon population therefore is \teq{\Omega_s \equiv 2\pi
(1-\cos\theta_c)}, so that \teq{\Omega_s\approx (\rns /r)^2} describes
the inverse square law when \teq{r\gg\rns}.  Departures from the
simplification in Eq.~(\ref{eq:cone_angle}) are required when taking
into account anisotropies that arise in radiative transfer models of
magnetic neutron star atmospheres (e.g. Zavlin, Shibanov \& Pavlov 1995;
P\'erez-Azor'n, Miralles \& Pons 2005; Mori, Ho \&  Wynn 2007), and
temperature non-uniformities in transitioning from polar to equatorial
zones; treatment of such refinements is deferred to future work. In this
section, we specialize to the dipole magnetic field, but other cases
such as a twisted dipole field (e.g. Thompson \& Beloborodov 2005) also
follow the logical developments below, with just routine geometrical
modifications. Also, in the analysis here, flat spacetime will be
presumed, so that general relativistic effects are neglected: photons
propagate in straight lines, and the photon energy and field morphology
are not modified by a stellar gravitational potential.  It is expected
that the general character of the results presented here would not
change appreciably with the accurate treatment of curved spacetime in
Schwarzschild or rotating metrics. Including general relativistic
effects such as magnetic field distortion and the enhancement of photon
temperatures and densities in the local inertial frame, and the
curvature of photon trajectories, would incur only moderate changes to
the shape and position of the cooling curves.

Considering interaction points remote from the magnetic axis, the local
magnetic field is in general at an angle $\thetaBr$ with respect to the
radial direction, which is easily found compactly using in the dipole
form of the magnetic field $\vec{B} = B_0\, r^{-3} ( 2\cos \Thetacol
\,\hat{r} +  \sin \Thetacol \,\hat{\theta} )/2 $ in terms of specified
colatitude \teq{\theta\equiv\Thetacol},
\begin{equation} 
   B  \; =\; \dover{B_0}{2 r^3} \sqrt{1+3\cos^2 \Thetacol} .
   \qquad , \qquad 
   \cos \thetaBr \;\equiv\; \dover{\vec{B} \cdot \hat{r}}{ \vert \vec{B} \vert} 
            \; =\; \dover{2 \cos \Thetacol}{\sqrt{1+3\cos^2\Thetacol}} 
 \label{eq:Br_form}
\end{equation}
The magnitude of the local magnetic field, central to the cross section
and cyclotron width calculations, and of critical import at resonance,
also depends on altitude and colatitude, and the surface polar magnetic
field, $B_0$. At varying locales, we may have \teq{\theta_c \leq
\thetaBr} or \teq{\theta_c \geq \thetaBr}, the former being realized at
high altitude or close to the magnetic equator, while the latter at low
altitudes or close to the magnetic axis. These delineate two algebraic
regimes in the ensuing analysis, corresponding to comparatively small or
large radii of curvature for the field lines. For more generalized,
non-dipolar magnetic field geometries, the principal modifications to
the subsequent developments can be introduced via alternative
expressions for \teq{B} and \teq{\thetaBr} to those in
Eq.~(\ref{eq:Br_form}).

\begin{figure*}[t]
\twofigureoutpdfadj{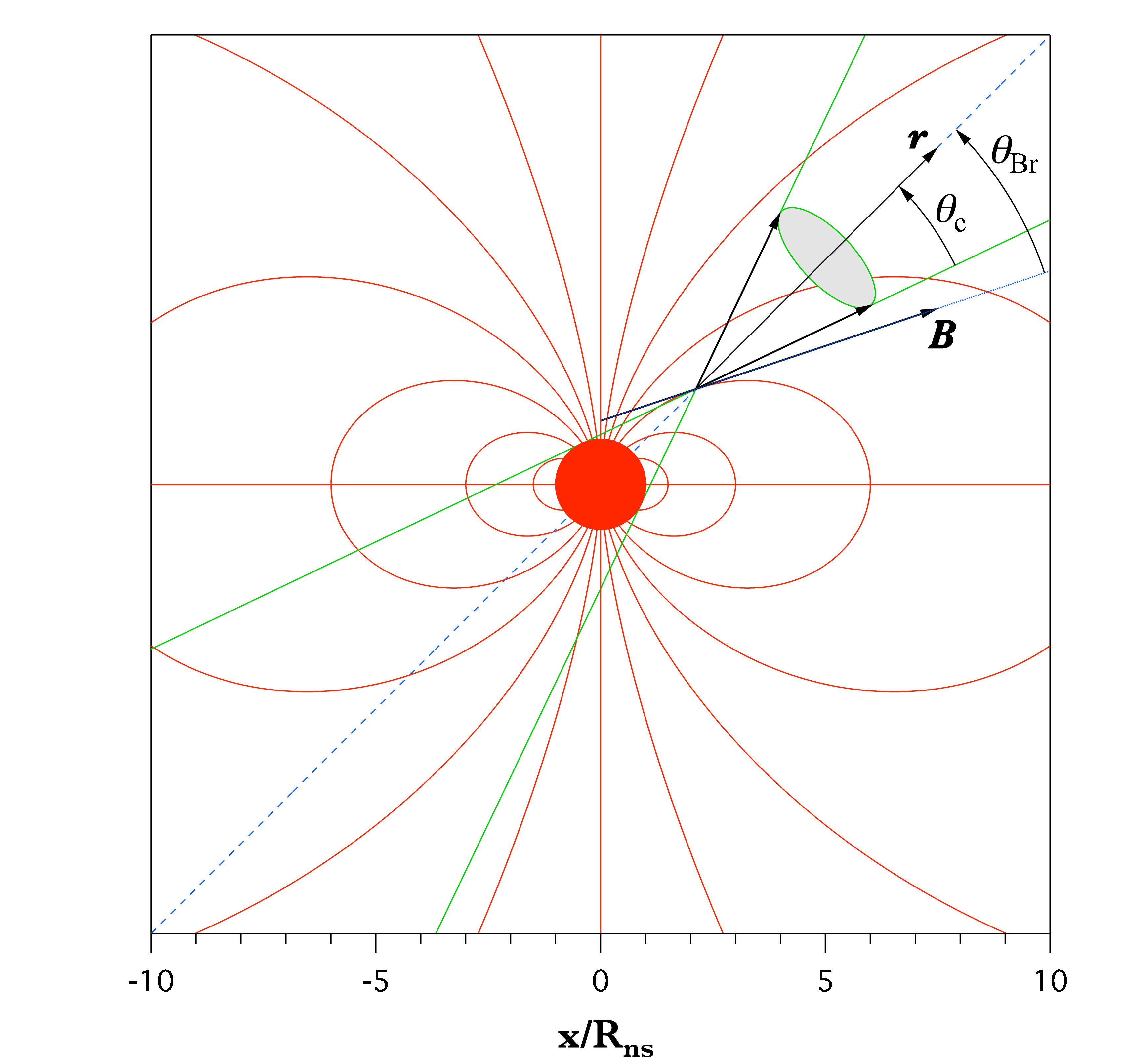}{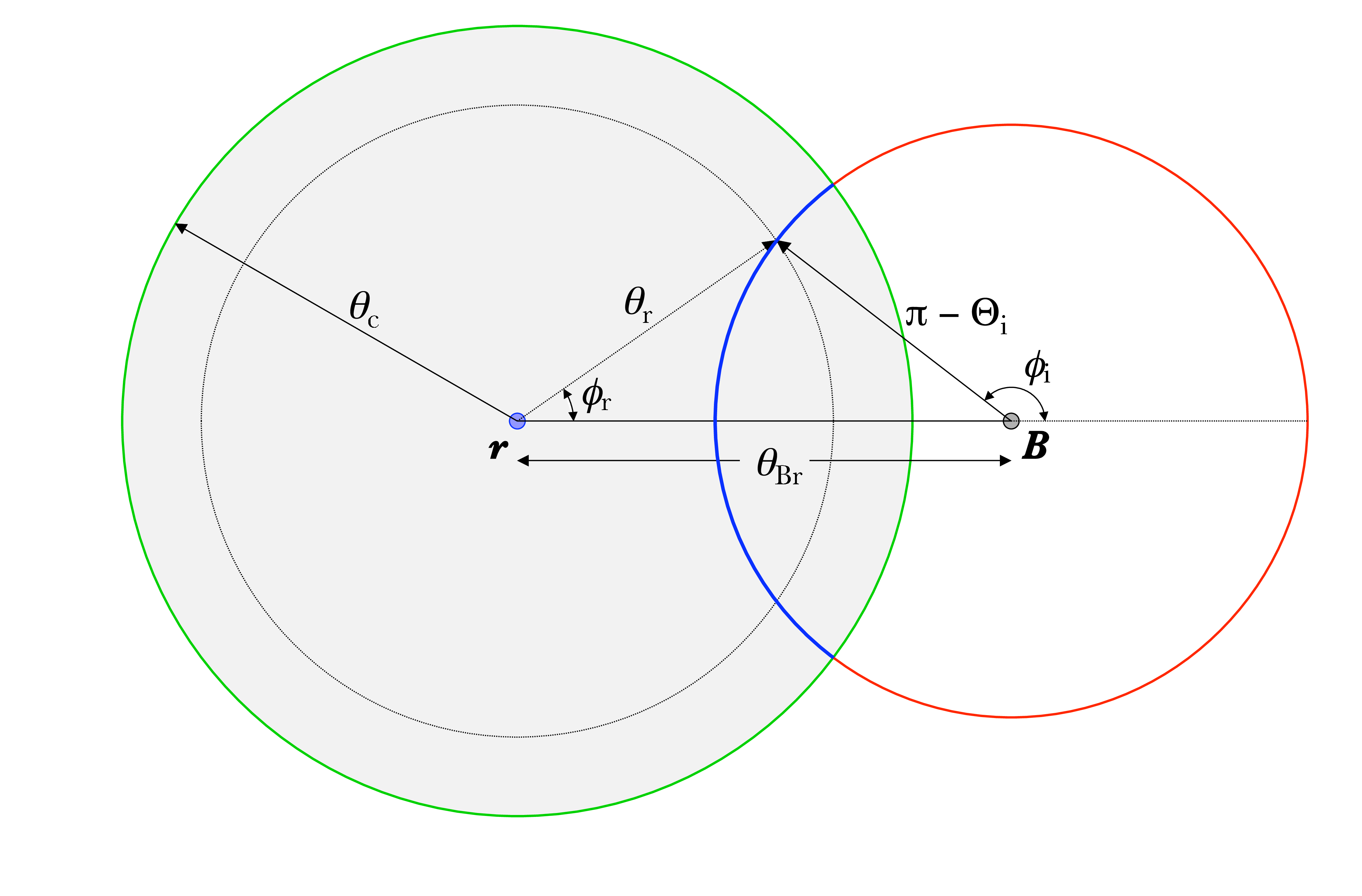}{
The geometry for Compton collisions at arbitrary altitudes and colatitudes in the
magnetosphere.  {\it Left panel:} Magnetospheric view depicting an interaction 
point located at \teq{r/\rns = 3} and a colatitude of \teq{\Thetacol = 45^{\circ}}. 
Here \teq{\thetaBr} lies outside the soft photon cone of opening 
angle \teq{\theta_c}.  The spatial scale is linear, in units of \teq{\rns}.
{\it Right panel:}  An outside blow-up perspective of the interaction cones, with 
the circles representing caps on angular cones with apexes at the 
interaction point \teq{{\vec r}}.  The left (shaded) circle defines the solid angle of the 
collimated soft photon cone above the stellar surface.  The unshaded circle 
at the right represents the cap of a cone, whose axis coincides with the local 
\teq{{\vec B}} vector, and whose opening angle \teq{\pi - \Theta_i} defines 
the incoming photon angle with respect to {\bf B} in the observer's frame.
The highlighted (blue) arc on this circle that lies within the soft photon cone's
cap signifies the azimuthal angles permitting interactions for a specified 
\teq{\Theta_i}.  Various angles in the depicted triangle
are related by spherical trigonometry according to Eq.~(\ref{eq:Br_sphere}).
 \label{fig:collision_geometry} } 
\end{figure*}

We wish to derive the \teq{f(\mu_i)} angular distribution function
(normalized to unity) at a given colatitude and altitude as a function
of the photon angle with respect to the magnetic field, \teq{\Theta_i =
\arccos \mu_i},  in the inertial frame of the star, but independent of
the azimuthal angles since the cross section is independent of such.
This involves the intersection of two spherical caps, one with a radial
axis and one along the magnetic field direction, and an integration over
the azimuthal angle with respect to the magnetic field; the global
geometry is displayed in the left panel of
Fig.~\ref{fig:collision_geometry}. Let \teq{\Theta_i} and \teq{\phi_i}
denote the spherical polar and azimuthal angles, respectively, with
respect to the magnetic field in the observer's frame. Similarly, let
\teq{\theta_r} and \teq{\phi_r} be the angles with respect to the radial
vector at some colatitude \teq{\theta}. These angles and the pertinent
geometry are displayed in Fig.~\ref{fig:collision_geometry} for the case
of outgoing electrons. The case of ingoing electrons differs from the
outgoing case by the simple mirroring of the angular distribution,
\teq{\pi- \Theta_i \rightarrow \Theta_i} in the ensuing algebra below.
From spherical trigonometry's laws of cosines and sines, we have the
following relations
\begin{eqnarray}
   -\cos \Theta_i &=& \cos \theta_r \cos \thetaBr + \sin \theta_r \sin \thetaBr \cos \phi_r \nonumber\\[4pt]
   \cos \theta_r &=& -\cos \thetaBr \cos \Theta_{i} - \sin \thetaBr \sin \Theta_{i} \cos \phi_i 
 \label{eq:Br_sphere}\\[3pt]
   \sin \theta_r &=& \frac{  \sin \Theta_i \sin \phi_i}{\sin \phi_r}\quad .\nonumber
\end{eqnarray}
Here $\theta_r$ is uniformly distributed in the range 
\teq{0 \leq \theta_r \leq \theta_c} reflecting the uniform brightness 
approximation of the surface emission, together with the constancy 
of light intensity during propagation in a flat spacetime.  It is easy to show 
that the Jacobian of the solid angle transformation is unity between the 
two spherical caps, i.e. $d\mu_r d\phi_r = d\mu_i d\phi_i$, signifying 
the conservation of volume elements under rotations. Thus the $f(\mu_i)$ 
dependence is imposed merely by geometrical restriction of the limits of 
integration over the azimuthal angles $\phi_i$.

These limits are specified by the two values of the azimuthal angle 
\teq{\phi_i} at the intersection of extremities to the spherical caps, 
corresponding to the ``threshold'' values \teq{\pm \phi_{rt}} of the 
azimuthal angle \teq{\phi_r} that are defined by the condition 
\teq{\theta_r = \theta_c} (see Fig.~\ref{fig:collision_geometry},
right panel), which has solutions provided that 
\teq{\vert \thetaBr-\theta_c\vert \leq \pi -\Theta_i \leq \thetaBr+\theta_c}.  
Using the spherical law of cosines, these values are expressed 
as functions of \teq{\Theta_i}:
\begin{equation}
   \phi_{rt} \; =\; \arccos \left[ \mu_{rt} \right] 
   \qquad , \qquad 
   \mu_{rt} \;\equiv\; 
   \dover{-\cos\Theta_i - \cos\theta_c\cos\thetaBr}{\sin\theta_c\sin\thetaBr}\quad ,
 \label{eq:phi_rt_def}
\end{equation}
while taking the appropriate branch cut for the inverse trigonometric 
function \teq{\arccos}.  The corresponding threshold angle \teq{\phi_{it}} 
is found in terms of \teq{\phi_{rt}} via the spherical law of sines,
\begin{equation}
   \sin \phi_{it} \; =\; \frac{\sin \theta_c}{\sin \Theta_i }\, \sqrt{ 1- \mu_{rt}^2} \quad ,
 \label{eq:phi_it_def}
\end{equation}
so that the threshold intersection \teq{\phi_{it}} is given by the 
appropriate definition of \teq{\arcsin} such that $ 0  \leq \phi_{it} \leq \pi$. 
After some routine algebra, we can express $\phi_{it}$ terms of $\mu_\pm$ 
for the case \teq{\thetaBr > \theta_c},
\begin{equation}
	\sin \phi_{it} \; =\;  \frac{1}{\sin \thetaBr}  \sqrt{ \left[ \frac{\mu_+ - \mu_i}{1 - \mu_i} \right] 
	\left[ \frac{\mu_i - \mu_-}{\mu_i + 1} \right] }  \quad , \quad \thetaBr \; >\; \theta_c\quad .
 \label{eq:phi_it_ident}
\end{equation}
Here the lower (\teq{\mu_-}) and upper (\teq{\mu_+}) limits to the \teq{\mu_i} integration 
in the cooling rate are restricted to the domain where \teq{f(\mu_i)} is non-vanishing,
and hence become functions of altitude and colatitude according to 
\begin{eqnarray}
   \mu_+ &=& -\cos (\thetaBr + \theta_c ) \nonumber \\[-3.5pt]
 \label{eq:mu_pm_ang}\\[-10.5pt]
   \mu_- &=& \left\{  \begin{array}{cl} 
                      -1 &\mbox{, \quad if } \thetaBr \leq \theta_c \\
   -\cos (\thetaBr - \theta_c ) &\mbox{, \quad if } \thetaBr > \theta_c
                               \end{array} \right.  \nonumber
\end{eqnarray}
for the case of outgoing electrons.  From these forms, one can deduce
that on the magnetic axis, \teq{(\mu_-,\mu_+) = (-1,-\mu_c)}, and at the
surface on the equator, \teq{(\mu_-,\mu_+) = (-1,1)}. It is clear that
for $\mu_- < \mu_i < \mu_+$, the argument of the square root in
Eq.~(\ref{eq:phi_it_ident}) is positive-definite, and so that this
equation can be inverted to generate a well-defined value of
$\phi_{it}$.  The value of \teq{\phi_{it}} controls the azimuthal
contribution to the angular distribution for a particular interaction
angle \teq{\pi -\Theta_i} with respect to the magnetic field direction.

\begin{figure*}[t]
\twofigureoutpdf{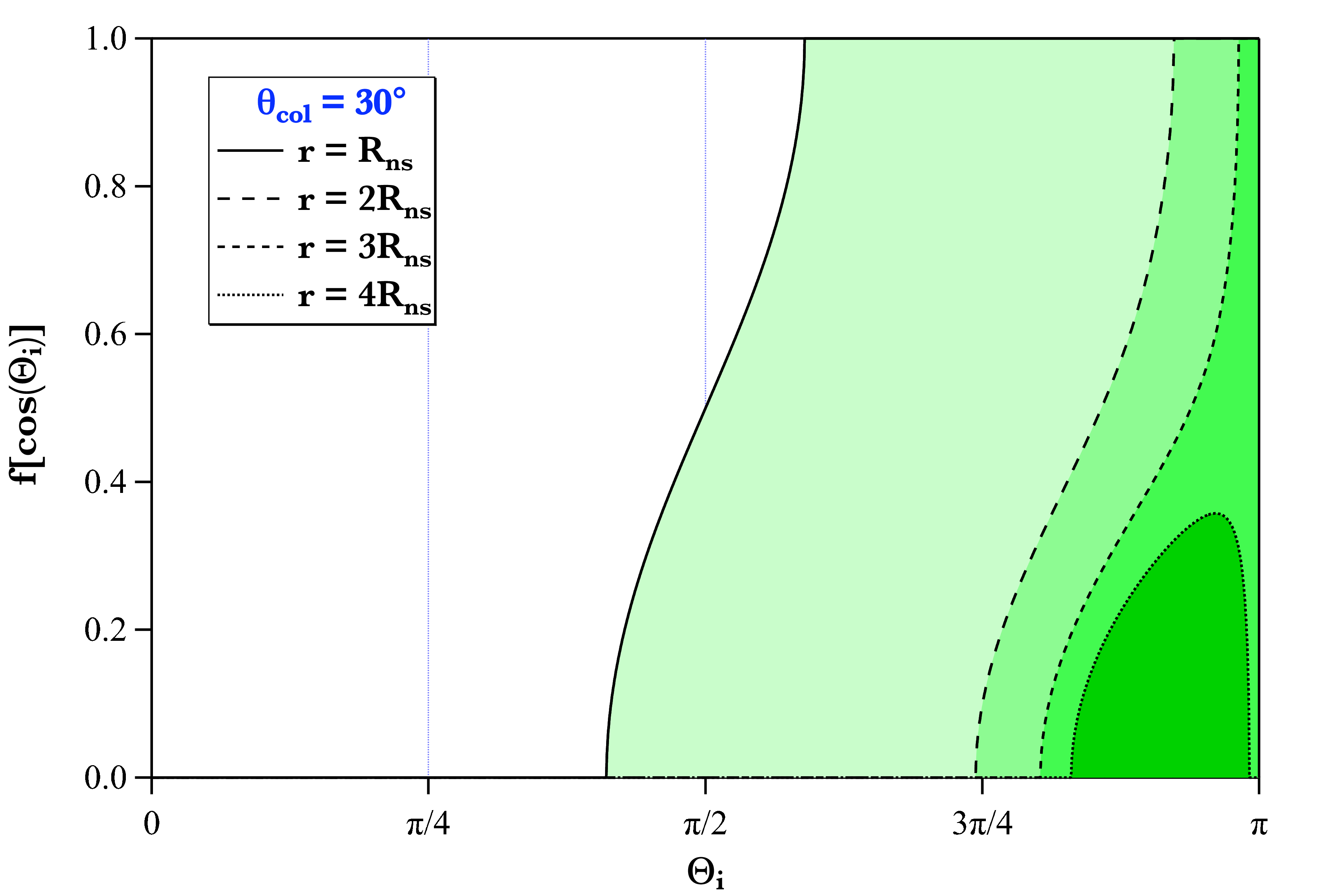}{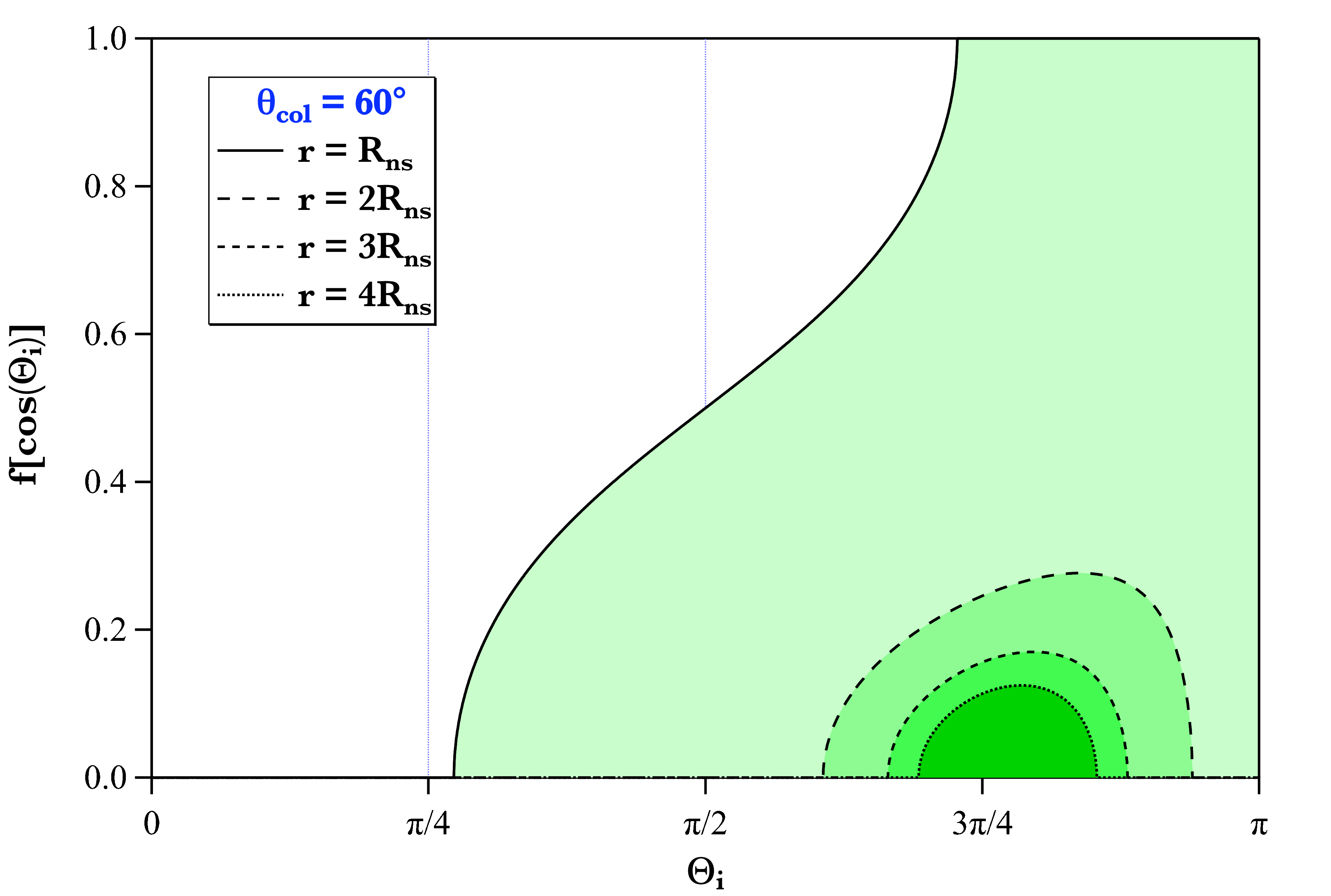}{
Shown here is the angular distribution function \teq{f(\mu_i)} for two
intermediate colatitudes \teq{\Thetacol} and four altitudes \teq{r}, as
labelled, for outgoing electrons. It is readily apparent that as the
photon cone becomes collimated at high altitudes, the phase space of
angles around the local magnetic field is both distorted and curtailed,
as highlighted for altitudes \teq{r/\rns = 1,2,3,4} by the solid,
long-dash, short-dash and dotted lines, respectively. Moreover, the
normalization of the areas under \teq{f(\mu_i)} in \teq{\mu_i} coordinates 
as a function of altitude yields \teq{1-\mu_c}, i.e., the inverse-square law 
to first order at high altitudes. As one moves to the equatorial regions 
(\teq{\Thetacol\sim 60^{\circ} - 90^{\circ}}), the ranges of angles sampled 
migrates towards a phase space in the neighborhood of \teq{\Theta_i = \pi/2}.
Distributions for ingoing electrons are simply obtained by invoking a
\teq{\Theta_i\to \pi - \Theta_i} transformation.
 \label{fig:ang_dist_3060} }      
\end{figure*}

Without loss of generality, the inversion of Eq.~(\ref{eq:phi_it_ident})
is established by restricting the range of the \teq{\arcsin x} function
to the interval \teq{[0,\, \pi /2]}.  The extremeties of this interval,
where \teq{\vert\sin\phi_{it}\vert =1}, correspond to a critical value
of \teq{\Theta_i}, denoted by \teq{\hat{\Theta}_i = \arccos
[\cos\theta_c/\cos\thetaBr ]}. This division point delineates two
branches to the inversion, encapsulated in the identity
\begin{equation}
 	\phi_{it} =  \left\{ 
	\begin{array} {cl}
 	\arcsin \left[ \dover{\sin \theta_c}{\sin \Theta_i }\, \sqrt{ 1- \mu_{rt}^2} \right]
	   &  \mbox{, \quad if $\pi- \Theta_i \leq \arccos \left[ \dover{\cos\theta_c}{\cos\thetaBr} \right] $} \\[12pt]
	\pi - \arcsin \left[ \dover{\sin \theta_c}{\sin \Theta_i }\, \sqrt{ 1- \mu_{rt}^2} \right] & \mbox{, \quad otherwise. } 
	\end{array}  \right. 
 \label{eq:phiitbranch}
\end{equation}
These branches span the intervals \teq{0\leq\phi_{it}\leq \pi/2} and 
\teq{\pi /2\leq\phi_{it}\leq \pi}, respectively.  Integrating over the 
azimuthal angles as restricted above, and normalizing to unity when at the magnetic 
pole at the surface (i.e. \teq{\Thetacol =0^{\circ}} and \teq{r=\rns}),
we obtain the angular distribution function for various cases,
\begin{equation}
 	f(\mu_i) =  \left\{ 
	\begin{array}{cl}
 	\! \! 1 - \dover{1}{\pi} \phi_{it} ( \Theta_i, \thetaBr, \theta_c )  
	   &  \mbox{, \quad if $\Theta_i \neq \pi \mbox{ and } \thetaBr \neq 0 $} \\[5pt]
	1 & \mbox{, \quad if } (\pi- \Theta_i \leq \theta_c \mbox{ and } \thetaBr = 0) 
	             \mbox{ or } (\Theta_i = \pi \mbox{ and } \thetaBr < \theta_c) \\[5pt]
	1/2 & \mbox{, \quad if } \Theta_i = \pi \mbox{ and } \thetaBr = \theta_c \\[5pt]
	0 & \mbox{, \quad if } (\Theta_i = \pi \mbox{ and } \thetaBr > \theta_c) 
	            \mbox{ or } \Theta_i < 0 
	\end{array}  \right. 
 \label{eq:ang_dist}
\end{equation}
with \teq{\cos \Theta_i = \mu_i \equiv [\omega_i / (\gamma_e \epsilon_s) -1]/\beta_e}
forging a connection to the scattering kinematics.  
Clearly  \teq{f(\mu_i)} vanishes on a domain that is a function of altitude and colatitude, 
a constraint transmitted through the values of \teq{\thetaBr} and \teq{\theta_c}. 
Angular distributions defined according to Eq.~(\ref{eq:ang_dist}) are
depicted in Fig.~\ref{fig:ang_dist_3060} and deployed in the computations below.
Only intermediate colatitudes, \teq{\Thetacol = 30^{\circ}} and \teq{\Thetacol = 60^{\circ}}, 
are illustrated.  The polar axis cases are simple top-hat functions, while the equatorial 
cases are quasi-circular bubbles that are concentrically declining in size at 
high altitudes (much like the \teq{\Thetacol = 60^{\circ}} case in Fig.~\ref{fig:ang_dist_3060}), 
and centered on and symmetric about \teq{\Theta_i=\pi/2}.  The declining size of the 
envelope loci with altitude reflects the dilution of the soft photon population with 
remoteness from the stellar surface.

Finally, the above derivation and analysis was performed for the case of
upward propagating electrons for \teq{0 \leq \Thetacol \leq \pi/2}. For
the case of ingoing electrons, the polarity of {\bf B} flips and
distribution is identical to the outgoing case for \teq{\pi/2 \leq
\Thetacol \leq\pi}. Moreover, the downward propagating distribution for
\teq{0 \leq \Thetacol \leq \pi/2} is simply a mirror of the upward case
in terms of \teq{\Theta_i}, i.e. \teq{\Theta_i \to \pi - \Theta_i},
along with some obvious redefinitions of \teq{\mu_\pm}, namely
\teq{\mu_- =\cos ( \thetaBr + \theta_c )}, and \teq{\mu_+ =\cos (
\thetaBr - \theta_c )} if \teq{\thetaBr > \theta_c}.  When \teq{\thetaBr
\leq\theta_c}, one sets \teq{\mu_+=1}.  These identities keep the
original definition of \teq{\thetaBr} from Eq.~(\ref{eq:Br_form}).

\begin{figure*}[t]
\fourfigureoutpdf{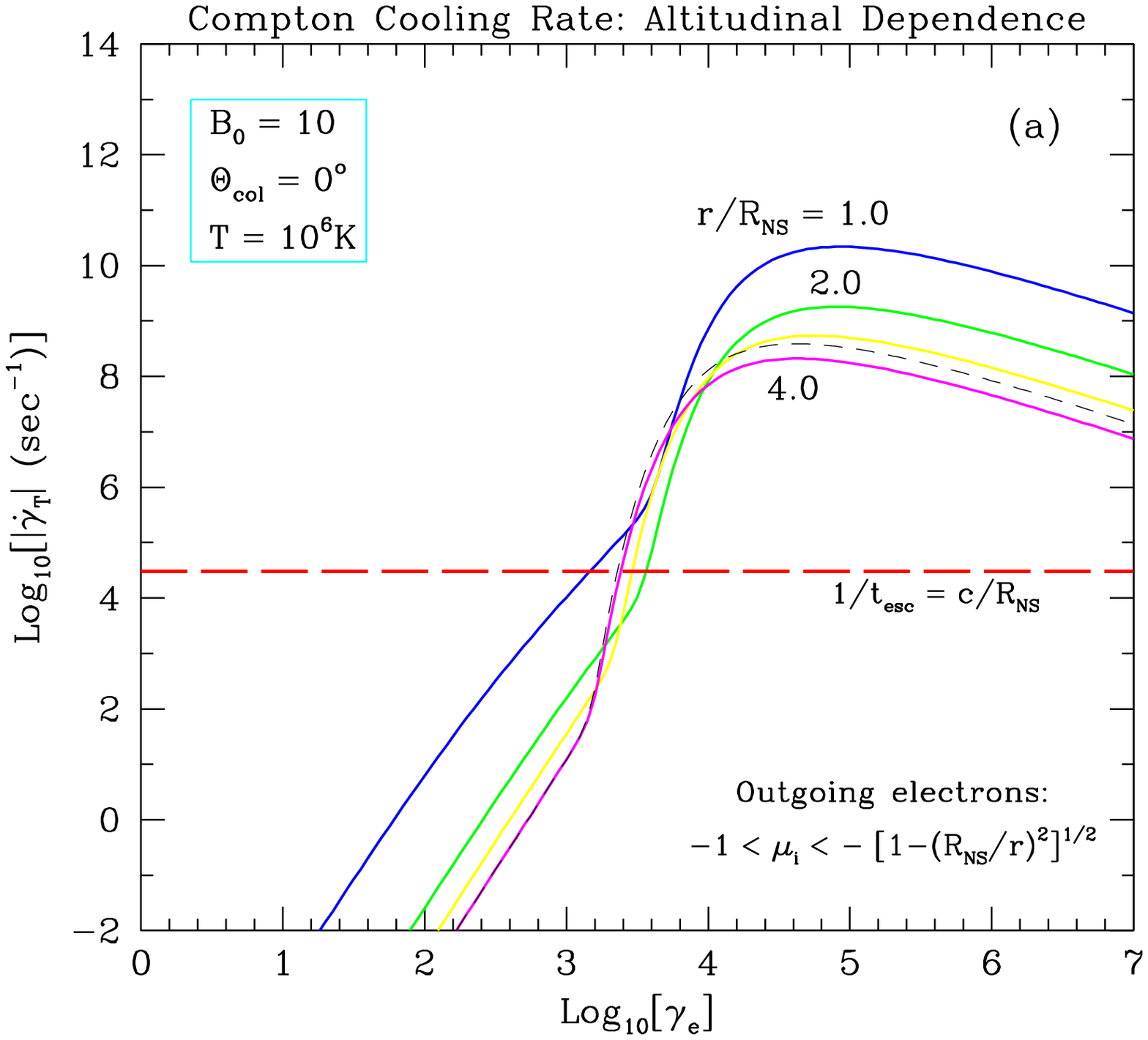}{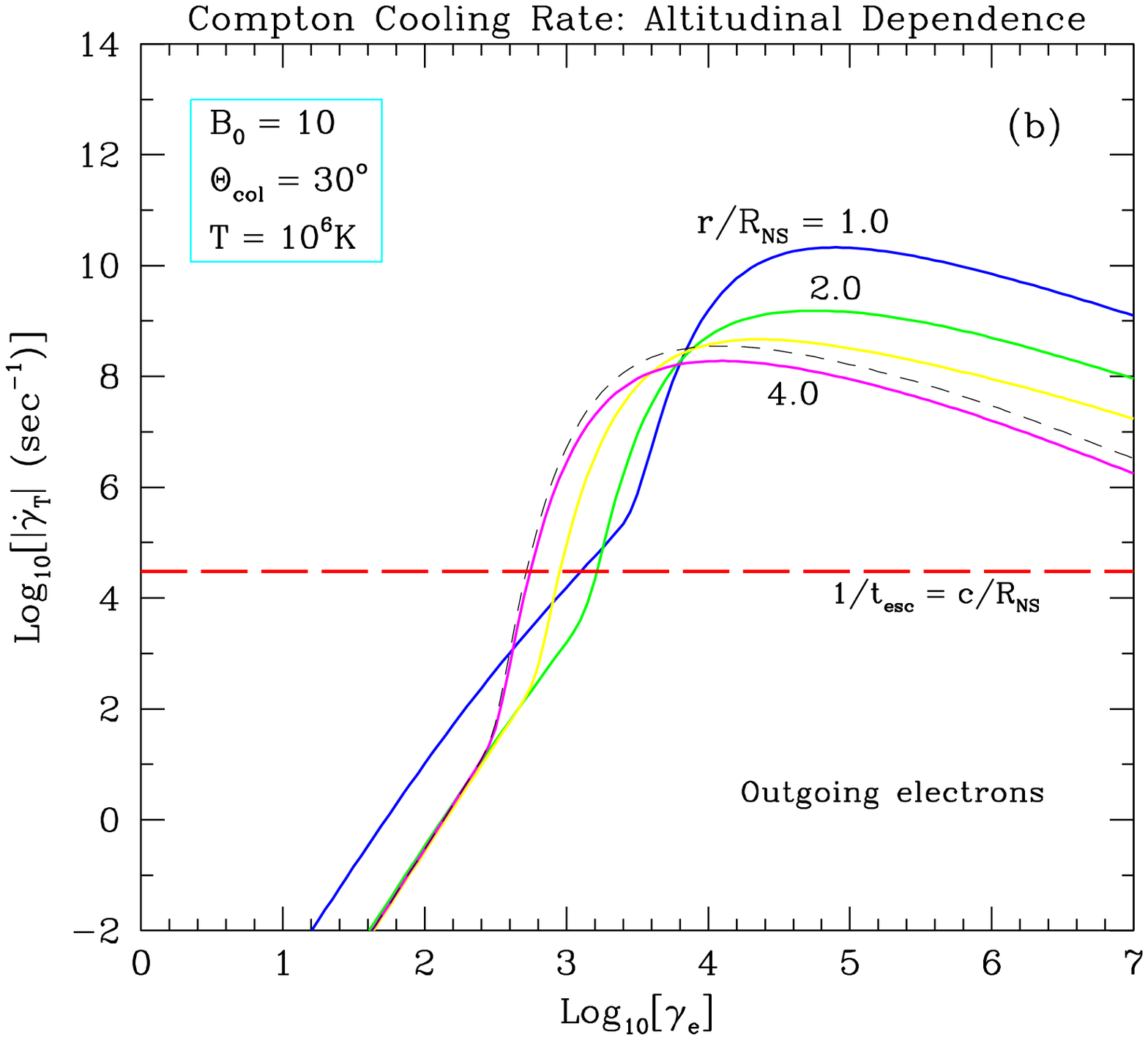}{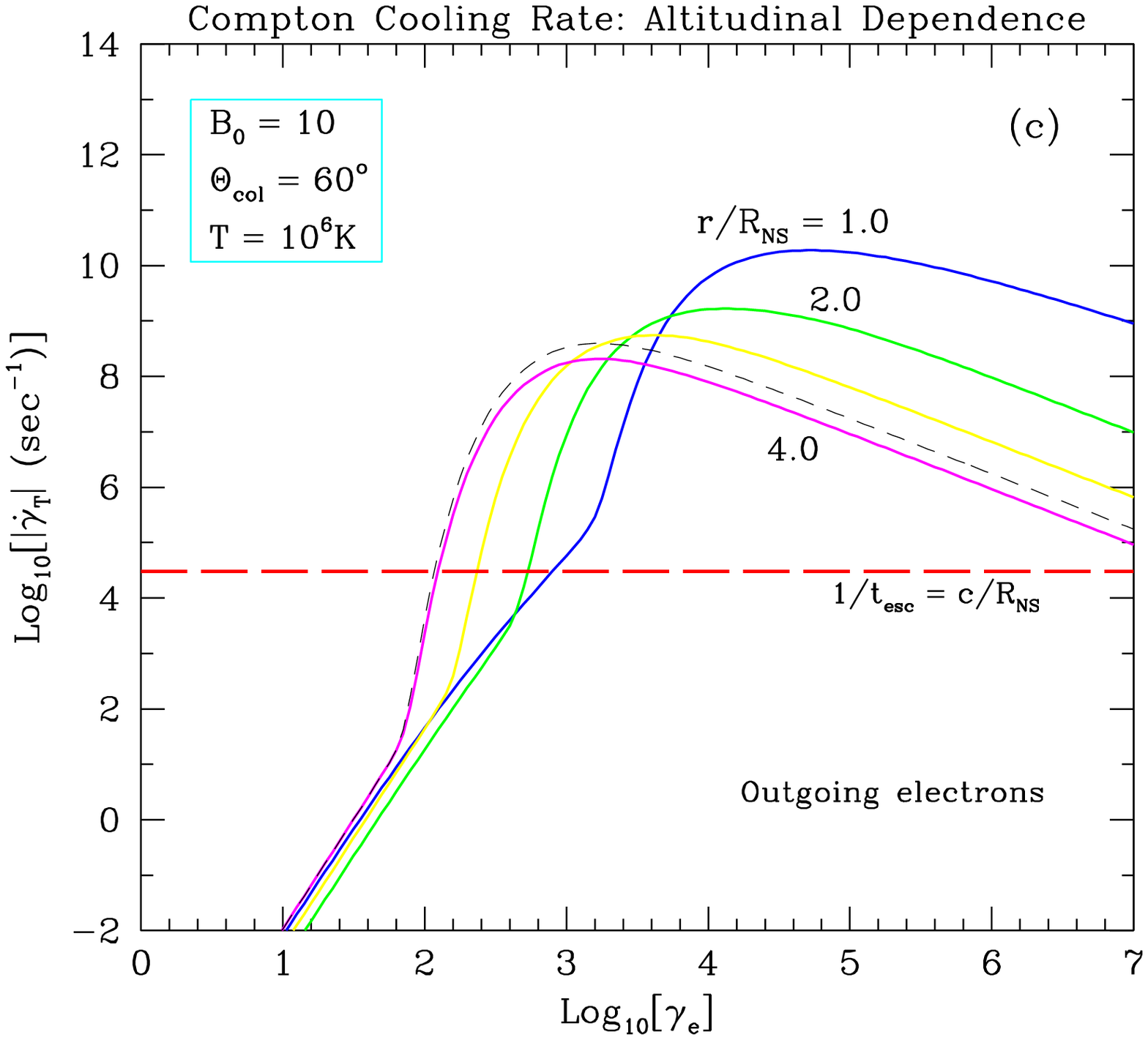}{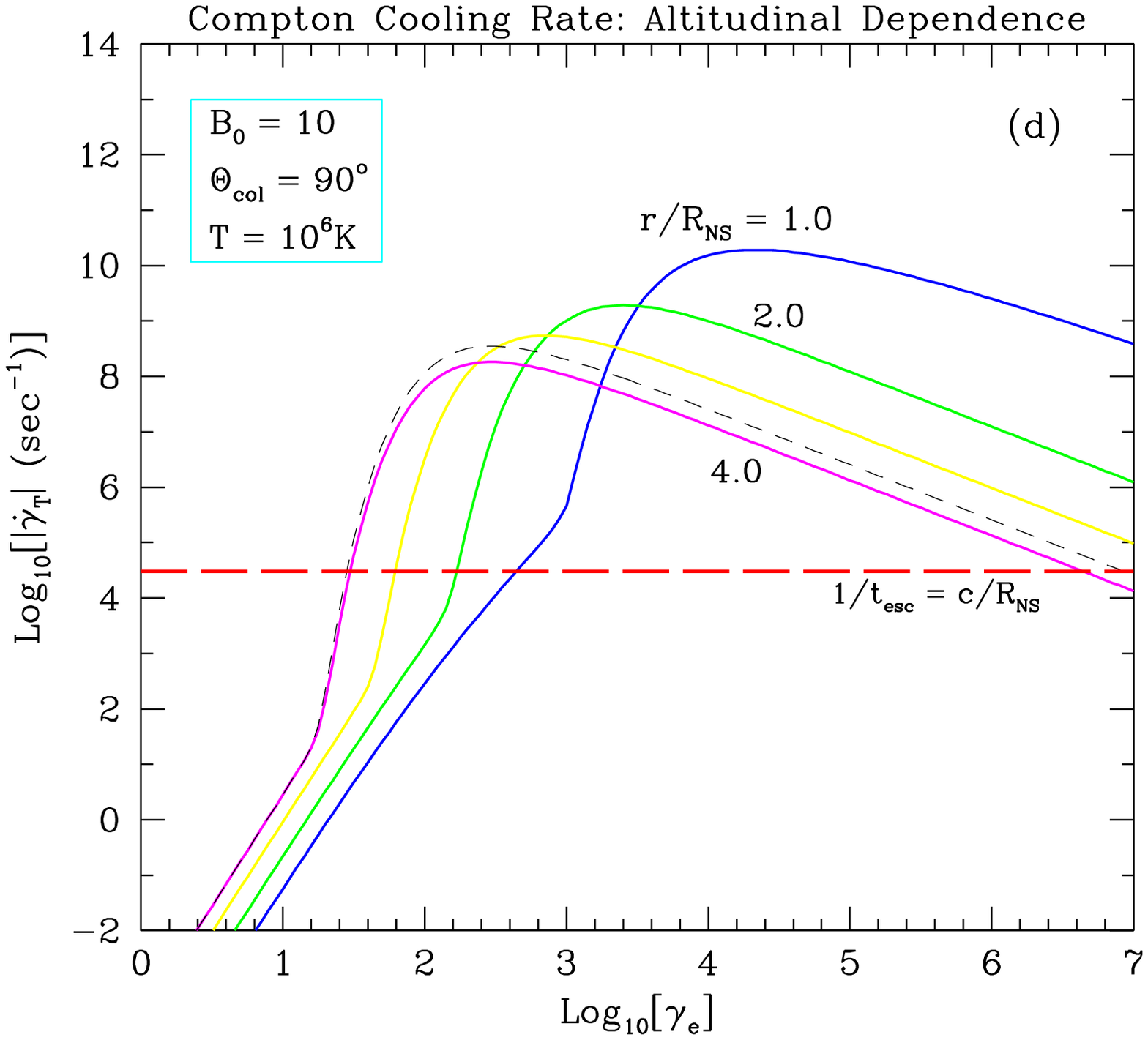}{
Resonant Compton cooling lengths for surface polar field strength
\teq{B_0=10}, as in Fig.~\ref{fig:resComp_th_ST_B10}, for interaction
locales at and above the surface (\teq{r/\rns =1} (blue), 2 (green), 3
(yellow, unlabelled), 4 (magenta), i.e., from top to bottom in the
resonant regime) and at different magnetic colatitudes: (a)
\teq{\Thetacol =0^{\circ}} (above the pole), (b) \teq{\Thetacol
=30^{\circ}}, (c) \teq{\Thetacol =60^{\circ}} and (d) \teq{\Thetacol
=90^{\circ}}, i.e. above the equator. The solid curves are for ST
evaluations, while the sole dashed curve in each panel is for JL rates
at \teq{r=4\rns}.  At lower altitudes, JL and ST cooling rates are
visually indistinguishable on this scale.  All curves are computed for
outgoing electrons, which at the equator becomes equivalent to inwardly
propagating \teq{e^-}.  The angular distributions for panels (b) and (c)
are displayed in Fig.~\ref{fig:ang_dist_3060}.
 \label{fig:cool_altitudes} } 
\end{figure*}

Numerical evaluations of the cooling rates at representative locales in
the magnetosphere are illustrated in Fig.~\ref{fig:cool_altitudes}.  The
rates were generally computed as fully three-dimensional integrals using
Eq.~(\ref{eq:cool_rate_final}) as a basis, inserting in the angular
distribution given by Eq.~(\ref{eq:ang_dist}), and then integrating over
the Planck spectrum in Eq.~(\ref{eq:Planck_spec}). The \teq{\Thetacol
=0^{\circ}}, \teq{r=\rns} case was computed as detailed in
Section~\ref{sec:therm_esoft}. Most curves are displayed for ST cases,
though each panel displays a single JL case for \teq{r=4\rns} for
comparison (JL computations at other altitudes are omitted in the
interest of clarity of the illustration).   The resonant portions of JL
cooling rates are closest to the depicted ST ones at the surface, where
the fields are the strongest, and exhibit the greatest departures from
ST values at high altitudes and therefore low local \teq{B} (but never
by more than a factor of two).  Below the resonant portions, the JL and
ST rates are, of course, identical. The general shapes and spectrum of
regimes are similar for all altitudes and colatitudes of the interaction
point: at low \teq{\gamma_e} the cooling is non-resonant and
approximately proportional to \teq{\gamma_e^4}, and at high Lorentz
factors, the cooling is resonant, sampling the Rayleigh-Jeans tail of
the Planckian, and scales roughly as \teq{1/\gamma_e} (i.e., modulo
logarithmic factors).  These global characteristics extend to both lower
and higher surface polar fields \teq{B_0}, though the specific values
for \teq{\gamma_e} for the onset of the resonant cooling, and the
normalizations of the cooling curves are modified accordingly, as
identified in Section~\ref{sec:cooling_gen}.

The principal differences in comparing curves for the various
colatitudes are imposed by the kinematic constraints due to the
\teq{\mu_{\pm}} values in Eq.~(\ref{eq:mu_pm_ang}) associated with the
angular distribution.  Above the magnetic pole, where photons and
electrons are predominantly chasing each other, altitudinal collimation
of the soft photons does not dramatically curtail the phase space for
accessibility to resonant interactions beyond the \teq{1/r^2} dilution
trend. There is significant compensation between the altitudinal decline
of the \teq{1+\beta_e\cos\Theta_i} factor in
Eq.~(\ref{eq:Lorentz_transform}) and the decline of \teq{B}, so that the
value of \teq{\gamma_e} marking the onset of resonant collisions does
not appreciably depend on \teq{r}. The normalization of the resonant
portion of the curves declines roughly as \teq{r^{-3}}, reflecting the
drop in the field strength at the locale of interaction. As the
colatitude increases, the reduction of \teq{\theta_c} not only
diminishes the number density of soft photons, but, for fixed
\teq{\gamma_e}, pushes the kinematic conditions that permit access to
the resonance deeper into the Rayleigh-Jeans tail.  This means that for
fixed \teq{\gamma_e}, there is both an impact of the altitude in
reducing the range of \teq{\mu_i} values that contribute to the cooling
rate, and a concomitant reduction in \teq{{\dot \gamma}_e} due to the
decline in the field strength.  Accordingly, these reinforcing effects
precipitate a decline with \teq{r} at the equator that is considerably
faster than that above the pole.  In addition, the effective pinning of
\teq{\Theta_i\approx \pi/2} in high altitude equatorial regions forces
the onset of resonant interactions to much lower Lorentz factors, being
approximately described by \teq{\gamma_e\propto B\propto r^{-3}}. This
trend is evident in Fig.~\ref{fig:cool_altitudes}d.

In all cases, polar, equatorial, or in between, the onset of resonant
cooling at \teq{30 < \gamma_e < 10^4} occurs on scales much shorter than
the neutron star light crossing time, and also the curvature emission
cooling timescale, at least for altitudes below around \teq{5 - 10\rns}
(i.e. coincident with the Compton resonaspheres enunciated in Baring \&
Harding 2007). This mirrors the discussion for surface polar locales at
the end of Section~\ref{sec:therm_esoft}. The implication is that for
moderately-scaled accelerating electric fields that permit resonant
Compton RRLA (the radiation-reaction limit), the maximum electron
Lorentz factors in the acceleration zone will be considerably lower for
electrodynamic potentials located in the equatorial regions as opposed
to polar zones.  The scattering kinematics then sets the maximum
dimensionless photon energy \teq{\erg_f} to be less than but around
\teq{\gamma_e} for Klein-Nishina interactions with strong recoil, or
more appropriately \teq{\gamma_e^2\Theta} for magnetic Thomson
collisions when \teq{B\lesssim 1} at higher altitudes.  Consequently,
one expects that the emergent spectrum from this process can potentially
extend to 1 GeV in polar emission geometries, such as is seen in some
high-field {\it Fermi} pulsars, but perhaps only to around 1 MeV in
equatorial acceleration locations.  This provides an interesting context
for resonant Compton upscattering models of magnetars, such as that
proposed by Baring \& Harding (2007): a quasi-equatorial RRLA picture at
altitudes above \teq{2-3\rns} should generate emission spectra that are
not too disparate from those needed to accommodate the Comptel upper
limits (e.g. Kuiper et al. 2006; den Hartog et al. 2008a,b; G\"otz et
al. 2006) at a few hundred keV to quiescent AXP and SGR hard X-ray tail
emission.  The detailed assessment of the viability of this picture
awaits an exploration of a combined photon emission and electron cooling
analysis in neutron star magnetospheres, including a foray into
scenarios invoking non-dipolar field morphologies (e.g. Thompson \&
Beloborodov 2005).  The development of time-dependent cooling of
electrons for such future models will build upon the foundations laid
and understanding presented in this paper.

\section{CONCLUSIONS}

This paper has presented a formulation for resonant Compton cooling
rates of ultrarelativistic electrons in magnetars and high-field pulsars
for arbitrary altitudes and colatitudes.  Such electrons can be
energized in the dynamic environments of these neutron stars, and their
cooling rates in collisions with thermal surface X-rays has never before
been explored using full QED scattering physics. We present both
numerical computations and analytic asymptotic forms of electron cooling
rates and reaction rates, using standard Boltzmann  equation
formulations in terms of the scattering differential cross section. The
evaluations of the cooling rates employ, for the first time, the
state-of-the-art Sokolov \& Ternov spin-dependent cross section for
magnetic Compton scattering.  This choice affords a more
physically-appropriate description of spin-dependent contributions to
the width of the cyclotron resonance, and generates significant
corrections relative to results obtained using the older Johnson \&
Lippmann cross section in this context, i.e. up to factors of 2 in
sub-critical fields.  Both the numerical work and the analytic
approximations will serve as useful tools for magnetic Compton
upscattering model development in the future.

It is demonstrated that the cooling rates scale with field as \teq{B^2}
in subcritical fields, but with a weaker dependence (\teq{\propto B}) in
supercritical field regimes, due to the Klein-Nishina and recoil
effects, and an ultra-quantum reduction of the cyclotron decay width.
Accordingly, a principal conclusion of this paper is that older magnetic
Thomson formulations will overestimate the cooling rates by orders of
magnitude if extrapolated to \teq{B\gg 1} cases.  This underpins the
major motivation for employing the full QED formalism developed herein.
Yet, as with previous magnetic Thomson cooling studies, it is found here
that distributing the soft photons with a thermal spectral energy
distribution, as is the case for surface X-ray seeds for scattering,
profoundly alters the collision kinematics accessed and the resulting
form of the electron cooling rates.   In such cases, the resonance is
always sampled above a certain Lorentz factor that is dependent on the
photon temperature, the interaction altitude and colatitude for a fixed
polar magnetic field. This enhances the cooling rates at Lorentz factors
above \teq{10^5} by orders of magnitude over those computed for
monoenergetic soft photons. Resonant Compton cooling is demonstrated to
be highly effective in cooling electrons; putative magnetospheric
acceleration of electrons can be halted effectively at lower altitudes
and Lorentz factors than typically occurs when using curvature
radiation.  Moroever, the cooling lengths are inferior to the neutron
star radius for \teq{10^3 < \gamma_e < 10^5} when the soft photon
temperature exceeds around \teq{3\times 10^5}K.  This renders the inner
magnetospheric, closed-field region a prime site for
Compton-cooling-limited non-thermal emission in magnetars.  In
particular, it is found that resonant cooling is accessed at lower
Lorentz factors in equatorial locales in a dipole field configuration,
suggesting that such a process might generate scattered photons only up
to energies of an MeV or so at high colatitudes; such a circumstance may
prove quite relevant to the hard X-ray tail quiescent emission seen in
magnetars.

These developments can be used in future investigations that
simultaneously treat the spatial and temporal evolution of photon and
electron distributions due to the resonant Compton process.  Their
general facility permits deployment in both kinetic equation or Monte
Carlo techniques.  Importing geometrical considerations and appropriate
acceleration/injection of electrons in such magnetospheric studies
should permit the detailed modeling of the hard X-ray tail spectra and
pulse profiles in observed in AXPs and SGRs, and also be pertinent to
gamma-ray emission from high-field pulsars such as PSR B1509-58.

\vskip 10pt
\acknowledgments 
We thank the referee for comments helpful to the polishing of the
manuscript, and are grateful to Alice Harding and Feryal \"Ozel for a
thorough reading of the paper and for providing numerous useful
suggestions for refining the presentation. We are also grateful for the
generous support of the National Science Foundation (grants AST-0607651
and PHY/DMR-1004811), and the NASA Astrophysics Theory Program 
through grants NNX06AI32G, NNX09AQ71G and NNX10AC59A.

\clearpage

\def\mn{M.N.R.A.S.}
\def\aassupp{{Astron. Astrophys. Supp.}}
\def\apss{{Astr. Space Sci.}}
\def\apj{ApJ}
\def\nat{Nature}
\def\aaps{{Astron. \& Astr. Supp.}}
\def\aap{{A\&A}}
\def\apjs{{ApJS}}
\def\sp{{Solar Phys.}}
\def\jgr{{J. Geophys. Res.}}
\def\jphysb{{J. Phys. B}}
\def\ssr{{Space Science Rev.}}
\def\araa{{Ann. Rev. Astron. Astrophys.}}
\def\nature{{Nature}}
\def\asr{{Adv. Space. Res.}}
\def\rmp{{Rev. Mod. Phys.}}
\def\prc{{Phys. Rev. C}}
\def\prd{{Phys. Rev. D}}
\def\pr{{Phys. Rev.}}


\appendix
\section{A: Analytic Reduction of Compton Cooling and Collision Rates in $\gamma_e\gg 1$ Domains}
 \label{sec:appendixA}
The expressions for the asymptotic resonant rates in supercritical fields that are 
summarized in Section~\ref{sec:cooling_gen}, and validate the numerical computations
therein, are derived in this Appendix.  For thermal photons of dimensionless 
temperature \teq{\Theta}, they correspond to the regime \teq{\gamma_e\Theta\gtrsim B}.
Yet, it must be noted that the core equations developed in this Appendix,
Eq.~(\ref{eq:cool_rate_genJL2_app}) for the Johnson \& Lippmann case,
and Eq.~(\ref{eq:cool_rate_genST1_app}) for the Sokolov \& Ternov formalism, 
are also applicable to non-resonant cases, but only under the \teq{\gamma_e\gg 1}
ultra-relativistic presumption.  Soft photon anisotropies at high altitudes and non-polar 
colatitudes are not encapsulated by the developments herein.

\subsection{A.1: Resonant Cooling Rates: Johnson \& Lippmann Formalism}
 \label{sec:JL_reduction}
The starting point is the full formulation in
Eq.~(\ref{eq:cool_rate_final}).  The resonant domain is
realized in the \teq{\gamma_e\gg 1} regime, for which the Jacobian in
Eq.~(\ref{eq:dergf_dcthetaf}) receives a leading order contribution of
\begin{equation}
   \biggl\vert \dover{\partial \erg_f}{\partial (\cos\theta_f) } \biggr\vert\,
   \;\approx\; \gamma_e\beta_e\omega_f\, \left[ 1 - \Psi \right]
   \quad ,\quad
   \Psi \; =\; \dover{(1-\mu )\, \omega_f\, (\omega_i - \omega_f\mu )}{
        2\omega_i - \omega_f-\zeta} \quad .
 \label{eq:dergf_dcthetaf_approx}
\end{equation}
The correction factor \teq{\Psi} is small in the limits \teq{B\to 0} and 
\teq{B\to\infty}, but contributes approximately 30-40\% in the near-critical field regime.
In addition, the relative velocity factor \teq{1-\beta_e\cos\theta_f} becomes
approximately \teq{\beta_e (1-\cos\theta_f)}.  Under these approximations, 
for isotropy of soft photons \teq{f(\mu_i)\to 1} on the cone (which is relaxed in
Section~\ref{sec:esoft_anisotropy}), the monoenergetic cooling rate simplifies to
\begin{equation}
   \dot{\gamma}_e \; \approx\; -  \dover{n_s\, c}{\mu_+-\mu_-}\, 
   \dover{1}{\gamma_e\erg_s^2}   \int_{\omega_-}^{\omega_+}
   \omega_i\, d\omega_i   \int_{-1}^1 d(\cos\theta_f )\, [1-\cos\theta_f]\,
   \omega_f \left[ 1 - \Psi \right] \, \dover{d\sigma}{d(\cos\theta_f) }
 \label{eq:cool_rate_approx1}
\end{equation}
As a reminder, here \teq{\omega_{\pm} = \gamma_e\erg_s\, (1+\beta_e\mu_{\pm})} defines the 
kinematic maximum and minimum of incoming photon energies in the ERF. 
The dominant contribution to the \teq{\omega_i} integration comes from
the resonance, so that the non-resonant terms in the JL cross section in 
Eqs.~(\ref{eq:sigma_unpol_JL}) 
can be neglected; it is not present in the approximate form for the 
ST cross section near resonance in Eq.~(\ref{eq:sigma_unpol_STapprox}).  
The Johnson \& Lippman differential cross section becomes
\begin{equation}
   \dover{d\sigJL}{d(\cos\theta_f)} \; \approx \; 
   \dover{3\sigt}{16} \, 
   \dover{ \omega_f^3\, e^{-\kappa} }{ \omega_i\, 
        [2\omega_i -\omega_f - \zeta \, ]}  \dover{\omega_f\, {\cal T}}{(\omega_i -B)^2 + (\Gamma /2)^2}
   \quad ,\quad
   \kappa \; =\; \dover{\omega_f^2\sin^2\theta_f}{2B}
 \label{eq:sigmaJL_unpol_app}
\end{equation}
for \teq{\zeta = \omega_i\omega_f  (1-\cos\theta_f)}.  The polarization-averaged
factor in the numerator is taken from Eq.~(\ref{eq:Tpar+Tperp}):
\begin{equation}
   \omega_f{\cal T} \; =\; 2 \omega_i 
                        - (1+\omega_i)\,\omega_f (1-\cos\theta_f)^2  \quad .
 \label{eq:wf_Tpar+Tperp}
\end{equation}
At this stage, the focus will be on analytic reduction of the the JL formulation.  

Using \teq{r=1/[1+\omega_i(1-\cos\theta_f)]}, 
a change of variables for the \teq{\theta_f} integration is employed, so that 
\teq{d(\cos\theta_f) = -dr/(r^2\omega_i)} with \teq{1/(1+2\omega_i) \leq r \leq 1}.  
Therefore, the scattering kinematics is described via
\begin{equation}
   \dover{\omega_f}{\omega_i}\; =\; \dover{2r}{1+\sqrt{1-q(r,\, \omega_i )}}\quad ,\quad
   q(r,\, \omega_i )\; =\; \dover{2}{\omega_i}\, (1-r)\, \Bigl[ (2\omega_i +1)r-1\Bigr]\quad .
 \label{eq:omega_omega_fin}
\end{equation}
The cooling rate reduces to
\begin{equation}
   \gammadotJL \; \approx\; - \dover{3}{16} \dover{n_s\, \sigt c}{\mu_+-\mu_-}\, 
   \dover{1}{\gamma_e\erg_s^2}  \int_{\omega_-}^{\omega_+}
   \dover{\omega_i^2\, d\omega_i}{(\omega_i -B)^2 + (\Gamma /2)^2} 
   \int_{1/(1+2\omega_i )}^1 dr\, \dover{(1-r)}{r^3} \,
   \left( \dover{\omega_f}{\omega_i}\right)^4\, 
   \dover{[1-\Psi ]\, \omega_f {\cal T}\, e^{-\kappa} }{2\omega_i-\omega_f-\zeta }\;\; ,
 \label{eq:cool_rate_approx_genJL1}
\end{equation}
The following identities then prove useful:
\begin{eqnarray}
   \omega_f\sin ^2\theta_f  &=& \dover{1}{r}\,\Bigl\{ 1-\sqrt{1-q} \Bigl\}\quad ,
   \nonumber\\
   \zeta &=&  \dover{2\omega_i\, (1-r)}{1+\sqrt{1-q}} \quad ,
 \label{eq:leq0_ident}\\
   \phi &=&  \dover{1-\sqrt{1-q}}{1+\sqrt{1-q}}\quad .
   \nonumber
\end{eqnarray}
Accordingly, the argument \teq{\kappa} of the exponential is purely a function
of \teq{q} at the peak of the resonance, and is most compactly expressed as  
\teq{\kappa =\phi\omega_i/B}.
It follows that \teq{r (2\omega_i-\omega_f-\zeta )= \omega_f\sqrt{1-q}} and
\teq{r\omega_f{\cal T} = 2\omega_i r - (1+\omega_i)\, [1-\sqrt{1-q}\, ]},
defining two useful identities that can compactly express the last factor in 
Eq.~(\ref{eq:cool_rate_approx_genJL1}).

The next step is to change to use \teq{\phi} as the angular integration variable,
since it proves more convenient than \teq{r}.  For this development,
the ERF incoming photon energy is expressed via the parameter
\begin{equation}
   z_{\omega}\; =\; \dover{1}{2} \biggl( \Phi + \dover{1}{\Phi} \biggr)
          \;\equiv\; 1 + \dover{1}{\omega_i}\quad , \quad
          \Phi\; =\; \dover{\sqrt{1+2\omega_i}-1}{\sqrt{1+2\omega_i}+1}\quad .
 \label{eq:zPhi_def}
\end{equation}
To facilitate the algebra, it proves convenient to effect the re-scaling
\teq{R=(1+z_{\omega})\, (1+\phi )r}, which spawns the identities
\begin{equation}
   \omega_i \; =\; \dover{1}{z_{\omega}-1}\quad ,\quad
   \omega_f \; = \; \dover{R}{z_{\omega}^2-1}\quad ,\quad
   \mu \; =\; z_{\omega} - \dover{z_{\omega}^2-1}{R}\, (1+\phi )\quad .
 \label{eq:Rscaling}
\end{equation}
The cooling rate can then be re-cast in the form
\begin{equation}
   \gammadotJL \; \approx\; - \dover{3}{8}\, \dover{n_s\, \sigt c}{\mu_+-\mu_-}\, 
   \dover{1}{\gamma_e\erg_s^2}  \int_{{\cal R}_-}^{{\cal R}_+} 
   \dover{e^{-\phi}\, dR}{(1+z_{\omega})^3}\, 
   \dover{1+\phi }{1-\phi}\, 
   \left({\cal R}_+ -R\right)\, 
   \Bigl\{ R - z_{\omega} (z_{\omega}+1)\phi  \Bigr\}\,
   \left[ 1 - \Psi \right]  \;\; ,
 \label{eq:cool_rate_approx4}
\end{equation}
for \teq{{\cal R}_{\pm}=(z_{\omega}\pm 1)(1+\phi )}.  We note also the identity
\begin{equation}
   \Psi\; =\; \dover{(1-\mu )\, r\, (\omega_i - \omega_f\mu )}{\sqrt{1-q}} 
   \; =\; \dover{ \left\{ {\cal R}_+ -R\right\}\, \left\{ (z_{\omega}-1){\cal R}_+ +1+z_{\omega} - Rz_{\omega} \right\}}{
          (1-\phi) (1+z_{\omega})^2} \quad .
 \label{eq:Psi_altform}
\end{equation}
Since \teq{q(r,\, \omega_i )} is quadratic in \teq{r}, there are two branches 
to the \teq{R\to\phi} transformation, defined by \teq{R=R_{\pm}},  where
\begin{equation}
   R_{\pm} \; =\; z_{\omega}\, (1+\phi) \pm \sqrt{1-2\phi z_{\omega} + \phi^2}  \quad .
 \label{eq:R_pm_def}
\end{equation}
Both branches, \teq{{\cal R}_- \leq R_-\leq z_{\omega} (1+\phi )} and
\teq{z_{\omega} (1+\phi ) \leq R_+\leq {\cal R}_+}, span the domain 
\teq{0\leq q\leq 2\omega_i /(1+2\omega_i )}, or equivalently, \teq{0\leq\phi\leq \Phi},
and for each we have 
\begin{equation}
   \left\vert \dover{dR}{d\phi} \right\vert \; =\;  
   \dover{1-\phi}{1+\phi}\,
   \dover{1+z_{\omega}}{\sqrt{1-2\phi z_{\omega} + \phi^2}}\quad .
 \label{eq:dr_dphi}
\end{equation}
After substantial cancellation from contributions from the two branches,
the algebra routinely yields a resonant asymptotic approximation
for the JL cooling rate in and near the resonance:
\begin{equation}
   \gammadotJL \; \approx\; - \dover{3}{4}\, \dover{n_s\, \sigt c}{\mu_+-\mu_-}\, 
   \dover{1}{\gamma_e\erg_s^2}   \int_{\omega_-}^{\omega_+}
   \dover{{\cal F}\left(z_{\omega},\, p\right) - {\cal G}\left(z_{\omega},\, p\right)}{
   (\omega_i -B)^2 + (\Gamma /2)^2} 
   \, \dover{\omega_i^4\, d\omega_i}{(1+2\omega_i)^2} 
   \quad ,\quad z_{\omega}\; =\; 1 + \dover{1}{\omega_i}\quad ,
 \label{eq:cool_rate_genJL2_app}
\end{equation}
for \teq{p=\omega_i/B\approx 1}, where the leading order term in the 
\teq{z_{\omega}\approx 1} and \teq{z_{\omega}\gg 1} domains has
the functional form
\begin{equation}
   {\cal F}(z,\, p)\; =\; \int_0^{\Phi (z)} \dover{ f(z,\, \phi ) \, e^{ -p\, \phi}\, d\phi }{
         \sqrt{1-2 z \phi + \phi^2}} 
   \quad ,\quad
   f(z,\, \phi )\; =\; z -1 + z (3- z) \phi - (1+z^2) \, \phi^2\quad .
 \label{eq:calFzpdef}
\end{equation}
Here, \teq{\Phi (z) = z - \sqrt{z^2-1}}.
The correction term corresponding to the \teq{\Psi} contribution is
\begin{equation}
   {\cal G}(z,\, p)\; =\; \dover{2 (z-1)}{(1+z)^2}
        \int_0^{\Phi (z)} \dover{ g(z,\, \phi ) \, e^{ -p\, \phi}\, d\phi }{ (1-\phi )
         \sqrt{1-2 z \phi + \phi^2}} \quad ,
 \label{eq:calGzpdef}
\end{equation}
where
\begin{equation}
   g(z,\, \phi ) \; =\; 2 z - \phi (1 - 2 z + 5 z^2) - \phi^2 (1 - 5 z + 3 z^2 - 3 z^3) 
      - \phi^3 (1 - 4 z + z^2 - 2 z^3) - \phi^4 (1 + z^2)\quad .
 \label{eq:gfn_def2}
\end{equation}
This separation is convenient, because it is trivial to show that 
\teq{{\cal G}(z,\, p)\to 0} in both the limits \teq{z\to 1^+} and \teq{z\to\infty}.
The integrals for \teq{{\cal F}} and \teq{{\cal G}} can be expressed in terms
of series of special functions, an analytic path that does not facilitate 
computation.  However, for numerical purposes, it
suffices to evaluate them as series in the \teq{(z-1)\ll 1} and \teq{z\gg 1}
limits, as outlined in Appendix B below.  Observe that Eq.~(\ref{eq:cool_rate_genJL2_app})
is reproduced in Eq.~(\ref{eq:cool_rate_genJL2}).

Numerical evaluation of Eq.~(\ref{eq:cool_rate_genJL2_app}) is somewhat 
cumbersome, and requires algorithmic intricacy and precision due to the presence of
the rapidly-varying Lorentz profile.  Since \teq{\Gamma/B\ll 1} for all 
domains of interest in this paper, 
the Lorentz profile in Eq.~(\ref{eq:sigma_unpol_JL}) can be approximated by a
delta function in \teq{\omega_i} space of identical normalization:
\begin{equation}
   \dover{1}{(\omega_i -B)^2 + (\Gamma/2)^2}\;\to\;
   \dover{2\pi}{\Gamma}\, \delta (\omega_i-B)\quad .
 \label{eq:resonance_approx}
\end{equation}
This standard mapping was adopted in Dermer (1990, for the specific non-relativistic
cyclotron decay case of \teq{\Gamma = 4\fsc B^2/3}) and Baring \& Harding (2007),
and renders the \teq{\omega_i} integration trivial.  Accordingly, one integral 
is removed from the computation, the integral that offers the greatest potential 
for numerical imprecision.  The result of this manipulation is
\begin{equation}
   \gammadotJL \; \approx\; - \dover{3\pi}{2}\, \dover{n_s\, \sigt c}{\mu_+-\mu_-}\, 
   \dover{B^2}{\gamma_e\erg_s^2\Gamma}  \, \dover{{\cal F}(z) - {\cal G}(z)}{(1+z)^2} 
   \quad ,\quad z\; =\; 1 + \dover{1}{B}\quad ,
 \label{eq:cool_rate_approx5}
\end{equation}
with the functional identification \teq{{\cal F}(z)\equiv {\cal F}(z,\, 1)}, 
\teq{{\cal G}(z)\equiv {\cal G}(z,\, 1)} representing the 
\teq{p=\omega_i/B=1} specialization.  In the high-field limit,
\teq{B\gg 1}, which is obtained via setting \teq{z\to 1}, we have 
\teq{{\cal G}(z)\to 0} and the integrand in Eq.~(\ref{eq:calFzpdef})
possesses no divergence as \teq{\phi\to\Phi\to 1}.  Then \teq{{\cal F}\to 2(1 - 2/e)},
giving
\begin{equation}
   \gammadotJL \; \approx\; - \dover{n_s\, \sigt c}{\mu_+-\mu_-}\, 
   \dover{3\pi B^2}{4\gamma_e\erg_s^2\Gamma}\,  \left( 1 - \dover{2}{e} \right)
   \quad ,\quad B\;\gg\; 1\quad .
 \label{eq:cool_rate_Bgg1_limit}
\end{equation}
Likewise, for \teq{B\ll 1}, \teq{z\gg 1} and \teq{\Phi\approx 1/(2z)}, yielding 
\teq{{\cal G}\to 0} and \teq{{\cal F}\to 2/3},
and the magnetic Thomson limit result becomes 
\begin{equation}
   \gammadotJL \; \approx\; - \dover{n_s\, \sigt c}{\mu_+-\mu_-}\, 
   \dover{\pi B^4}{\gamma_e\erg_s^2\Gamma}
   \quad ,\quad B\;\ll\; 1\quad .
 \label{eq:cool_rate_Bll1_limit}
\end{equation}
When substituting \teq{\Gamma\to 4\fsc B^2/3} for the ``non-relativistic'' cyclotron decay width,
this result generates Eq.~(\ref{eq:gammadot_res}), and is commensurate with the leading-order, high-\teq{\gamma_e} 
contribution from Eq.~(24) of Dermer (1990), specifically for mono-energetic soft photons.
We note in concluding this reduction of the Johnson and Lippmann resonant formalism
that Eqs.~(\ref{eq:cool_rate_genJL2_app}) and~(\ref{eq:cool_rate_approx5}) can 
be routinely integrated over a Planck spectrum for soft photons to yield a separable 
temperature factor contributing to the thermal, resonant cooling rate; this aspect is 
outlined in Section~\ref{sec:therm_esoft}.

\subsection{A.2: Resonant Cooling Rates: Sokolov \& Ternov Formulation}
 \label{sec:ST_reduction}
\begin{figure*}[t]
\figureoutpdf{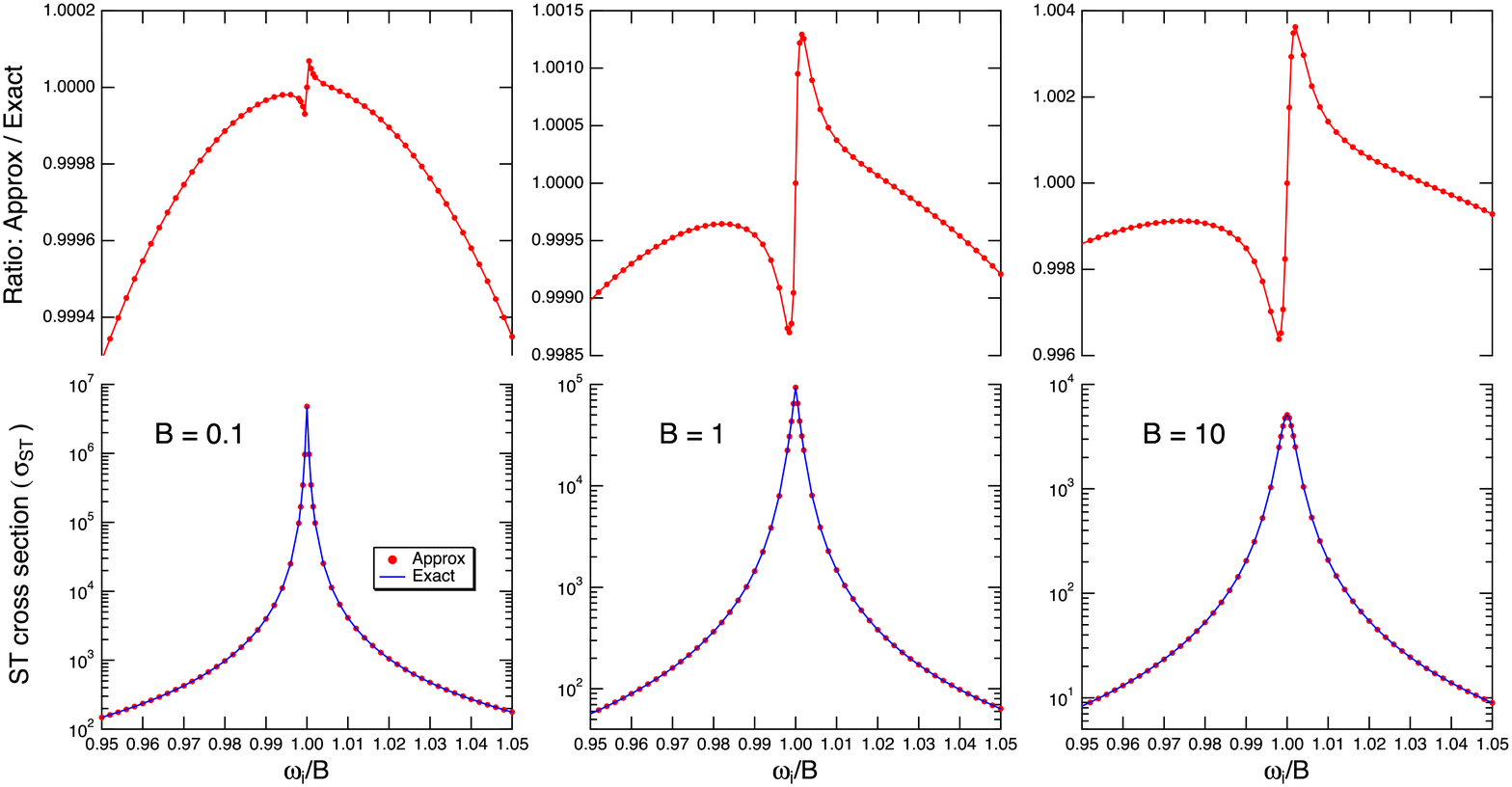}{7.0}{0.0}{-0.2}{
The total cross section for the Sokolov and Ternov (ST) QED formulation of
magnetic Compton scattering, for the field strengths \teq{B=0.1, 1, 10}, as 
labelled.  The panels focus on a small range \teq{0.95\leq \omega_i/B\leq 1.05} 
centered on the cyclotron resonance.  The numerical evaluation of the 
exact cross section derived in Gonthier et al. (2011) is presented as the
solid curves in the lower panels, and includes the spin-dependent 
cyclotron decay widths in the resonant denominators.  The approximation
in Eq.~(\ref{eq:sigmaST_unpol_approx}), derived in Gonthier et al. (2011),
when integrated over \teq{\theta_f} produces the points on the lower panels.
The ratio of the approximate cross section to the exact one is 
illustrated as points connected by lines in the upper three panels, clearly indicating the excellent 
precision of the approximation in Eq.~(\ref{eq:sigmaST_unpol_approx})
in this energy range near the peak and in the wings of the cyclotron resonance.
 \label{fig:ST_csect} } 
\end{figure*}

For Sokolov and Ternov states, the analytic manipulations proceed using an almost 
identical protocol.   Eq.~(\ref{eq:sigmaJL_unpol_app}) is replaced 
by the spin-dependent form for
the approximate differential cross section at and near resonance:
\begin{equation}
   \dover{d\sigST}{d(\cos\theta_f)} \; = \; 
   \dover{3\sigt}{64} \, 
   \dover{ \omega_f^2\, e^{-\kappa} }{ \omega_i\,  [ 2\omega_i-\omega_f-\zeta]\, \eperp^3}
                  \sum_{s=\pm 1} \dover{(\eperp + s)^2\, \Lambda_s
                         }{(\omega_i -B)^2 + (\Gamma_s /2)^2}   \quad ,
  \label{eq:sigmaST_unpol_approx}
\end{equation}
for \teq{\kappa = \omega_f^2\sin^2\theta_f/[2B]} and
\begin{equation}
   \Lambda_s\; =\; \left( 2\eperp -s \right)\, \omega_f^2 {\cal T} 
             + s\, \eperp^2\, (\omega_i-\omega_f)\quad .
 \label{eq:Lambda_s_def2}
\end{equation}
Here the quantum number \teq{s} labels the intermediate electron's spin state,
yielding the two Lorentz profiles that are described by the widths
\begin{equation}
   \Gamma_s \; =\; (\eperp + s) \, \dover{1+B}{\eperp}\,\Gamma
   \quad ,\quad 
   \eperp\; =\; \sqrt{1+2 B}\quad .
 \label{eq:ST_widths2}
\end{equation}
Again, \teq{\Gamma} is the same spin-averaged cyclotron width/decay rate 
as is used in the JL formulation.  Before proceeding with the analytic 
manipulations, a brief description of the origin of the approximation in 
Eq.~(\ref{eq:sigmaST_unpol_approx}) is offered.  A full derivation,
too lengthy to include here,
is presented in Gonthier et al. (2011, in preparation), a paper on the physics 
of magnetic Compton scattering focusing on the use of Sokolov and Ternov 
electron spin states.  The starting point for that work is Eq. ~(3.25) of 
Sina (1996), with the S-matrix elements in the ST formulation being 
encapsulated in Eqs.~(3.15) and~(3.16) of Sina's (1996) thesis.  
Gonthier et al. (2011) specialized this formal result to the 
case of incoming photons propagating along {\bf B}, and developed 
and simplified the algebra for the differential cross section.  Near the resonance, 
the approximation \teq{\omega_i\approx B} was invoked, for efficacy,
to condense the numerators of the squares of the S-matrix elements.
The result was Eq.~(\ref{eq:sigmaST_unpol_approx}).  A numerical 
illustration of the total cross section for the ST formulation is presented in 
Fig.~(\ref{fig:ST_csect}), exhibiting both the exact cross section (curves), and the 
integration over \teq{\theta_f} of the 
approximation in Eq.~(\ref{eq:sigmaST_unpol_approx}) in the bottom 
panels (as points).  The upper panels in the Figure highlight the ratio of the approximation 
to the exact computation, and clearly demonstrate that the approximation is 
accurate to a fraction of a percent for the fields \teq{B=0.1, 1, 10}.

Now, in terms of the development algebra, the \teq{ \omega_f^2 {\cal T}} piece 
has already been handled in the JL resonance asymptotics, yielding
an integral involving the \teq{{\cal F}(z_{\omega},\, p) - {\cal G}(z_{\omega},\, p)} factor.
The remaining piece possesses an integrand proportional to \teq{\omega_i-\omega_f},
and can be manipulated in a similar fashion through the changes of variables 
\teq{\theta_f\to\phi}, amounting to a replacement of the factor 
\teq{R - z_{\omega} (z_{\omega}+1)\phi} in Eq.~(\ref{eq:cool_rate_approx4})
by one proportional to \teq{(1+\phi)\, (R-z-1)}.  This algebraic rearrangement is 
routine.  Progress can be expedited by extracting out the \teq{{\cal F}-{\cal G}} piece that reproduces
half the JL result, i.e. provides the dominant contribution to the \teq{B\ll 1} limit.
The end result for the Sokolov and Ternov cooling rate
in and near the resonance can be cast in the form, summarized in Eq.~(\ref{eq:cool_rate_genST1}),
\begin{equation}
   \gammadotST \; \approx\; - \dover{3}{16}\, \dover{n_s\, \sigt c}{\mu_+-\mu_-}\, 
   \dover{1}{\gamma_e\erg_s^2}
   \sum_{s=\pm 1}  \int_{\omega_-}^{\omega_+}
   \dover{(s\eperp + 1)\, \Upsilon_s \left(z_{\omega},\, p\right)}{
   (\omega_i -B)^2 + (\Gamma_s /2)^2} 
   \, \dover{\omega_i^4\, d\omega_i}{(1+2\omega_i)^2} 
   \;\; ,\quad z_{\omega}\; =\; 1 + \dover{1}{\omega_i}\quad ,
 \label{eq:cool_rate_genST1_app}
\end{equation}
for \teq{p=\omega_i/B\approx 1}, and
\begin{equation}
   \Upsilon_s \left(z,\, p\right)\; =\; \left\{
   \left( \dover{s}{\eperp} +1 \right)\, \Bigl\lbrack{\cal F}(z,\, p) - {\cal G}(z, \, p) \Bigr\rbrack
    + \left( \dover{s}{\eperp} -1 \right) \, \Bigl\lbrack {\calFST}(z, \, p) 
                 - {\calGST}(z, \, p) \Bigr\rbrack \right\}\quad .
 \label{eq:Upsilon_def_app}
\end{equation}
This path of manipulation leads to the identification of two new integral functions:
\begin{eqnarray}
   \calFST (z,\, p) & = & (z+1)
       \int_0^{\Phi (z)} \dover{\phi (1-\phi)\, e^{-p\, \phi}}{\sqrt{1-2 z\phi + \phi^2}}\, d\phi\nonumber\\[-5.5pt]
 \label{eq:calFG_ST_def_app}\\[-5.5pt]
   \calGST (z, \, p) & = &  2\, \dover{z-1}{z+1} \int_0^{\Phi (z)} 
       \dover{\phi^2\, (z+2\phi z - \phi^2)\, e^{-p\, \phi}}{(1-\phi )\, \sqrt{1-2 z\phi + \phi^2}}\, d\phi\quad .\nonumber
\end{eqnarray}
The incorporation of spin-dependence necessitates a total of two integrals expressing 
the angular integrations, which in this paper are each divided into differences of terms 
(\teq{{\cal F}-{\cal G}} and \teq{{\cal F}_{\Delta}-{\cal G}_{\Delta}}) that isolate the 
mathematical character of the contributions to the Jacobian in Eq.~(\ref{eq:dergf_dcthetaf_approx}).
As with the JL resonant analytics above, for numerical purposes, it
suffices to evaluate each of these four functions as series in the \teq{(z-1)\ll 1} and \teq{z\gg 1}
limits, a development outlined below. 

Again, one may form a more compact analytic approximation to the resonant ST 
contribution by appealing to the fact that \teq{\Gamma/B\ll 1} for all 
domains of interest in this paper.
The Lorentz profile in Eq.~(\ref{eq:sigma_unpol_JL}) is then again approximated by a
delta function in \teq{\omega_i} space via Eq.~(\ref{eq:resonance_approx}).
Then the \teq{p\to 1} limit can be taken, the \teq{\omega_i} integration is trivial, and 
the two-dimensional integral in 
Eq.~(\ref{eq:cool_rate_genST1_app}) reduces to
\begin{equation}
   \gammadotST \; \approx\; - \dover{3\pi}{4}\, \dover{n_s\, \sigt c}{\mu_+-\mu_-}\, 
   \dover{1}{\gamma_e\erg_s^2\Gamma}  \, \dover{B^2}{(1+z)^2}\, 
   \Bigl\{  \left\lbrack {\cal F}(z) - {\cal G}(z) \right\rbrack
    +  \left\lbrack {\calFST}(z) - {\calGST}(z) \right\rbrack  \Bigr\}
   \quad ,\quad z\; =\; 1 + \dover{1}{B} \quad ,
 \label{eq:cool_rate_ST_tot_v2}
\end{equation}
again with the functional identification \teq{{\cal F}(z)\equiv {\cal F}(z,\, 1)}, 
\teq{{\cal G}(z)\equiv {\cal G}(z,\, 1)}, etc., representing the 
\teq{p=\omega_i/B=1} specialization.  Observe that the magnetic field-dependent
factors here are just represented by the \teq{\calRST (B)} function defined in 
Eq.~(\ref{eq:calRST_def}).  Observe also that Eq.~(\ref{eq:cool_rate_ST_tot_v2})
can be derived in an alternative fashion by asserting the Eq.~(\ref{eq:resonance_approx}) 
correspondence early on, and then following the same changes of variables
for the angular integration.  This amounts to employing the spin-averaged 
ST differential cross section
\begin{equation}
   \dover{d\sigST}{d(\cos\theta_f)} \; \approx \; 
   \dover{3\pi\sigt}{16\Gamma} \, 
   \dover{ \omega_f^2\, e^{-\kappa} \, \delta (\omega_i -B)\, {\cal S} }{ 
              \omega_i\, (1+\omega_i) \,  [ 2\omega_i-\omega_f-\zeta] }\, 
              \quad ,
  \label{eq:sigmaST_unpol_approx2}
\end{equation}
for
\begin{equation}
   {\cal S}\; =\; \dover{1}{2\eperp^2} \sum_{s=\pm 1} (\eperp + s)\, \Lambda_s 
   \; =\; \left( 2 - \dover{1}{\eperp^2} \right)\,  \omega_f^2 {\cal T} 
             + (\omega_i-\omega_f)\quad .
 \label{eq:calSdef}
\end{equation}
This is slightly more involved than the corresponding JL form, which is obtained by 
substituting \teq{{\cal S}\to 2 (1+B) \omega_f^2 {\cal T}}.  Note that an identical 
alternative protocol for derivation applies to the JL resonant asymptotic result
in Eq.~(\ref{eq:cool_rate_approx5}).

The high field limit of the ST resonant rate is readily obtained.   Setting \teq{z\to 1}, we have 
\teq{{\cal F}(z)\to 2(1-2/e)}, as before, and 
\teq{{\cal F}_{\Delta} \to 2(1-2/e)} also.  The \teq{{\cal G}(z)} and \teq{{\cal G}_{\Delta}(z)} terms 
contribute zero in this limit, yielding
\begin{equation}
    \gammadotST \; \approx\; - \dover{n_s\, \sigt c}{\mu_+-\mu_-}\, 
   \dover{3\pi B^2}{4\gamma_e\erg_s^2\Gamma}\,  \left( 1 - \dover{2}{e} \right)
   \quad ,\quad B\;\gg\; 1\quad ,
 \label{eq:STcool_rate_Bgg1_limit}
\end{equation}
a result identical to that in Eq.~(\ref{eq:cool_rate_Bgg1_limit}) for the 
JL cross section formalism.  In the \teq{B\ll 1} limit, since \teq{z\gg 1} and 
\teq{\Phi\approx 1/(2z)}, again the \teq{{\cal G}(z)} and \teq{{\cal G}_{\Delta}(z)} 
terms can be neglected, indicating that the \teq{\omega_i-\omega_f} term
is of order \teq{O(1/z)}.   Similarly, the 
\teq{{\cal F}_{\Delta}(z)} contribution is small.   Then, since
\teq{{\cal F}(z)\to 2/3} when \teq{z\to\infty}, the Thomson limit result is 
\begin{equation}
    \gammadotST \; \approx\; - \dover{n_s\, \sigt c}{\mu_+-\mu_-}\, 
   \dover{\pi B^4}{2\gamma_e\erg_s^2\Gamma}
   \quad ,\quad B\;\ll\; 1\quad ,
 \label{eq:STcool_rate_Bll1_limit}
\end{equation}
i.e. half that of the Johnson \& Lippman value in Eq.~(\ref{eq:cool_rate_Bll1_limit}).


\subsection{A.3: Compton Upscattering Collisional Rates}
 \label{sec:coll_reduction}
Now the focus turns to {\it collision rates}, i.e. inverse lifetimes for Compton
upscattering collisions that are employed in the computation of mean energy 
losses for the electrons that are considered in Section~\ref{sec:mean_energy}.
These are mathematically very similar to the cooling rates, being just devoid 
of an energy factor \teq{\erg_f-\erg_i\approx \erg_f} in the integrand.  Accordingly,
the reaction rate can quickly be written down by introducing a factor
\teq{-1/\erg_f=-(1+z_{\omega})/\gamma_e/({\cal R}_+-R)}
into Eq.~(\ref{eq:cool_rate_approx4}) for the JL case.  This reduces the order of the 
polynomial factors in the integrand by one.  The manipulations then 
parallel those for the cooling rates, and the end result is of a very similar 
form.  For JL states, one arrives at
\begin{equation}
   \dover{1}{\tauJL} \; \approx\; \dover{3}{4}\, \dover{n_s\, \sigt c}{\mu_+-\mu_-}\, 
   \dover{1}{\gamma_e^2\erg_s^2}   \int_{\omega_-}^{\omega_+}
   \dover{{\cal F}_{\tau}\left(z_{\omega},\, p\right) - {\cal G}_{\tau}\left(z_{\omega},\, p\right)}{
   (\omega_i -B)^2 + (\Gamma /2)^2} 
   \, \dover{\omega_i^3\, d\omega_i}{1+2\omega_i} 
   \quad ,\quad z_{\omega}\; =\; 1 + \dover{1}{\omega_i}\quad ,
 \label{eq:coll_rate_JL_app}
\end{equation}
where again, \teq{p=\omega_i/B}.  Now we have two new functions expressing 
the integrals over photon scattering angles in the ERF:
\begin{equation}
   {\cal F}_{\tau}(z,\, p)\; =\; \int_0^{\Phi (z)} \dover{ f_{\tau}(z,\, \phi ) \, e^{ -p\, \phi}\, d\phi }{
         \sqrt{1-2 z \phi + \phi^2}} 
   \quad ,\quad
   f_{\tau}(z,\, \phi )\; =\; z (1 - z \phi ) \quad ,
 \label{eq:calFtauzpdef}
\end{equation}
where again, \teq{\Phi (z) = z - \sqrt{z^2-1}}.
The correction term corresponding to the \teq{\Psi} contribution is
\begin{equation}
   {\cal G}_{\tau}(z,\, p)\; =\; \dover{z-1}{(1+z)^2}
        \int_0^{\Phi (z)} \dover{ g_{\tau}(z,\, \phi ) \, e^{ -p\, \phi}\, d\phi }{ (1-\phi )
         \sqrt{1-2 z \phi + \phi^2}} \quad ,
 \label{eq:calGtauzpdef}
\end{equation}
with
\begin{equation}
   g_{\tau}(z,\, \phi ) \; =\; 2 z - \phi (1 - z + 4 z^2) + \phi^2 z (5 + z + 2 z^2) 
      - \phi^3 (1 + z^2) \quad .
 \label{eq:gtaufn_def}
\end{equation}
This separation yields a dominance of \teq{{\cal F}_{\tau}} in both the limits 
\teq{z\to 1^+} and \teq{z\to\infty}, for which it is trivial to show that 
\teq{{\cal G}_{\tau}(z,\, p)\to 0}.
The reduction proceeds along similar lines for the ST states, resulting in
\begin{equation}
   \dover{1}{\tauST} \; \approx\; \dover{3}{16}\, \dover{n_s\, \sigt c}{\mu_+-\mu_-}\, 
   \dover{1}{\gamma_e^2\erg_s^2}
   \sum_{s=\pm 1}  \int_{\omega_-}^{\omega_+}
   \dover{(s\eperp + 1)\, \Upsilon^{\tau}_s \left(z_{\omega},\, p\right)}{
   (\omega_i -B)^2 + (\Gamma_s /2)^2} 
   \, \dover{\omega_i^3\, d\omega_i}{1+2\omega_i} 
   \;\; ,\quad z_{\omega}\; =\; 1 + \dover{1}{\omega_i}\quad ,
 \label{eq:coll_rate_ST_app}
\end{equation}
for \teq{p=\omega_i/B\approx 1}, and
\begin{equation}
   \Upsilon^{\tau}_s \left(z,\, p\right)\; =\; \left\{
   \left( \dover{s}{\eperp} +1 \right)\, \Bigl\lbrack{\cal F}_{\tau}(z,\, p) - {\cal G}_{\tau}(z, \, p) \Bigr\rbrack
    + \left( \dover{s}{\eperp} -1 \right) \, \Bigl\lbrack {\calFtauST}(z, \, p) 
                 - {\calGtauST}(z, \, p) \Bigr\rbrack \right\}\quad .
 \label{eq:Upsilontau_def_app}
\end{equation}
This arrangement seeds the identification of two new integral functions:
\begin{eqnarray}
   \calFtauST (z,\, p) & = & 
       \int_0^{\Phi (z)} \dover{(1-z\phi )\, e^{-p\, \phi}}{\sqrt{1-2 z\phi + \phi^2}}\, d\phi\nonumber\\[-5.5pt]
 \label{eq:calFGtau_STtau_def_app}\\[-5.5pt]
   \calGtauST (z, \, p) & = &  \dover{z-1}{z+1} \int_0^{\Phi (z)} 
       \dover{\phi^2 (1+2z-\phi )\, e^{-p\, \phi}}{(1-\phi )\, \sqrt{1-2 z\phi + \phi^2}}\, d\phi\quad .\nonumber
\end{eqnarray}
Analytics and evaluation of these functions are addressed in Appendix B.

As with the cooling rates, it is possible to form more compact analytic approximations to the 
resonant contributions by invoking the delta function in \teq{\omega_i} space to the
Lorentz profile via Eq.~(\ref{eq:resonance_approx}), since \teq{\Gamma/B\ll 1} for all 
fields.  Then the \teq{p\to 1} limit can be taken, the \teq{\omega_i} integrations are trivial, and 
the two-dimensional integral in Eq.~(\ref{eq:coll_rate_JL_app}) collapses to
\begin{equation}
   \dover{1}{\tauJL} \; \approx\;  \dover{3\pi}{2}\, \dover{n_s\, \sigt c}{\mu_+-\mu_-}\, 
   \dover{B^2}{\gamma_e^2\erg_s^2\Gamma}  \, \dover{{\cal F}_{\tau}(z) - {\cal G}_{\tau}(z)}{1+z} 
   \quad ,\quad z\; =\; 1 + \dover{1}{B}\quad ,
 \label{eq:coll_rate_JLdelta_app}
\end{equation}
and likewise for the ST form in Eq.~(\ref{eq:coll_rate_ST_app}):
\begin{equation}
   \dover{1}{\tauST} \; \approx\;  \dover{3\pi}{4}\, \dover{n_s\, \sigt c}{\mu_+-\mu_-}\, 
   \dover{1}{\gamma_e^2\erg_s^2\Gamma} 
   \dover{B^2}{1+z}\, \Bigl\{  \left\lbrack {\cal F}_{\tau}(z) - {\cal G}_{\tau}(z) \right\rbrack
    +  \left\lbrack {\calFtauST}(z) - {\calGtauST}(z) \right\rbrack  \Bigr\}  \quad  .
 \label{eq:coll_rate_STdelta_app}
\end{equation}
Using the fact that \teq{{\cal F}_{\tau}\to 2/3} as \teq{z\to\infty} yields the dominant 
contribution for \teq{B\ll 1}, and \teq{{\cal F}_{\tau}\to 1-1/e} and \teq{\calFtauST\to 1-1/e} 
as \teq{z\to 1^+} dominate the \teq{B\gg 1} case, one arrives at the asymptotic forms
\begin{equation}
   \dover{1}{\tauJL} \; \approx\;  \dover{n_s\, \sigt c}{\mu_+-\mu_-}\, 
   \dover{\pi B^2}{\gamma_e^2\erg_s^2\Gamma} 
   \cases{ \vphantom{\Bigl(} B \quad &, $\;\;B\,\ll\, 1\;\;$,\cr
    \dover{3}{4} \left( 1 - \dover{1}{e} \right) \quad &, $\;\;B\,\gg\, 1\;\;$,\cr}
 \label{eq:coll_rate_JLdelta_asymp}
\end{equation}
and
\begin{equation}
   \dover{1}{\tauST} \; \approx\;  \dover{n_s\, \sigt c}{\mu_+-\mu_-}\, 
   \dover{\pi B^2}{\gamma_e^2\erg_s^2\Gamma} 
   \cases{ \dover{B}{2} \quad &, $\;\;B\,\ll\, 1\;\;$,\cr
    \dover{3}{4} \left( 1 - \dover{1}{e} \right) \quad &, $\;\;B\,\gg\, 1\;\;$.\cr}
 \label{eq:coll_rate_STdelta_asymp}
\end{equation}
As with the cooling rates, the collision rates for the JL and ST formulations 
differ by a factor of two in the highly-subcritical field limit, but are 
identical in ultra-quantum domains, \teq{B\gg 1}, as expected.
Combining these properties automatically gives a fractional 
mean energy loss per collision that is independent of the cross
section employed in both the \teq{B\ll 1} and \teq{B\gg 1} limits.

In the non-resonant regime, either Eq.~(\ref{eq:coll_rate_JL_app})
or Eq.~(\ref{eq:coll_rate_ST_app}) can be used to derive the
asymptotic collision rate appropriate to \teq{\omega_+\ll B} cases.
Remembering that they are both posited in the \teq{\gamma_e\gg 1}
approximation, an extra factor of \teq{1/\beta_e} must be 
reintroduced to capture the more general nature of the integral
over Eq.~(\ref{eq:scatt_spec}).
Considering the JL formulation, as \teq{z_{\omega}\to\infty} 
the \teq{{\cal G}_{\tau}} term becomes insignificant, and 
\teq{{\cal F}_{\tau}\to 2/3}, resulting in an integrand that 
is proportional to \teq{\omega_i^3/B^2}.  This formally includes 
only the resonant term; the non-resonant term in the cross
section contributes an identical value in this limit (deducible,
for example, from the presentation in Section~\ref{sec:cooling_Thom}), 
so that a factor of two is introduced.  The integral over \teq{\omega_i} is then trivial,
producing a dependence \teq{1/\tau_e\propto \erg_s^2}.  This can
then be routinely integrated over the Planck spectrum in 
Eq.~(\ref{eq:Planck_spec}) to assemble the limiting non-resonant result
\begin{equation}
   \dover{1}{\tau_e}\; \approx\; 
   \dover{6\zeta (5)\Omega_s}{\pi^2}\, \dover{ \sigt c}{\lambar^3} \; \dover{\gamma_e^2}{\beta_e}
   \, \dover{\Theta^5}{B^2}  \; \Bigl\{ (1+\beta_e\mu_+)^4 - (1+\beta_e\mu_-)^4 \Bigr\} 
   \quad ,\quad   \gamma_e\Theta (1+\beta_e\mu_+)\; \ll\;  B\quad .
 \label{eq:collrate_belowres_th}
\end{equation}
The subscripts JL and ST are suppressed because this 
form applies to either formalism.  Generating a rate proportional 
to \teq{\gamma_e^2} is naturally expected given that the
cross section in the ERF scales as \teq{\omega_i^2},
with \teq{\omega_i\propto\gamma_e}.

\section{B: Evaluation of the Angular Integrals}
 \label{sec:angular_int}

The cooling and collisional rates developed in Appendix A subsume the 
angular integration into a set of relatively compact integrals.  This Appendix 
hones these integrals by expressing them in terms of four key integrals 
that are chosen to afford stability of numerical evaluation.  Specifically, 
the \teq{z\approx 1} range can be problematic in some of the integrals due
to singularities in the integrands.  The first step is 
to isolate identities for the more numerically pernicious integrals in terms
of the more benign ones.  Consider first \teq{{\cal F}(z, \, p)} and \teq{\calFST (z, \, p)}.
Both are fairly routine to compute in the \teq{z\to 1^+} limit since the singularity 
at \teq{\phi =\Phi\approx 1} in the denominator is negated by an emerging root
in the numerator.  Define a class of integrals
\begin{equation}
   {\cal I}_{\nu}(z,\, p)\; =\; \int_0^{\Phi (z)} e^{-p\, \phi} \, \left(1-2 z\phi + \phi^2\right)^{\nu/2}\, d\phi\quad .
 \label{eq:calI_def}
\end{equation}
Then rearrangement of the quadratic factor in  the numerator of the integrands
can be employed to establish the identity
\begin{equation}
   {\cal F}(z, \, p)\; =\; z\, \calFST (z, \, p) + (z-1) \, {\cal I}_1(z,\, p)\quad .
 \label{eq:calF_ident}
\end{equation}
All three of these integrals are particularly stable to numerical integration, 
and we opt to use \teq{\calFST} and \teq{{\cal I}_1} as a basis for the 
determination of the JL and ST rates.

The integrals \teq{{\cal G}(z, \, p)} and \teq{\calGST (z, \, p)} are more 
pathological around \teq{z\approx 1}, and since \teq{\calGST (z, \, p)}
possess a more compact integrand, we seek a relation between 
the two functions.   In analogy with the \teq{{\cal F}(z, \, p)} case, we
define another class of integrals
\begin{equation}
   {\cal J}_{\nu}(z,\, p)\; =\; \int_0^{\Phi (z)} \dover{e^{-p\,\phi}}{1-\phi} 
   \, \left(1-2 z\phi + \phi^2\right)^{\nu/2}\, d\phi\quad ,
 \label{eq:calJ_def}
\end{equation}
which are introduced because the integral for \teq{{\cal G}} cannot be 
expressed simply in terms of the \teq{{\cal I}_{\nu}} functions.
The integrand for \teq{{\cal G}}
can be efficaciously simplified using the identification of a class of 
polynomials that result from useful perfect derivatives:
\begin{equation}
   p_k(z,\, \phi ) \; =\; e^{p\,\phi} \sqrt{1-2 z \phi + \phi^2}\,
   \dover{d}{d\phi} \left\{ \phi^k e^{-p\,\phi} \sqrt{1-2 z \phi + \phi^2}\, \right\}\quad .
 \label{eq:pk_def}
\end{equation}
%
Because \teq{1-2 z \Phi + \Phi^2=0}, these readily 
generate the following useful integral identities:
\begin{equation}
  \int_0^{\Phi} \dover{e^{ -\phi}\, p_0(z,\, \phi )}{\sqrt{1-2 z \phi + \phi^2}} \; d\phi \; =\; -1
  \quad ,\quad
  \int_0^{\Phi} \dover{e^{ -\phi}\, p_n(z,\, \phi )}{\sqrt{1-2 z \phi + \phi^2}} \; d\phi \; =\; 0
  \quad \hbox{for}\quad n \geq 1\quad .
 \label{eq:phi_integ_ident}
\end{equation}
These are used to reduce the order of the polynomial \teq{g(z,\, \phi )} 
in Eq.~(\ref{eq:gfn_def2}).  The same can be done for the quartic 
in the numerator of \teq{\calGST (z,\, p)}.  This process incurs an
increase in the polynomial order in the parameter \teq{z}.
Rather than solve each explicitly, which effectively incorporates \teq{{\cal I}_{-1}} terms,  
it is expedient to form a linear combination of \teq{{\cal G}}, \teq{\calGST}, \teq{{\cal J}_1} and 
\teq{{\cal I}_1} and solve for the coefficients as a linear algebra problem, demanding identity 
in all powers of \teq{\phi}.  The result is quickly obtained:
\begin{eqnarray}
    && \left[2 + p + 2 (p-1) z \right] (z+1)^2\, {\cal G}(z,\, p) \; =\; 
              \left[ 2 + (p-2) z + (3 p + 2) z^2 + 2 (p-1) z^3\right] \, (z+1)\, \calGST(z, \, p)\nonumber\\[4pt]
    && \qquad\qquad\qquad  + 2 (z-1) \left[ - 3 - 2 p + (p+10) z + 9 (p-1) z^2 - 2 (p-1) z^3 \right] \, {\cal J}_1(z)
  \label{eq:calG_ident_v2}\\[4pt]
    && \qquad\qquad\qquad   - 2 (z-1)^2 \left[ 4 + 3 p + 3 p z - 2 (p-1) z^2\right] \, {\cal I}_1(z) +  2(z-1)^2  \quad .\nonumber
\end{eqnarray}
Through this relation, and~ Eq.~(\ref{eq:calF_ident}), we can express the cooling rates 
in terms of the basis integrals \teq{\calFST}, \teq{\calGST}, \teq{{\cal J}_1} and 
\teq{{\cal I}_1}, and these four functions form the anchors of our numerical evaluations.
The same protocol can be adopted for the collisional rates so that no new functions 
need to be computed.  The relevant identities are 
\begin{eqnarray}
   (1+2z)\, {\cal F}_{\tau} (z,p) & = & z\, (1+2 z)\, {\cal F}_{\tau\Delta}(z,p)
         \; =\; z \Bigl\{ 1+ {\cal F}_{\Delta} + (1-p+z){\cal I}_1 \Bigr\} \nonumber\\
   (1+2z)  (1+z)^2 {\cal G}_{\tau} (z, p) & = & z\, (z+1)\, (1 + {\cal F}_\Delta) 
         -  (1 + 2 z) \, \Bigl\{ z \left[  z ( z-2 ) + 4 \right] -1 \Bigr\} \, {\cal J}_1 \nonumber\\[-6.5pt]
  \label{eq:calFGtau_ident}\\[-6.5pt]
        && \quad -  \left[ 1 + z (p + (-3 + p) z - 2 z^3) \right] {\cal I}_1\nonumber\\[2pt]
   (1+z)\, (1+2z) \, {\cal G}_{\tau\Delta}(z,p) & = & 
          1+ {\cal F}_{\Delta} +(2z^2-p){\cal I}_1 - z\, (2z+1){\cal J}_1 \quad .\nonumber
\end{eqnarray}
These suffice to evaluate the reaction rates in Eqs.~(\ref{eq:coll_rate_JL_app})
and~(\ref{eq:coll_rate_ST_app}) in terms of the same four basis integrals.
It should be remarked that \teq{{\cal G}_{\Delta}} is not more stable, numerically,
in the neighborhood of \teq{z=1} than \teq{{\cal G}}, \teq{{\cal G}_{\tau}} or 
\teq{{\cal G}_{\tau\Delta}}; it is preferred only because it is analytically more compact.

The procedure with evaluating all four integrals is to change variables to 
\teq{x=1-\phi/\Phi} and then
form a Taylor series expansion of the exponential in the integrand in terms 
of the parameter \teq{p\Phi}.  The result is a convergent series, 
rapidly so in the limit of \teq{p\Phi\ll 1}, and collectively, or term-by-term,
each of the integrals assumes a form expressible in terms of elementary 
functions:
\begin{equation}
   \left( A + B \arctan (\sqrt{\Phi}) + C  \log_e \left[\dover{1 + \Phi}{1-\Phi}\right] \right) e^{-p\, \Phi}
 \label{eq:integral_forms}
\end{equation}
where \teq{A, B} and \teq{C} are functions of \teq{p} and \teq{\Phi(z)}. The \teq{B} terms
are zero for the \teq{{\cal F}_{\Delta}} and \teq{{\cal I}_1} integrals.  This method 
is computationally faster than a full numerical evaluation of the integrals, which 
slows down when tailoring the algorithm to achieve suitable precision near \teq{z\approx 1}.
Using the notation \teq{q=(1-\Phi^2)/\Phi^2}, we start with \teq{{\cal I}_1},
and manipulate it as follows:
\begin{equation}
   {\cal I}_{1}(z,p) \; \equiv\; \Phi^2 e^{-p \Phi} \int_0^{1} e^{p \Phi x} \sqrt{ x \left( x + q\right) } \, dx 
   \; =\;  e^{-p \Phi}  \sum_{n=0}^\infty \dover{( p \Phi )^n}{n!} \, i_n\quad , 
 \label{eq:calI1_series}
\end{equation}
for series coefficients
\begin{equation}
   i_n \; =\; \Phi^2 \int_0^{1} x^{n+1/2} \sqrt{x+q}\,  dx
 \label{eq:i_n_def}
\end{equation}
that can be expressed as special cases of the standard hypergeometric 
function using identity 3.197.8 of Gradshteyn \& Ryzhik (1980).
For integer \teq{n}, the \teq{i_n} are analytically tractable in terms of elementary functions,
and the result can be expressed as a recurrence relation that is readily obtained by
an integration by parts technique:
\begin{equation}
   i_n \; =\; \dover{1}{2+n} \left[  \dover{1}{\Phi} - \dover{1-\Phi^2}{\Phi^2}\, \left(n+ \dover{1}{2} \right) i_{n-1} \right]
   \;\; ,\quad
   i_0 \; =\; \dover{1 +\Phi ^2}{4\Phi}
      - \dover{(1-\Phi^2)^2}{8\Phi^2} \log_e \left[\dover{1+\Phi }{1-\Phi }\right]\;\;  .
 \label{eq:in_recurr}
\end{equation}
In this way, retaining up to the \teq{n=3} term of the series in Eq.~(\ref{eq:calI1_series})
generated an analytic approximation,
which, at \teq{p=1}, is accurate to \teq{< 0.85 \%} for \teq{z\geq 1}.  We also observe that 
\teq{{\cal I}_1} can be expressed as a series of Legendre functions of the second kind,
\teq{Q_{\nu}(z)}, in a manner similar to that in Eq.~(\ref{eq:I1B_BGH05_alt}), but that this
offers no advantage of increased computational facility.

The \teq{{\cal J}_1} integral is treated in a similar fashion, and includes terms with
the \teq{\arctan} form.  The same change of variables yields
\begin{equation}
   {\cal J}_{1}(z,p) \; \equiv\; \Phi^2 e^{-p \Phi} \int_0^{1} e^{p \Phi x} 
   \dover{\sqrt{ x \left( x + q\right) } }{1-\Phi + \Phi x}\, dx 
   \; =\;  e^{-p \Phi}  \sum_{n=0}^\infty \dover{( p \Phi )^n}{n!} \, j_n\quad , 
 \label{eq:calJ1_series}
\end{equation}
where the series coefficients are
\begin{equation}
   j_n \; =\; \Phi^2 \int_0^{1} \dover{x^{n+1/2} \sqrt{x+q}}{1-\Phi + \Phi x}\,  dx\quad . 
 \label{eq:j_n_def}
\end{equation}
A partial fractions manipulation of the denominator in the integrand then 
quickly yields a recurrence relation for the \teq{j_n}:
\begin{equation}
   j_n \; =\; \dover{1}{\Phi} i_{n-1} - \dover{1-\Phi}{\Phi}  j_{n-1}  \;\; ,\quad
   j_0 \; =\; 1 - \dover{2(1-\Phi )}{\sqrt{\Phi}} \, \arctan \sqrt{\Phi} 
      + \dover{(1-\Phi)^2}{2\Phi} \log_e \left[\dover{1+\Phi }{1-\Phi }\right]\;\;  .
\end{equation}
Truncating the series in Eq.~(\ref{eq:calJ1_series}) at three terms is sufficient 
for the purposes of this paper: at \teq{p=1}, this is accurate to \teq{< 0.6 \%} 
for \teq{z\geq 1}.

The \teq{{\cal F}_{\Delta}} integral is handled similarly.  The presence of more 
powers of \teq{\phi} in the numerator of the integrand lengthens the algebra somewhat,
but the manipulations involve partial fractions and integration by parts.  In 
this way, one arrives at
\begin{equation}
   {\cal F}_{\Delta}(z,p) \; \equiv\; \dover{(1+\Phi )^2}{2}\, e^{-p \Phi} \int_0^{1} e^{p \Phi x} 
   \, (1-x)\, \dover{1-\Phi + \Phi x}{\sqrt{ x \left( x + q\right) } }\, dx 
   \; =\;  e^{-p \Phi}  \sum_{n=0}^\infty \dover{( p \Phi )^n}{n!} \, f_n\quad , 
 \label{eq:calFDelta_series}
\end{equation}
where the series coefficients are
\begin{eqnarray}
   f_n &=& \dover{(1+\Phi )^2}{2}
   \int_0^{1} x^{n-1/2}\, (1-x)\, \dover{1-\Phi + \Phi x}{\sqrt{ \left( x + q\right) } }\,  dx\nonumber\\[-5.5pt]
 \label{eq:f_n_def}\\[-5.5pt]
   &=& \dover{ 3 + 4 n (2+n) - (1+2 n) \Phi (1-\Phi ) - 3 \Phi ^3 }{2 (2 n+1) \, \Phi \, (1-\Phi )} i_n 
        - \dover{2 n +1- \Phi }{(2 n+1)\,  (1-\Phi ) }\quad ,\nonumber
 \end{eqnarray}
thereby making use of the \teq{i_n} recurrence relation.  Accordingly, the \teq{{\cal F}_{\Delta}}
integral exhibits the same elementary function structure as the \teq{{\cal I}_1} integral.
Truncating the series at \teq{n=3} generates, at \teq{p=1}, an accuracy of  
\teq{< 0.25 \%} for \teq{z\geq 1}.

Finally, the \teq{{\cal G}_{\Delta}} integral can be expressed as 
\begin{eqnarray}
   {\cal G}_{\Delta}(z,p) &\equiv & \dover{\Phi (1-\Phi )^2}{(1+\Phi )^2}\, e^{-p \Phi} \int_0^{1} e^{p \Phi x} 
   \, \dover{(1-x)^2 (1+\Phi^2 + 2 (1-x) \Phi + 2(1-x)x \Phi^3 )}{(1-\Phi + \Phi x)\, \sqrt{ x \left( x + q\right) } }\, dx\nonumber\\[-5.5pt]
 \label{eq:calGDelta_series}\\[-5.5pt]
   &=&  e^{-p \Phi}  \sum_{n=0}^\infty \dover{( p \Phi )^n}{n!} \, g_n\quad , \nonumber
\end{eqnarray}
with 
\begin{equation}
   g_n \; =\; \dover{\Phi (1-\Phi )^2}{(1+\Phi )^2} \int_0^{1} x^{n-1/2}
   \, \dover{(1-x)^2 (1+\Phi^2 + 2 (1-x) \Phi + 2(1-x)x \Phi^3 )}{(1-\Phi + \Phi x)\, \sqrt{ \left( x + q\right) } }\, dx\quad .
 \label{eq:gn_def}
\end{equation}
If we define
\begin{equation}
   l_n \; =\; \dover{\Phi (1-\Phi )^2}{(1+\Phi )^2} \int_0^{1} 
   \, \dover{x^{n-1/2}}{(1-\Phi + \Phi x)\, \sqrt{ \left( x + q\right) } }\, dx\quad ,
 \label{eq:ln_def}
\end{equation}
then
\begin{equation}
   g_n \; =\; (1+\Phi)^2 l_n + 2(\Phi^3-\Phi^2-3\Phi -1) l_{n+1}  
   + (1+6\Phi + \Phi^2 - 6\Phi^3) l_{n+2} + 2 \Phi ( 3\Phi^2 -1) l_{n+3}  - 2 \Phi^3 l_{n+4}
 \label{eq:gn_ident}
\end{equation}
with a recurrence relation for the \teq{l_n} in terms of the \teq{j_n} and the \teq{i_n}:
\begin{equation}
   l_n \; =\;  \dover{\Phi \, (1-\Phi )}{(1+\Phi )^2}\, \left\{   
  j_{n-1} - \dover{2(n+1)\,\Phi}{1-\Phi^2}\, i_{n-1} + \dover{2\Phi^2}{1-\Phi^2} \right\} \;\; ,\quad
   l_0 \; =\; \dover{2(1-\Phi )\, \Phi^{3/2} }{(1+\Phi )^2 } \, \arctan \sqrt{\Phi} \;\;  .
\end{equation}
The result is, for \teq{p\approx 1}, that retaining terms to 
second order in \teq{p} gives an approximation for 
\teq{{\cal G}_\Delta} that is accurate to \teq{< 0.1 \%} 
for all \teq{z\geq 1}.

\end{document}